\documentclass[useAMS,usenatbib]{mn2e}  
\usepackage{epsfig}  
\usepackage{times}  
  
\newcommand{\placeTabone}{
\begin{table*}
\caption{Parameters of the sample galaxies. The columns show the following: 
(2) morphological classification from RC3, except for 
    ESO-LV~5340200 (NASA/IPAC Extragalactic Database); 
(3) heliocentric systemic velocity of the galaxy derived at centre of 
    symmetry of the rotation curve of the gas (this paper). The 
    typical error on systemic velocity is $\Delta V_\odot = 10$ \kms ; 
(4) distance obtained as $V_0/H_0$ with $H_0 = 75$ km s$^{-1}$ 
    Mpc$^{-1}$ and $V_0$ the systemic velocity derived from $V_\odot$ 
    corrected to the CMB reference frame assuming the dipole direction 
    by \citet{Fixsen1996}; 
(5) major-axis position angle from ESO-LV; 
(6) inclination derived as $\cos^2 i = (q^2-q^2_0 )/(1-q^2_0)$. The 
    observed axial ratio $q$ was taken from ESO-LV, except for 
    ESO-LV~4460170 \citep{Palunas2000} and ESO-LV~2060140 
    \citep{McGaugh2001}. The intrinsic flattening $q_0 = 0.11$ was 
    assumed following \citet{Guthrie1992}; 
(7) radius of the 25 $B-$mag arcsec$^{-2}$ isophote derived as $D_{25}/2$  
    with $D_{25}$ from ESO-LV; 
(8) total observed red magnitude from ESO-LV catalog; 
(9) central velocity dispersion of the stellar component (this paper);  
(10) galaxy circular velocity (this paper);  
(11) radial extension of the ionized-gas rotation curve in units of 
    $R_{25}$ (this paper);  
(12) radial extension of the stellar rotation curve in units of 
    $R_{25}$ (this paper).} 
\begin{center} 
\begin{small} 
\begin{tabular}{ll rrr ccc rr cc} 
\hline 
\multicolumn{1}{c}{Galaxy} & 
\multicolumn{1}{c}{Type} & 
\multicolumn{1}{c}{$V_\odot$} & 
\multicolumn{1}{c}{$D$} & 
\multicolumn{1}{c}{PA} & 
\multicolumn{1}{c}{$i$} & 
\multicolumn{1}{c}{$R_{25}$} & 
\multicolumn{1}{c}{$M_R$} & 
\multicolumn{1}{c}{$\sigma_{\rm c}$} & 
\multicolumn{1}{c}{$V_{\rm c}$} & 
\multicolumn{1}{c}{$R_{\rm g}/R_{25}$} & 
\multicolumn{1}{c}{$R_\star/R_{25}$} \\ 
\multicolumn{1}{c}{} & 
\multicolumn{1}{c}{} & 
\multicolumn{1}{c}{(\kms)} & 
\multicolumn{1}{c}{(Mpc)} & 
\multicolumn{1}{c}{($^\circ$)} & 
\multicolumn{1}{c}{($^\circ$)} & 
\multicolumn{1}{c}{(arcsec)} & 
\multicolumn{1}{c}{(mag)} & 
\multicolumn{1}{c}{(\kms)} & 
\multicolumn{1}{c}{(\kms)} & 
\multicolumn{1}{c}{} & 
\multicolumn{1}{c}{} \\ 
\multicolumn{1}{c}{(1)} & 
\multicolumn{1}{c}{(2)} & 
\multicolumn{1}{c}{(3)} & 
\multicolumn{1}{c}{(4)} & 
\multicolumn{1}{c}{(5)} & 
\multicolumn{1}{c}{(6)} & 
\multicolumn{1}{c}{(7)} & 
\multicolumn{1}{c}{(8)} & 
\multicolumn{1}{c}{(9)} & 
\multicolumn{1}{c}{(10)} & 
\multicolumn{1}{c}{(11)} & 
\multicolumn{1}{c}{(12)} \\ 
\hline 
ESO-LV~1860550 & Sab(r)?    & 4640 &  60.1 & 105 & 63 & 40 & 13.39 &  $91.7\pm2.0$ &    $235\pm11$ & 1.0 & 0.6\\  
ESO-LV~2060140 & SABc(s)    & 4648 &  60.5 & 174 & 39 & 35 & 13.74 &  $54.3\pm2.0$ & $141.0\pm4.5$ & 1.4 & 1.4\\  
ESO-LV~2340130 & Sbc        & 4703 &  60.9 & 172 & 69 & 42 & 13.43 &  $64.1\pm2.0$ &    $194\pm19$ & 1.2 & 0.5\\  
ESO-LV~4000370 & SBcd(s) pec& 3016 &  37.5 &  68 & 50 & 46 & 13.64 &  $42.0\pm2.5$ & $128.7\pm1.8$ & 1.0 & 1.5\\  
ESO-LV~4880490 & SBdm(s)    & 1790 &  25.0 & 136 & 67 & 46 & 14.38 &  $48.2\pm2.5$ & $102.7\pm8.2$ & 1.2 & 0.9\\  
ESO-LV~5340200 & Sa:        &17320 & 226.7 & 167 & 46 & 19 & 14.97 & $153.9\pm7.1$ &    $297\pm11$ & 1.4 & 0.5\\  
\hline 
\end{tabular} 
\end{small} 
\label{tab:sample_lsb} 
\end{center} 
\end{table*} 
}

\newcommand{\placeTabtwo}{
\begin{table}
\begin{tiny}
\caption{Log of spectroscopic observations of the sample galaxies. The 
columns show the following: 
(2) observing run; 
(3) slit position: MJ = major axis, MN = minor axis; 
(4) slit position angle; 
(5) number and exposure time of the single exposures; 
(6) total exposure time.}  
\begin{center} 
\begin{tabular}{lc lr lc} 
\hline 
\multicolumn{1}{c}{Galaxy} & 
\multicolumn{1}{c}{Run} & 
\multicolumn{1}{c}{Position} & 
\multicolumn{1}{c}{PA} & 
\multicolumn{1}{c}{Single Exp. Time} & 
\multicolumn{1}{c}{Tot. Exp. Time} \\ 
\multicolumn{1}{c}{} & 
\multicolumn{1}{c}{} & 
\multicolumn{1}{c}{} & 
\multicolumn{1}{c}{($^\circ$)} & 
\multicolumn{1}{c}{(s)} & 
\multicolumn{1}{c}{(h)} \\ 
\multicolumn{1}{c}{(1)} & 
\multicolumn{1}{c}{(2)} & 
\multicolumn{1}{c}{(3)} & 
\multicolumn{1}{c}{(4)} & 
\multicolumn{1}{c}{(5)} & 
\multicolumn{1}{c}{(6)} \\ 
\hline 
ESO-LV~1860550 & 2 & MJ$-30^\circ$ &  75  & $3\times2750$             & 2.3 \\    
               & 2 & MN$-30^\circ$ & 165  & $2\times2750$             & 1.5 \\    
ESO-LV~2060140 & 3 & MJ            & 175  & $3\times2620+1\times2600$ & 2.9 \\    
               & 3 & MN            &  85  & $1\times2600$             & 0.7 \\    
ESO-LV~2340130 & 2 & MJ$+16^\circ$ &   8  & $3\times2750$             & 2.3 \\    
               & 2 & MN$+16^\circ$ &  98  & $2\times2750$             & 1.5 \\    
ESO-LV~4000370 & 2 & MJ$+44^\circ$ & 112  & $3\times2750$             & 2.3 \\    
               & 2 & MN$+44^\circ$ &  22  & $2\times2750$             & 1.5 \\    
ESO-LV~4880490 & 3 & MJ            &  46  & $3\times2750$             & 2.3 \\    
               & 3 & MN            & 136  & $3\times2750$             & 2.3 \\    
ESO-LV~5340200 & 2 & MJ            & 167  & $3\times3050$             & 2.5 \\  
               & 2 & MN            &  77  & $1\times2750$             & 0.8 \\  
               & 2 & MJ$+26^\circ$ &  13  & $1\times2400$             & 0.7 \\  
               & 2 & MN$+26^\circ$ & 103  & $1\times2750$             & 0.8 \\  
\hline  
\end{tabular}   
\end{center}  
\end{tiny}
\label{tab:speclog_lsb}  
\end{table}  
}

\newcommand{\placeTabthree}{
\begin{table}
\caption{Log of photometric observations of the sample galaxies. 
The columns show the following: 
(2) observing run; 
(3) filter; 
(4) number and exposure time of the single exposures; 
(5) seeing FWHM of the combined image.} 
\begin{center} 
\begin{tabular}{lc lr c} 
\hline 
\multicolumn{1}{c}{Galaxy} & 
\multicolumn{1}{c}{Run} & 
\multicolumn{1}{c}{Filter} & 
\multicolumn{1}{c}{Single Exp. Time} & 
\multicolumn{1}{c}{FWHM} \\ 
\multicolumn{1}{c}{} & 
\multicolumn{1}{c}{} & 
\multicolumn{1}{c}{} & 
\multicolumn{1}{c}{(s)} & 
\multicolumn{1}{c}{(arcsec)} \\ 
\multicolumn{1}{c}{(1)} & 
\multicolumn{1}{c}{(2)} & 
\multicolumn{1}{c}{(3)} & 
\multicolumn{1}{c}{(4)} & 
\multicolumn{1}{c}{(5)} \\ 
\hline                        
ESO-LV~1860550 & 2 & Gunn $z$ &  $3\times50$ & 0.9 \\   
ESO-LV~2060140 & 3 & Gunn $z$ & $6\times100$ & 0.7 \\   
ESO-LV~2340130 & 2 & Gunn $z$ &  $5\times50$ & 1.0 \\   
ESO-LV~4000370 & 2 & Gunn $z$ &  $3\times50$ & 1.4 \\ 
ESO-LV~4880490 & 3 & Gunn $z$ & $9\times100$ & 0.8 \\ 
ESO-LV~5340200 & 2 & Gunn $z$ &  $7\times50$ & 1.1 \\ 
\hline 
\end{tabular} 
\end{center} 
\label{tab:imalog_lsb} 
\end{table}  
}

\newcommand{\placeTabfour}{
\begin{table} 
\caption{Log of spectroscopic observations of the kinematical template stars. 
The columns show the following: 
(2) spectral type from SIMBAD; 
(3) blue apparent magnitude from SIMBAD; 
(4) observing run; 
(5) exposure time; 
(6) dispersion of the auto-correlation function.}  
\begin{center} 
\begin{tabular}{lc rc rc} 
\hline 
\multicolumn{1}{c}{Star} & 
\multicolumn{1}{c}{Sp. Type} & 
\multicolumn{1}{c}{$m_B$} & 
\multicolumn{1}{c}{Run} & 
\multicolumn{1}{c}{Exp. Time} & 
\multicolumn{1}{c}{$\sigma$} \\ 
\multicolumn{1}{c}{} & 
\multicolumn{1}{c}{} & 
\multicolumn{1}{c}{(mag)} & 
\multicolumn{1}{c}{} & 
\multicolumn{1}{c}{(s)} & 
\multicolumn{1}{c}{(\kms)} \\ 
\multicolumn{1}{c}{(1)} & 
\multicolumn{1}{c}{(2)} & 
\multicolumn{1}{c}{(3)} & 
\multicolumn{1}{c}{(4)} & 
\multicolumn{1}{c}{(5)} & 
\multicolumn{1}{c}{(6)} \\ 
\hline                         
SAO  99192 & G0 &  9.84 & 3 &  7 & 43 \\   
SAO 119387 & K0 & 10.64 & 1 &  8 & 65 \\   
SAO 119458 & G0 & 10.49 & 3 & 12 & 50 \\  
SAO 123779 & K0 &  8.08 & 1 &  1 & 51 \\    
SAO 137138 & G5 & 10.47 & 3 & 12 & 47 \\  
SAO 137330 & G8 & 10.78 & 1 &  5 & 49 \\   
\hline  
\end{tabular}  
\end{center} 
\label{tab:templog}  
\end{table}  
}

\newcommand{\placeTabfive}{
\begin{table*} 
\begin{tiny}
\caption{Photometric parameters of the bulge and disc in the sample galaxies. 
The columns show the following: 
(2) effective surface brightness of the bulge;  
(3) effective radius of the bulge;  
(4) shape parameter of the bulge;  
(5) axial ratio of the bulge isophotes;  
(6) position angle of the bulge major axis;  
(7) central surface brightness of the disc;  
(8) scale length of the disc;  
(9) axial ratio of the disc isophotes;  
(10) position angle of the disc major axis;  
(11) bulge-to-total luminosity ratio. 
} 
\begin{tabular}{lcrccrcrcrc} 
\hline 
\multicolumn{1}{c}{Galaxy} & 
\multicolumn{1}{c}{$\mu_{\rm e}$} & 
\multicolumn{1}{c}{$r_{\rm e}$} & 
\multicolumn{1}{c}{$n$} & 
\multicolumn{1}{c}{$q_{\rm b}$} & 
\multicolumn{1}{c}{PA$_{\rm b}$} & 
\multicolumn{1}{c}{$\mu_0$} & 
\multicolumn{1}{c}{$h$} & 
\multicolumn{1}{c}{$q_{\rm d}$} & 
\multicolumn{1}{c}{PA$_{\rm d}$} & 
\multicolumn{1}{c}{$B/T$} \\ 
\multicolumn{1}{c}{ } & 
\multicolumn{1}{c}{(mag/arcsec$^2$)} & 
\multicolumn{1}{c}{(arcsec)} & 
\multicolumn{1}{c}{} & 
\multicolumn{1}{c}{} & 
\multicolumn{1}{c}{($^{\circ}$)} & 
\multicolumn{1}{c}{(mag/arcsec$^2$)} & 
\multicolumn{1}{c}{(arcsec)} & 
\multicolumn{1}{c}{} & 
\multicolumn{1}{c}{($^{\circ}$)} & 
\multicolumn{1}{c}{} \\ 
\multicolumn{1}{c}{(1)} & 
\multicolumn{1}{c}{(2)} & 
\multicolumn{1}{c}{(3)} & 
\multicolumn{1}{c}{(4)} & 
\multicolumn{1}{c}{(5)} & 
\multicolumn{1}{c}{(6)} & 
\multicolumn{1}{c}{(7)} & 
\multicolumn{1}{c}{(8)} & 
\multicolumn{1}{c}{(9)} & 
\multicolumn{1}{c}{(10)} & 
\multicolumn{1}{c}{(11)} \\ 
\hline 
ESO-LV~1860550 & $21.16\pm0.04$ & $ 9.7\pm0.3$ & $2.16\pm0.05$ & $0.51\pm0.01$ & $126\pm3$ & $20.72\pm0.03$ & $12.5\pm0.3$ & $0.25\pm0.01$ &  $111\pm3$ & 0.68\\       
ESO-LV~2060140 & $22.27\pm0.04$ & $ 3.1\pm0.1$ & $1.64\pm0.04$ & $0.76\pm0.02$ & $ 18\pm1$ & $21.67\pm0.03$ & $17.9\pm0.5$ & $0.80\pm0.02$ &  $  4\pm1$ & 0.04\\       
ESO-LV~2340130 & $20.62\pm0.05$ & $ 7.5\pm0.3$ & $1.40\pm0.04$ & $0.54\pm0.01$ & $ 72\pm2$ & $21.00\pm0.04$ & $12.7\pm0.4$ & $0.30\pm0.01$ &  $ 77\pm2$ & 0.66\\       
ESO-LV~4000370 & $22.36\pm0.05$ & $ 4.2\pm0.2$ & $0.59\pm0.01$ & $0.58\pm0.01$ & $107\pm3$ & $21.63\pm0.04$ & $22.5\pm0.7$ & $0.54\pm0.01$ &  $140\pm4$ & 0.03\\ 
ESO-LV~4880490 & $23.48\pm0.04$ & $18.2\pm0.6$ & $1.43\pm0.03$ & $0.32\pm0.01$ & $ 83\pm2$ & $23.33\pm0.03$ & $36.6\pm0.8$ & $0.56\pm0.01$ &  $ 88\pm2$ & 0.21\\ 
ESO-LV~5340200 & $20.44\pm0.06$ & $ 3.0\pm0.6$ & $1.96\pm0.07$ & $0.62\pm0.01$ & $ 10\pm1$ & $22.94\pm0.05$ & $15.2\pm0.7$ & $0.50\pm0.02$ &  $178\pm6$ & 0.56\\   
\hline 
\end{tabular} 
\label{tab:parameters_lsb}  
\end{tiny}
\end{table*} 
}

\newcommand{\placeTabsix}{
\begin{table} 
\caption{Central surface brightness of the LSB discs. 
The columns show the following: 
(2) effective radius of the galaxy from ESO-LV;
(3) mean color outside $R_{\rm e}$ from ESO-LV;
(4) central blue surface brightness of the disc corrected for inclination.
} 
\begin{tabular}{lccc} 
\hline 
\multicolumn{1}{c}{Galaxy} & 
\multicolumn{1}{c}{$R_{\rm e}$} & 
\multicolumn{1}{c}{$B-R$} & 
\multicolumn{1}{c}{$\mu_{0,B}^0$} \\ 
\multicolumn{1}{c}{} & 
\multicolumn{1}{c}{(arcsec)} & 
\multicolumn{1}{c}{(mag)} & 
\multicolumn{1}{c}{(mag/arcsec$^2$)} \\
\multicolumn{1}{c}{(1)} & 
\multicolumn{1}{c}{(2)} & 
\multicolumn{1}{c}{(3)} & 
\multicolumn{1}{c}{(4)} \\ 
\hline 
ESO-LV~1860550 & 15 & 1.23 & 22.81 \\       
ESO-LV~2060140 & 26 & 1.14 & 23.08 \\       
ESO-LV~2340130 & 18 & 1.10 & 23.22 \\       
ESO-LV~4000370 & 28 & 0.97 & 23.08 \\ 
ESO-LV~4880490 & 29 & 0.53 & 24.88 \\ 
ESO-LV~5340200 & 11 & 0.88 & 24.22 \\   
\hline 
\end{tabular} 
\label{tab:discs_lsb}  
\end{table} 
}

\newcommand{\placeTabseven}{
\begin{table} 
\caption{Stellar kinematics of the sample galaxies. The complete table is published in the electronic issue.} 
\begin{tiny} 
\begin{tabular}{rrrrrrrrrrr}
\multicolumn{1}{c}{Name} &
\multicolumn{1}{c}{PA} &
\multicolumn{1}{c}{r} &
\multicolumn{1}{c}{v} &
\multicolumn{1}{c}{$\Delta$v} &
\multicolumn{1}{c}{$\sigma$} &
\multicolumn{1}{c}{$\Delta\sigma$} &
\multicolumn{1}{c}{h$_3$} &
\multicolumn{1}{c}{$\Delta$h$_3$} &
\multicolumn{1}{c}{h$_4$} &
\multicolumn{1}{c}{$\Delta$h$_4$}\\
\multicolumn{1}{c}{ESO-LV } &
\multicolumn{1}{c}{$^\circ$} &
\multicolumn{1}{c}{``} &
\multicolumn{1}{c}{km/s} &
\multicolumn{1}{c}{km/s} &
\multicolumn{1}{c}{km/s} &
\multicolumn{1}{c}{km/s} &
\multicolumn{1}{c}{} &
\multicolumn{1}{c}{} &
\multicolumn{1}{c}{} &
\multicolumn{1}{c}{}\\
\multicolumn{1}{c}{(1)} &
\multicolumn{1}{c}{(2)} &
\multicolumn{1}{c}{(3)} &
\multicolumn{1}{c}{(4)} &
\multicolumn{1}{c}{(5)} &
\multicolumn{1}{c}{(6)} &
\multicolumn{1}{c}{(7)} &
\multicolumn{1}{c}{(8)} &
\multicolumn{1}{c}{(9)} &
\multicolumn{1}{c}{(10)} &
\multicolumn{1}{c}{(11)}\\
\hline
1860550 & 75 & -10.02 &  102.5 & 6.0 &  54.8 & 2.9 & -0.004 & 0.064 & -0.075 & 0.133\\
1860550 & 75 & -16.75 &  106.7 & 7.4 &  63.8 & 4.3 & -0.033 & 0.051 & -0.080 & 0.075\\
1860550 & 75 & -25.68 &  129.4 & 8.4 &  53.6 & 9.4 &      0 & 0     &   0          0  \\
\multicolumn{1}{c}{...} &
\multicolumn{1}{c}{...} &
\multicolumn{1}{c}{...} &
\multicolumn{1}{c}{...} &
\multicolumn{1}{c}{...} &
\multicolumn{1}{c}{...} &
\multicolumn{1}{c}{...} &
\multicolumn{1}{c}{...} &
\multicolumn{1}{c}{...} &
\multicolumn{1}{c}{...} &
\multicolumn{1}{c}{...}\\
\end{tabular}
\end{tiny} 
NOTE: columns are (1) ESO-LV galaxy number; (2) Position angle on the
sky; (3) distance from the galaxy center; (4-5) velocity and 1-$\sigma$
uncertain; (6-7) velocity dispersion and 1-$\sigma$ uncertain;
(8-9) h$_3$ and 1-$\sigma$ uncertain; (10-11) h$_4$ and 1-$\sigma$
uncertain.

\label{tab:starkin_lsb} 
\end{table} 
}

\newcommand{\placeTabeight}{
\begin{table} 
\caption{Ionized-gas kinematics of the sample galaxies. The complete table is published in the electronic issue.} 
\begin{tabular}{rrrrrrrr}
\multicolumn{1}{c}{Name} &
\multicolumn{1}{c}{PA} &
\multicolumn{1}{c}{r} &
\multicolumn{1}{c}{v} &
\multicolumn{1}{c}{$\Delta$v} &
\multicolumn{1}{c}{$\sigma$} &
\multicolumn{1}{c}{$\Delta\sigma_-$} &
\multicolumn{1}{c}{$\Delta\sigma_+$}\\

\multicolumn{1}{c}{ESO-LV} &
\multicolumn{1}{c}{$^\circ$} &
\multicolumn{1}{c}{``} &
\multicolumn{1}{c}{km/s} &
\multicolumn{1}{c}{km/s} &
\multicolumn{1}{c}{km/s} &
\multicolumn{1}{c}{km/s} &
\multicolumn{1}{c}{km/s}\\
\multicolumn{1}{c}{(1)} &
\multicolumn{1}{c}{(2)} &
\multicolumn{1}{c}{(3)} &
\multicolumn{1}{c}{(4)} &
\multicolumn{1}{c}{(5)} &
\multicolumn{1}{c}{(6)} &
\multicolumn{1}{c}{(7)} &
\multicolumn{1}{c}{(8)} \\

\hline
1860550 & 75 & -0.75 & 23.8 & 5.5 & 73.7 & 4.9 & 4.8\\
1860550 & 75 &  0.25 &  7.4 & 5.7 & 75.1 & 5.5 & 5.4\\
1860550 & 75 &  1.25 &-29.4 & 6.2 & 61.5 & 7.0 & 6.8\\
\multicolumn{1}{c}{...} &
\multicolumn{1}{c}{...} &
\multicolumn{1}{c}{...} &
\multicolumn{1}{c}{...} &
\multicolumn{1}{c}{...} &
\multicolumn{1}{c}{...} &
\multicolumn{1}{c}{...} &
\multicolumn{1}{c}{...}\\
\end{tabular}
NOTE: columns are (1) ESO-LV galaxy number; (2) Position angle on the
sky; (3) distance from the galaxy center; (4-5) velocity and 1-$\sigma$
uncertain; (6-8) velocity dispersion and 1-$\sigma$ lower and upper
uncertain following the scheme:
$\sigma_{-\Delta\sigma_-}^{+\Delta\sigma_+}$

\label{tab:gaskin_lsb} 
\end{table} 
}

\newcommand{\placeTabAone}{
\begin{table*} 
\caption{As in Tab.~\ref{tab:sample_lsb} but for the HSB galaxies. The
inclination of ESO-LV~4460170 is from \citet{Palunas2000}.} 
\begin{center} 
\begin{small} 
\begin{tabular}{ll rrr ccc rr cc} 
\hline 
\multicolumn{1}{c}{Galaxy} & 
\multicolumn{1}{c}{Type} & 
\multicolumn{1}{c}{$V_\odot$} & 
\multicolumn{1}{c}{$D$} & 
\multicolumn{1}{c}{PA} & 
\multicolumn{1}{c}{$i$} & 
\multicolumn{1}{c}{$R_{25}$} & 
\multicolumn{1}{c}{$M_R$} & 
\multicolumn{1}{c}{$\sigma_{\rm c}$} & 
\multicolumn{1}{c}{$V_{\rm c}$} & 
\multicolumn{1}{c}{$R_{\rm g}/R_{25}$} & 
\multicolumn{1}{c}{$R_\star/R_{25}$} \\ 
\multicolumn{1}{c}{} & 
\multicolumn{1}{c}{} & 
\multicolumn{1}{c}{(\kms)} & 
\multicolumn{1}{c}{(Mpc)} & 
\multicolumn{1}{c}{($^\circ$)} & 
\multicolumn{1}{c}{($^\circ$)} & 
\multicolumn{1}{c}{(arcsec)} & 
\multicolumn{1}{c}{(mag)} & 
\multicolumn{1}{c}{(\kms)} & 
\multicolumn{1}{c}{(\kms)} & 
\multicolumn{1}{c}{} & 
\multicolumn{1}{c}{} \\ 
\multicolumn{1}{c}{(1)} & 
\multicolumn{1}{c}{(2)} & 
\multicolumn{1}{c}{(3)} & 
\multicolumn{1}{c}{(4)} & 
\multicolumn{1}{c}{(5)} & 
\multicolumn{1}{c}{(6)} & 
\multicolumn{1}{c}{(7)} & 
\multicolumn{1}{c}{(8)} & 
\multicolumn{1}{c}{(9)} & 
\multicolumn{1}{c}{(10)} & 
\multicolumn{1}{c}{(11)} & 
\multicolumn{1}{c}{(12)} \\ 
\hline 
ESO-LV~1890070 & SABbc(rs)  & 2973 &  37.5 &  18 & 49 & 90 & 11.63 &  $91.3\pm2.0$ & $185.9\pm6.9$ & 1.0 & 1.0\\  
ESO-LV~4460170 & (R)SBb(s)  & 4193 &  58.9 & 155 & 54 & 54 & 12.22 & $133.6\pm2.0$ &    $196\pm33$ & 1.2 & 1.3\\  
ESO-LV~4500200 & SBbc(s):   & 2247 &  31.6 & 154 & 30 & 58 & 11.27 & $112.4\pm2.4$ &    $245\pm35$ & 1.0 & 1.0\\ 
ESO-LV~5140100 & SABc(s):   & 2857 &  40.4 &  50 & 36 & 50 & 11.62 &  $60.2\pm4.0$ &    $181\pm15$ & 0.8 & 1.2\\ 
\hline 
\end{tabular} 
\end{small} 
\label{tab:sample_hsb} 
\end{center} 
\end{table*} 
}

\newcommand{\placeTabAtwo}{
\begin{table}
\caption{As in Tab.~\ref{tab:imalog_lsb} but for the HSB galaxies.} 
\begin{center} 
\begin{tabular}{lc lr c} 
\hline 
\multicolumn{1}{c}{Galaxy} & 
\multicolumn{1}{c}{Run} & 
\multicolumn{1}{c}{Filter} & 
\multicolumn{1}{c}{Single Exp. Time} & 
\multicolumn{1}{c}{FWHM} \\ 
\multicolumn{1}{c}{} & 
\multicolumn{1}{c}{} & 
\multicolumn{1}{c}{} & 
\multicolumn{1}{c}{(s)} & 
\multicolumn{1}{c}{(arcsec)} \\ 
\multicolumn{1}{c}{(1)} & 
\multicolumn{1}{c}{(2)} & 
\multicolumn{1}{c}{(3)} & 
\multicolumn{1}{c}{(4)} & 
\multicolumn{1}{c}{(5)} \\ 
\hline                        
ESO-LV~1890070 & 1 & none     &  $4\times20$ & 0.8 \\   
ESO-LV~4460170 & 1 & none     &  $4\times20$ & 1.4 \\ 
ESO-LV~4500200 & 1 & none     &  $2\times20$ & 1.1 \\ 
ESO-LV~5140100 & 1 & none     &  $4\times20$ & 1.1 \\ 
\hline 
\end{tabular} 
\end{center} 
\label{tab:imalog_hsb} 
\end{table}  
}

\newcommand{\placeTabAthree}{
\begin{table}  
\begin{tiny} 
\caption{As in Tab.~\ref{tab:speclog_lsb} but for the HSB galaxies.} 
\begin{center} 
\begin{tabular}{lc lr lc} 
\hline 
\multicolumn{1}{c}{Galaxy} & 
\multicolumn{1}{c}{Run} & 
\multicolumn{1}{c}{Position} & 
\multicolumn{1}{c}{PA} & 
\multicolumn{1}{c}{Single Exp. Time} & 
\multicolumn{1}{c}{Tot. Exp. Time} \\ 
\multicolumn{1}{c}{} & 
\multicolumn{1}{c}{} & 
\multicolumn{1}{c}{} & 
\multicolumn{1}{c}{($^\circ$)} & 
\multicolumn{1}{c}{(s)} & 
\multicolumn{1}{c}{(h)} \\ 
\multicolumn{1}{c}{(1)} & 
\multicolumn{1}{c}{(2)} & 
\multicolumn{1}{c}{(3)} & 
\multicolumn{1}{c}{(4)} & 
\multicolumn{1}{c}{(5)} & 
\multicolumn{1}{c}{(6)} \\ 
\hline 
ESO-LV~1890070 & 1 & MJ            &  18  & $3\times2400$             & 2.0 \\    
               & 1 & MN            & 108  & $2\times2900$             & 1.6 \\    
ESO-LV~4460170 & 1 & MJ            & 155  & $3\times2400$             & 2.0 \\    
               & 1 & MN            &  65  & $2\times3000$             & 1.7 \\  
ESO-LV~4500200 & 1 & MJ            & 154  & $3\times2400$             & 2.0 \\    
               & 1 & MN            &  64  & $3\times2400$             & 2.0 \\    
ESO-LV~5140100 & 1 & MJ            &  50  & $3\times2520+2\times2400$ & 4.8 \\  
               & 1 & MN            & 140  & $1\times2400$             & 0.7 \\  
\hline  
\end{tabular}   
\end{center}  
\label{tab:speclog_hsb}  
\end{tiny} 
\end{table}  
}

\newcommand{\placeTabAfour}{
\begin{table*} 
\begin{tiny}
\caption{As in Tab.~\ref{tab:parameters_lsb} but for the HSB galaxies.} 
\begin{tabular}{lcrccrcrcrc} 
\hline 
\multicolumn{1}{c}{Galaxy} & 
\multicolumn{1}{c}{$\mu_{\rm e}$} & 
\multicolumn{1}{c}{$r_{\rm e}$} & 
\multicolumn{1}{c}{$n$} & 
\multicolumn{1}{c}{$q_{\rm b}$} & 
\multicolumn{1}{c}{PA$_{\rm b}$} & 
\multicolumn{1}{c}{$\mu_0$} & 
\multicolumn{1}{c}{$h$} & 
\multicolumn{1}{c}{$q_{\rm d}$} & 
\multicolumn{1}{c}{PA$_{\rm d}$} & 
\multicolumn{1}{c}{$B/T$} \\ 
\multicolumn{1}{c}{ } & 
\multicolumn{1}{c}{(mag/arcsec$^2$)} & 
\multicolumn{1}{c}{(arcsec)} & 
\multicolumn{1}{c}{} & 
\multicolumn{1}{c}{} & 
\multicolumn{1}{c}{($^{\circ}$)} & 
\multicolumn{1}{c}{(mag/arcsec$^2$)} & 
\multicolumn{1}{c}{(arcsec)} & 
\multicolumn{1}{c}{} & 
\multicolumn{1}{c}{($^{\circ}$)} & 
\multicolumn{1}{c}{} \\ 
\multicolumn{1}{c}{(1)} & 
\multicolumn{1}{c}{(2)} & 
\multicolumn{1}{c}{(3)} & 
\multicolumn{1}{c}{(4)} & 
\multicolumn{1}{c}{(5)} & 
\multicolumn{1}{c}{(6)} & 
\multicolumn{1}{c}{(7)} & 
\multicolumn{1}{c}{(8)} & 
\multicolumn{1}{c}{(9)} & 
\multicolumn{1}{c}{(10)} & 
\multicolumn{1}{c}{(11)} \\ 
\hline 
ESO-LV~1890070 & $19.24\pm0.03$ & $ 3.4\pm0.1$ & $0.99\pm0.02$ & $0.71\pm0.01$ & $178\pm4$ & $20.32\pm0.03$ & $31.1\pm0.8$ & $0.46\pm0.01$ &  $  3\pm1$ & 0.09\\       
ESO-LV~4460170 & $18.97\pm0.03$ & $ 3.0\pm0.1$ & $1.23\pm0.03$ & $0.85\pm0.02$ & $177\pm4$ & $20.11\pm0.03$ & $19.4\pm0.5$ & $0.51\pm0.01$ &  $  1\pm1$ & 0.19\\ 
ESO-LV~4500200 & $18.01\pm0.03$ & $ 3.3\pm0.1$ & $1.79\pm0.04$ & $0.70\pm0.01$ & $179\pm4$ & $19.03\pm0.03$ & $17.3\pm0.4$ & $0.67\pm0.01$ &  $  4\pm1$ & 0.19\\ 
ESO-LV~5140100 & $20.19\pm0.03$ & $ 2.2\pm0.1$ & $0.93\pm0.02$ & $0.91\pm0.02$ & $171\pm4$ & $20.45\pm0.03$ & $27.1\pm0.7$ & $0.78\pm0.02$ &  $175\pm5$ & 0.02\\ 
\hline 
\end{tabular} 
\label{tab:parameters_hsb}  
\end{tiny}
\end{table*} 
}

\newcommand{\placeTabAfive}{
\begin{table}  
\caption{As in Tab.~\ref{tab:discs_lsb} but for the HSB galaxies.} 
\begin{tabular}{lccc} 
\hline 
\multicolumn{1}{c}{Galaxy} & 
\multicolumn{1}{c}{$R_{\rm e}$} & 
\multicolumn{1}{c}{$B-R$} & 
\multicolumn{1}{c}{$\mu_{0,B}^0$} \\ 
\multicolumn{1}{c}{} & 
\multicolumn{1}{c}{(arcsec)} & 
\multicolumn{1}{c}{(mag)} & 
\multicolumn{1}{c}{(mag/arcsec$^2$)} \\
\multicolumn{1}{c}{(1)} & 
\multicolumn{1}{c}{(2)} & 
\multicolumn{1}{c}{(3)} & 
\multicolumn{1}{c}{(4)} \\
\hline 
ESO-LV~1890070 & 48 & 0.89 & 21.67\\       
ESO-LV~4460170 & 24 & 1.24 & 21.92\\ 
ESO-LV~4500200 & 23 & 1.68 & 20.87\\ 
ESO-LV~5140100 & 42 & 1.16 & 21.84\\ 
\hline 
\end{tabular} 
\label{tab:discs_hsb}  
\end{table} 
}

\newcommand{\placeTabAsix}{
\begin{table} 
\caption{Stellar kinematics of the HSB galaxies. The complete table is published in the electronic issue.} 
\begin{tiny} 
\begin{tabular}{rrrrrrrrrrr}
\multicolumn{1}{c}{Name} &
\multicolumn{1}{c}{PA} &
\multicolumn{1}{c}{r} &
\multicolumn{1}{c}{v} &
\multicolumn{1}{c}{$\Delta$v} &
\multicolumn{1}{c}{$\sigma$} &
\multicolumn{1}{c}{$\Delta\sigma$} &
\multicolumn{1}{c}{h$_3$} &
\multicolumn{1}{c}{$\Delta$h$_3$} &
\multicolumn{1}{c}{h$_4$} &
\multicolumn{1}{c}{$\Delta$h$_4$}\\
\multicolumn{1}{c}{ESO-LV } &
\multicolumn{1}{c}{$^\circ$} &
\multicolumn{1}{c}{``} &
\multicolumn{1}{c}{km/s} &
\multicolumn{1}{c}{km/s} &
\multicolumn{1}{c}{km/s} &
\multicolumn{1}{c}{km/s} &
\multicolumn{1}{c}{} &
\multicolumn{1}{c}{} &
\multicolumn{1}{c}{} &
\multicolumn{1}{c}{}\\
\multicolumn{1}{c}{(1)} &
\multicolumn{1}{c}{(2)} &
\multicolumn{1}{c}{(3)} &
\multicolumn{1}{c}{(4)} &
\multicolumn{1}{c}{(5)} &
\multicolumn{1}{c}{(6)} &
\multicolumn{1}{c}{(7)} &
\multicolumn{1}{c}{(8)} &
\multicolumn{1}{c}{(9)} &
\multicolumn{1}{c}{(10)} &
\multicolumn{1}{c}{(11)}\\
\hline
1890070 & 108 & -94.50 & -139.4 & 25.5 & 117.7 & 22.8 &  0.00 & 0.00 &  0.00 & 0.00\\
1890070 & 108 & -70.64 & -131.4 &  8.5 &  63.5 &  7.0 & -0.01 & 0.09 & -0.07 & 0.13\\
1890070 & 108 & -62.55 & -127.8 &  6.4 &  58.4 &  4.6 &  0.00 & 0.07 & -0.06 & 0.12\\
\multicolumn{1}{c}{...} &
\multicolumn{1}{c}{...} &
\multicolumn{1}{c}{...} &
\multicolumn{1}{c}{...} &
\multicolumn{1}{c}{...} &
\multicolumn{1}{c}{...} &
\multicolumn{1}{c}{...} &
\multicolumn{1}{c}{...} &
\multicolumn{1}{c}{...} &
\multicolumn{1}{c}{...} &
\multicolumn{1}{c}{...}\\
\end{tabular}
\end{tiny} 
NOTE: columns are (1) ESO-LV galaxy number; (2) Position angle on the
sky; (3) distance from the galaxy center; (4-5) velocity and 1-$\sigma$
uncertain; (6-7) velocity dispersion and 1-$\sigma$ uncertain;
(8-9) h$_3$ and 1-$\sigma$ uncertain; (10-11) h$_4$ and 1-$\sigma$
uncertain.

\label{tab:starkin_hsb} 
\end{table} 
}

\newcommand{\placeTabAseven}{
\begin{table} 
\caption{Ionized-gas kinematics of the HSB galaxies. The complete table is published in the electronic issue.}
\begin{tabular}{rrrrrrrr}
\multicolumn{1}{c}{Name} &
\multicolumn{1}{c}{PA} &
\multicolumn{1}{c}{r} &
\multicolumn{1}{c}{v} &
\multicolumn{1}{c}{$\Delta$v} &
\multicolumn{1}{c}{$\sigma$} &
\multicolumn{1}{c}{$\Delta\sigma_-$} &
\multicolumn{1}{c}{$\Delta\sigma_+$}\\

\multicolumn{1}{c}{ESO-LV} &
\multicolumn{1}{c}{$^\circ$} &
\multicolumn{1}{c}{``} &
\multicolumn{1}{c}{km/s} &
\multicolumn{1}{c}{km/s} &
\multicolumn{1}{c}{km/s} &
\multicolumn{1}{c}{km/s} &
\multicolumn{1}{c}{km/s}\\
\multicolumn{1}{c}{(1)} &
\multicolumn{1}{c}{(2)} &
\multicolumn{1}{c}{(3)} &
\multicolumn{1}{c}{(4)} &
\multicolumn{1}{c}{(5)} &
\multicolumn{1}{c}{(6)} &
\multicolumn{1}{c}{(7)} &
\multicolumn{1}{c}{(8)} \\

\hline
1890070 &  108 &  -125.30 & -151.6 &   6.2&   13.6 &  13.6  & 92.2\\
1890070 &  108 &  -123.30 & -135.3 &   5.8&   22.7 &   7.0  &  7.0\\
1890070 &  108 &  -121.30 & -139.2 &   5.4&   25.9 &   4.5  &  4.5\\
\multicolumn{1}{c}{...} &
\multicolumn{1}{c}{...} &
\multicolumn{1}{c}{...} &
\multicolumn{1}{c}{...} &
\multicolumn{1}{c}{...} &
\multicolumn{1}{c}{...} &
\multicolumn{1}{c}{...} &
\multicolumn{1}{c}{...}\\
\end{tabular}
NOTE: columns are (1) ESO-LV galaxy number; (2) Position angle on the
sky; (3) distance from the galaxy center; (4-5) velocity and 1-$\sigma$
uncertain; (6-8) velocity dispersion and 1-$\sigma$ lower and upper
uncertain following the scheme:
$\sigma_{-\Delta\sigma_-}^{+\Delta\sigma_+}$

\label{tab:gaskin_hsb} 
\end{table} 
}

\newcommand{\placeFigone}{
\begin{figure}  
\begin{center}  
\centering\epsfig{figure=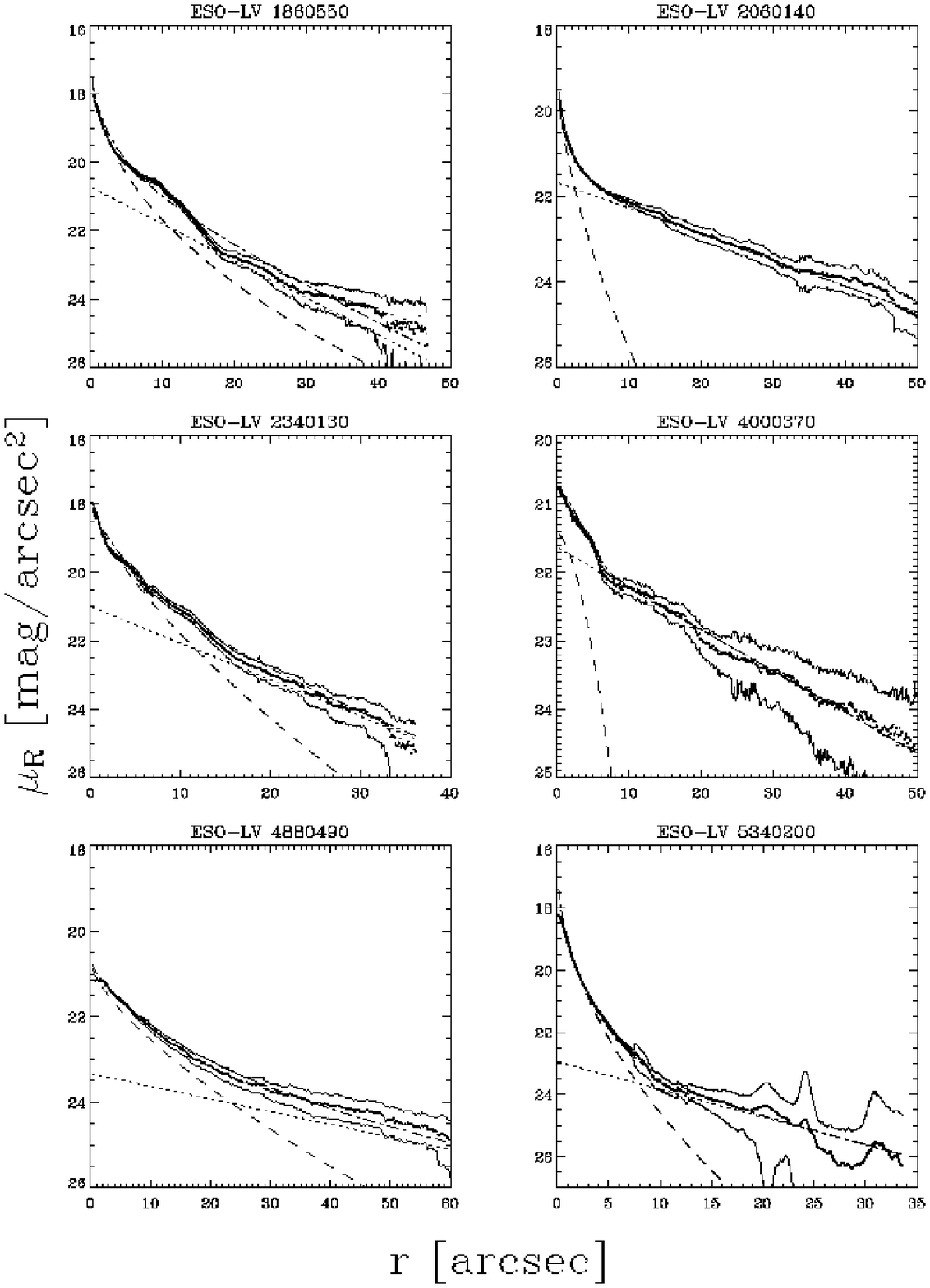, width=9cm}  
\caption{Surface-brightness radial profile and photometric 
  decomposition of the sample galaxies. 
  The radial profiles were obtained by fitting ellipses to the
  isophotes of the image of the galaxy (thick continuous line), model
  bulge (dashed line), model disc (dotted line), and bulge-disc model
  (dot-dashed line). All the profiles are given as a function of the
  semimajor axes of the fitting ellipses. The bulge and disc
  contributions are shown before the convolution with PSF. The surface
  brightness of the bulge-disc model takes into account for PSF.  The
  two thin continuous lines show the $\pm1\sigma$ confidence region
  from the ellipse fit of the galaxy image.}
\label{fig:decomposition_lsb}  
\end{center}  
\end{figure}  
}

\newcommand{\placeFigtwo}{
\begin{figure}  
\begin{center}  
 \begin{minipage}[b]{.999\linewidth}  
  \centering\epsfig{figure=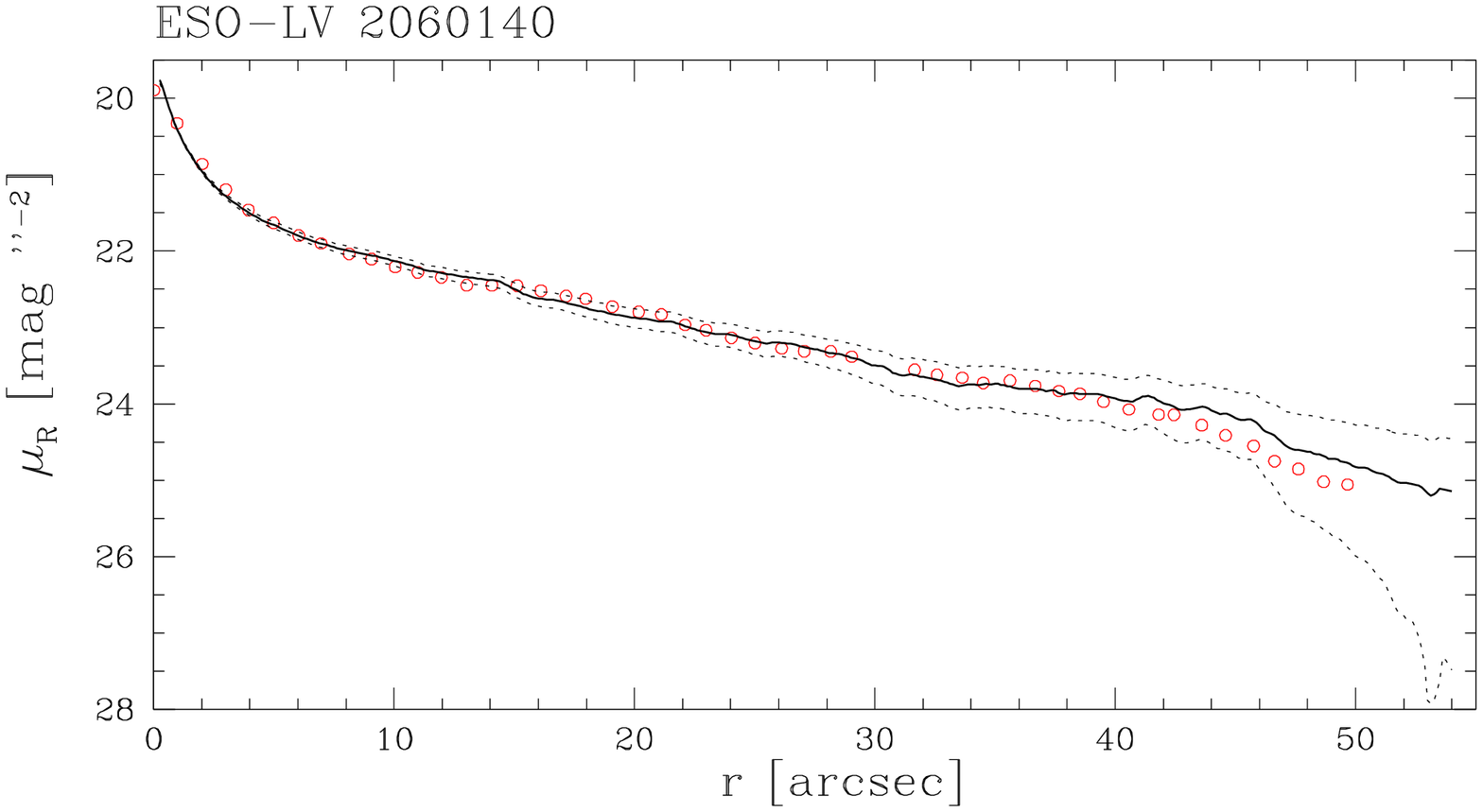,width=8.6cm, angle=0}  
 \end{minipage}  
\caption{Comparison between the elliptically-averaged radial profiles 
of surface brightness of ESO-LV~2060140 measured in this work 
(continuous line, with the $\pm1\sigma$ confidence region indicated by 
the dotted lines) and by \citet[circles]{Beijersbergen1999}.} 
\label{fig:imacomparison}  
\end{center}  
\end{figure}  
}

\newcommand{\placeFigthree}{
\begin{figure}  
\begin{center}  
\begin{minipage}[b]{.99\linewidth}  
\centering\epsfig{figure=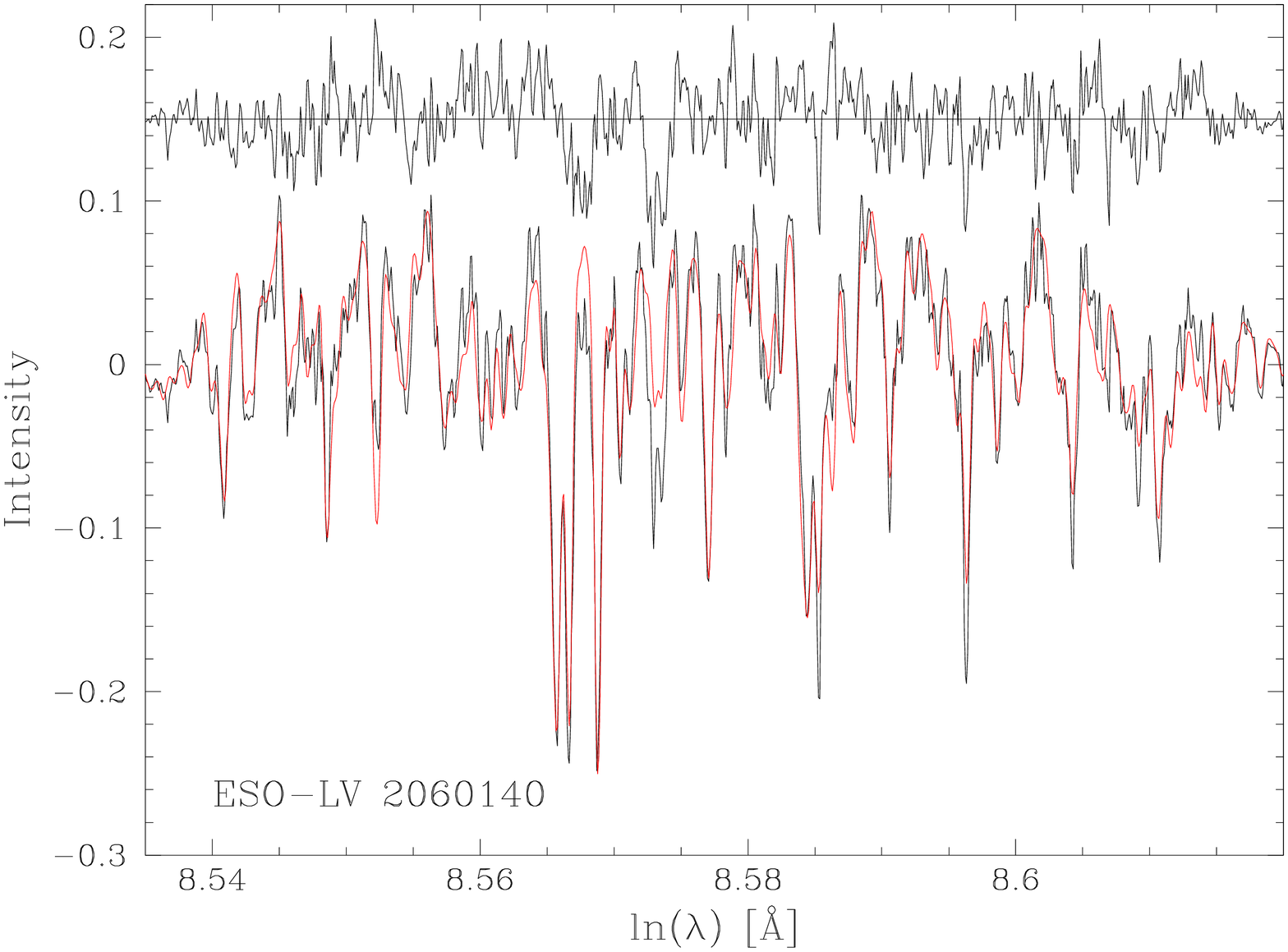, width=6cm, angle=270}  
\end{minipage}   
\begin{minipage}[b]{.99\linewidth}  
\centering\epsfig{figure=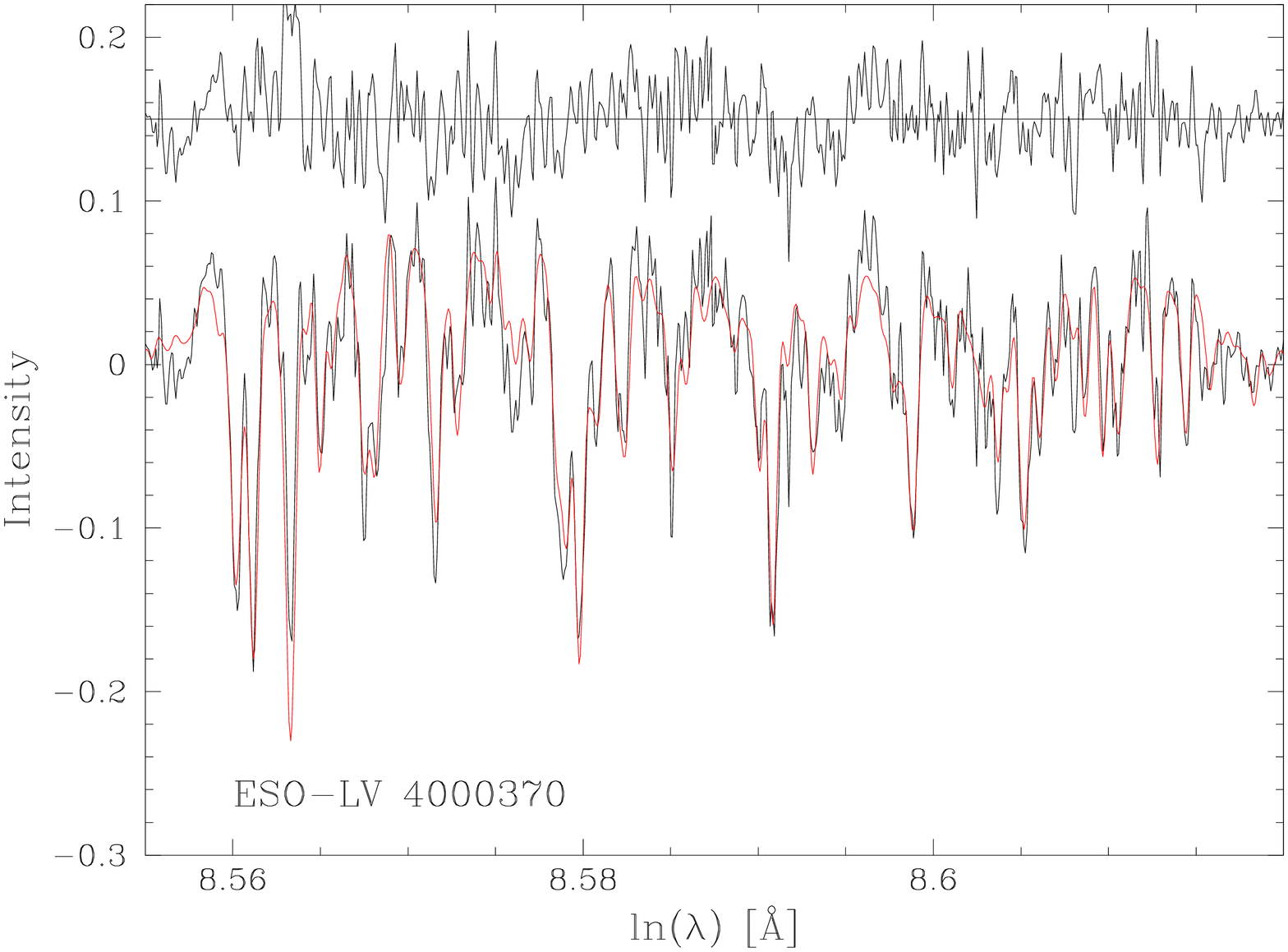, width=6cm, angle=270}  
\end{minipage}  
\begin{minipage}[b]{.99\linewidth}  
\centering\epsfig{figure=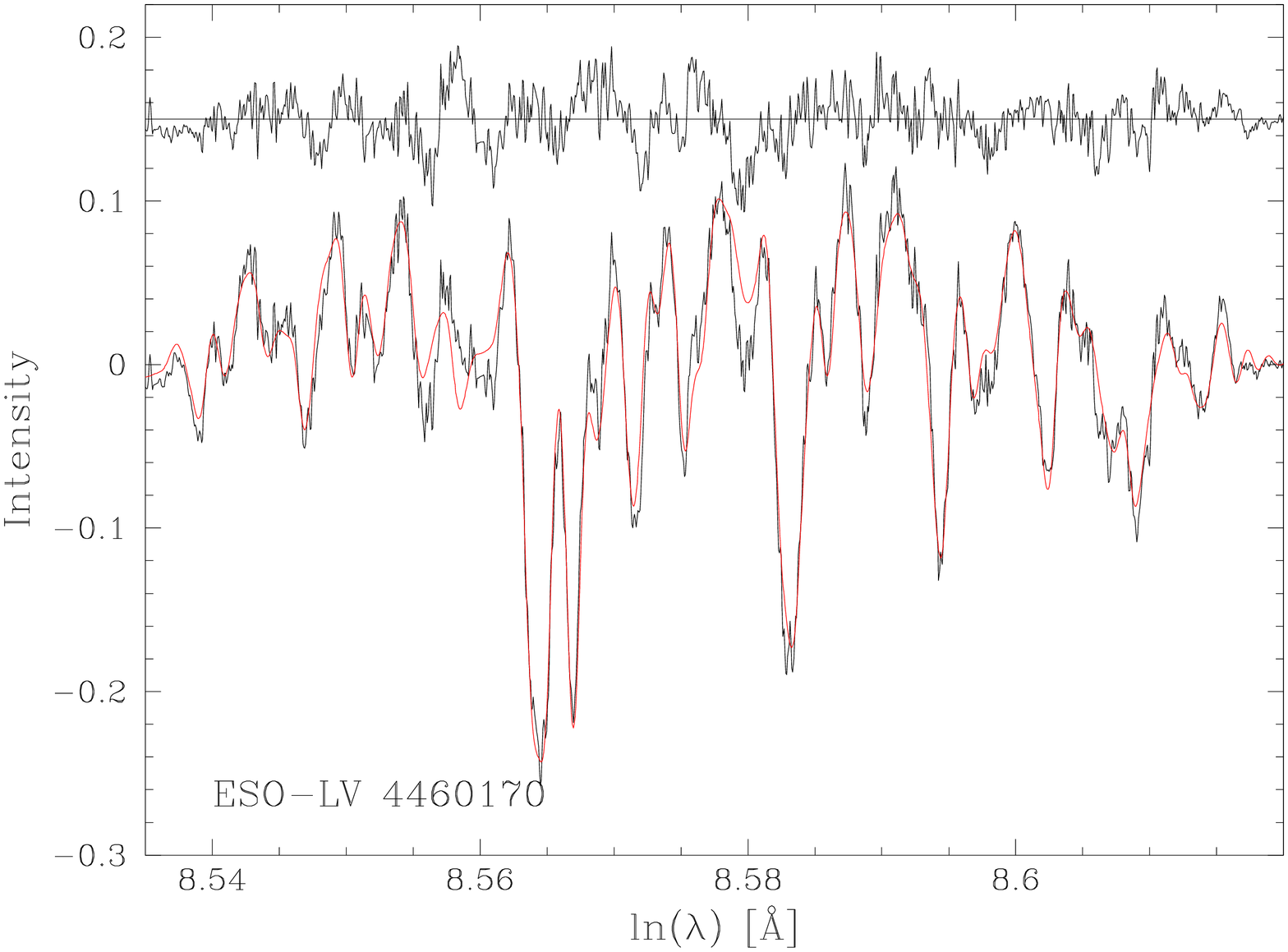, width=6cm, angle=270}  
\end{minipage}  
\caption{Examples of the central spectra (continuum line) for three 
  sample galaxies. They are compared with the template spectra (dotted
  line) convolved with their corresponding LOSVDs. The galaxy and
  template spectra have been continuum-subtracted and tapered at the
  ends with a cosine bell function. Residuals are plotted in the upper
  part of each panel and have false zero points for viewing
  convenience.  ESO-LV~2060140 (upper panel) was observed in
  run 3 and fitted with SAO~99192. ESO-LV~4000370 (central
  panel) and ESO-LV~4460170 (see App. \ref{sec:hsb}, lower
  panel) were observed in runs 2 and 1, respectively. They were fitted
  with SAO~137330.}
\label{fig:tempmism}  
\end{center}  
\end{figure}  
}

\newcommand{\placeFigfour}{
\begin{figure}  
\begin{center}  
\begin{minipage}[b]{.99\linewidth}  
\centering\epsfig{figure=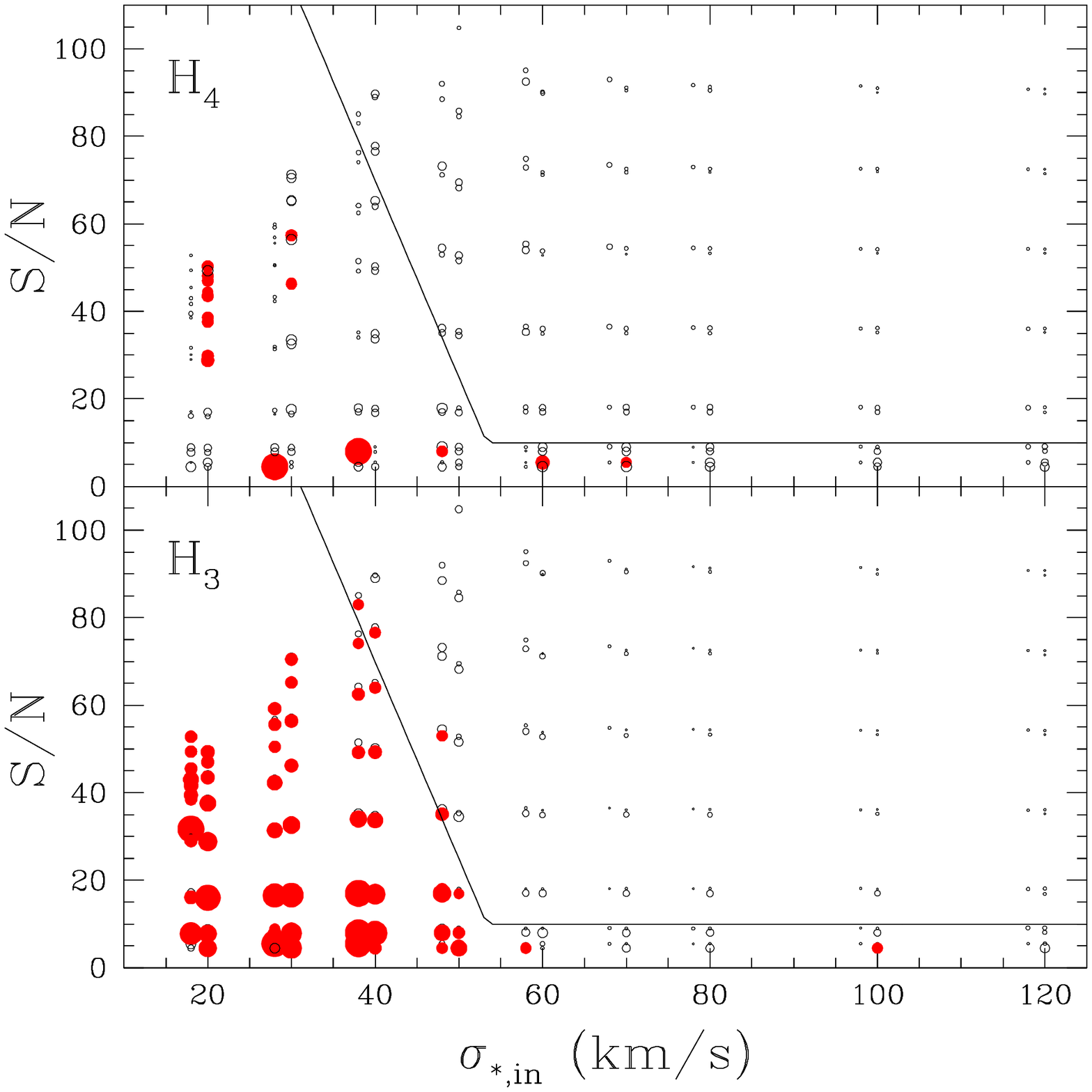, width=8cm, angle=0}  
\end{minipage}   
\caption{Reliability of the $h_3$ and $h_4$ measurements from Monte 
  Carlo simulations. Upper panel: The value of $|\Delta h_4|$ obtained 
  for $h_3=0$ as a function of $\sigma_{\rm \star, in}$ and $S/N$. The 
  values obtained for $h_4=0$ are shifted from those obtained with 
  $h_4=-0.1$ for viewing convenience. Symbol size is proportional to 
  the measured value. Open and filled symbols correspond to $|\Delta 
  h_4|$ smaller and larger than 0.1, respectively. The minimum $S/N$ 
  ratios to obtain reliable measurements of $h_4$ for the different 
  $\sigma_{\rm \star, in}$ are connected with a continuous line. Lower 
  panel: As in the upper panel but for $|\Delta h_3|$.} 
\label{fig:sigmaSNlimit}  
\end{center}  
\end{figure}  
}

\newcommand{\placeFigfive}{
\begin{figure}  
\begin{center}  
\begin{minipage}[b]{.99\linewidth}  
\centering\epsfig{figure=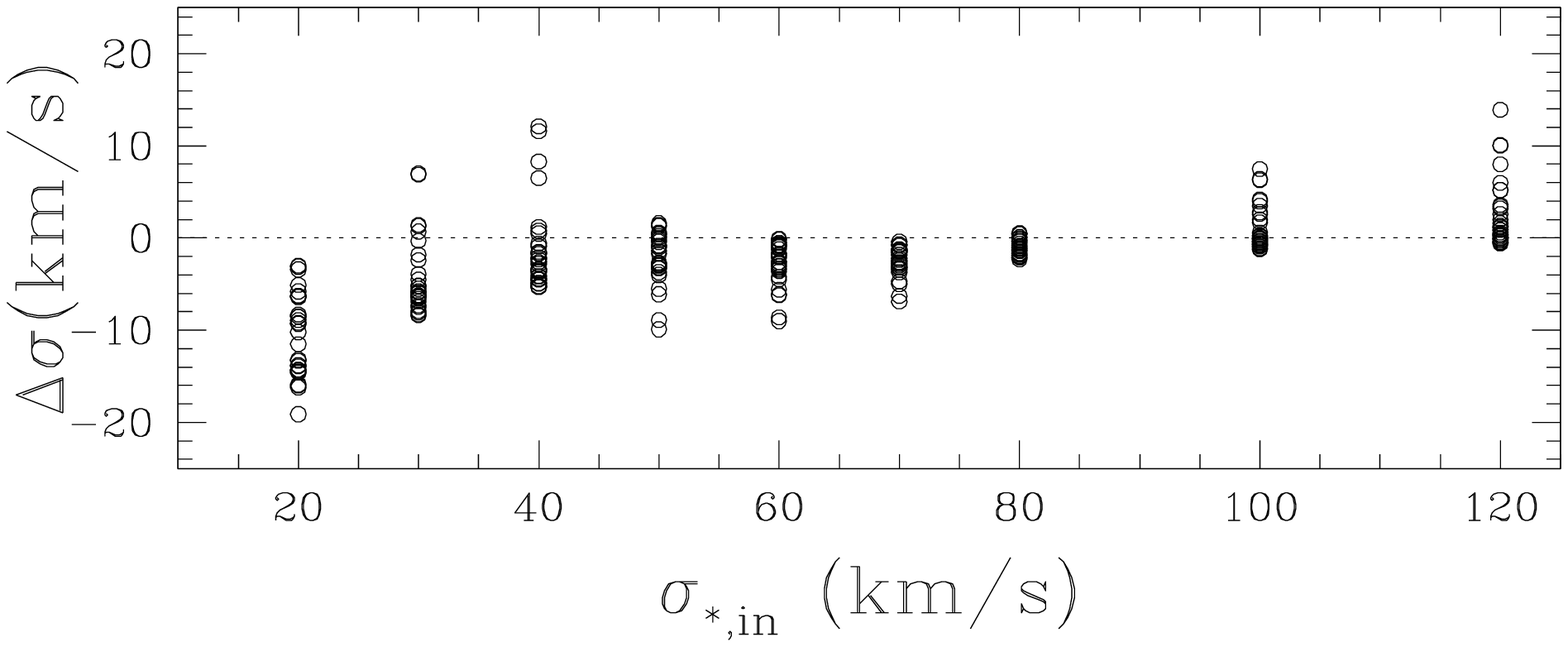, width=8cm, angle=0}  
\end{minipage}   
\caption{Reliability of $\sigma_\star$ measurements from Monte 
  Carlo simulations. The values of $\Delta\sigma_\star$ is plotted as 
  a function of $\sigma_\star$ only for the artificial spectra with 
  reliable measurements of $h_3$ and $h_4$.} 
\label{fig:sigmatest}  
\end{center}  
\end{figure}  
}

\newcommand{\placeFigsix}{
\begin{figure*}  
\begin{center}  
 \begin{minipage}[b]{.49\linewidth}  
  \centering\epsfig{figure=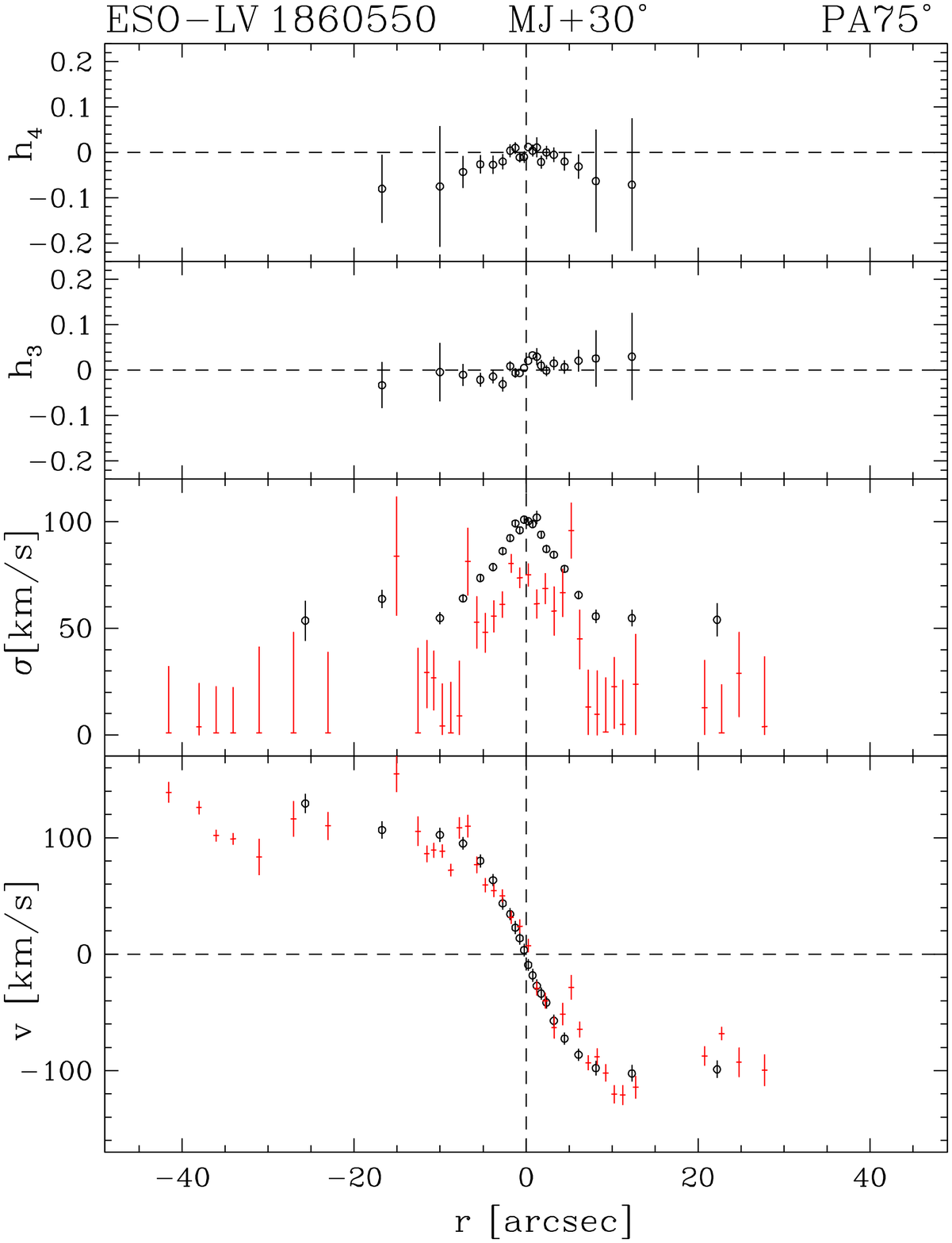,width=\linewidth}  
 \end{minipage} \hfill  
 \begin{minipage}[b]{.49\linewidth}  
  \centering\epsfig{figure=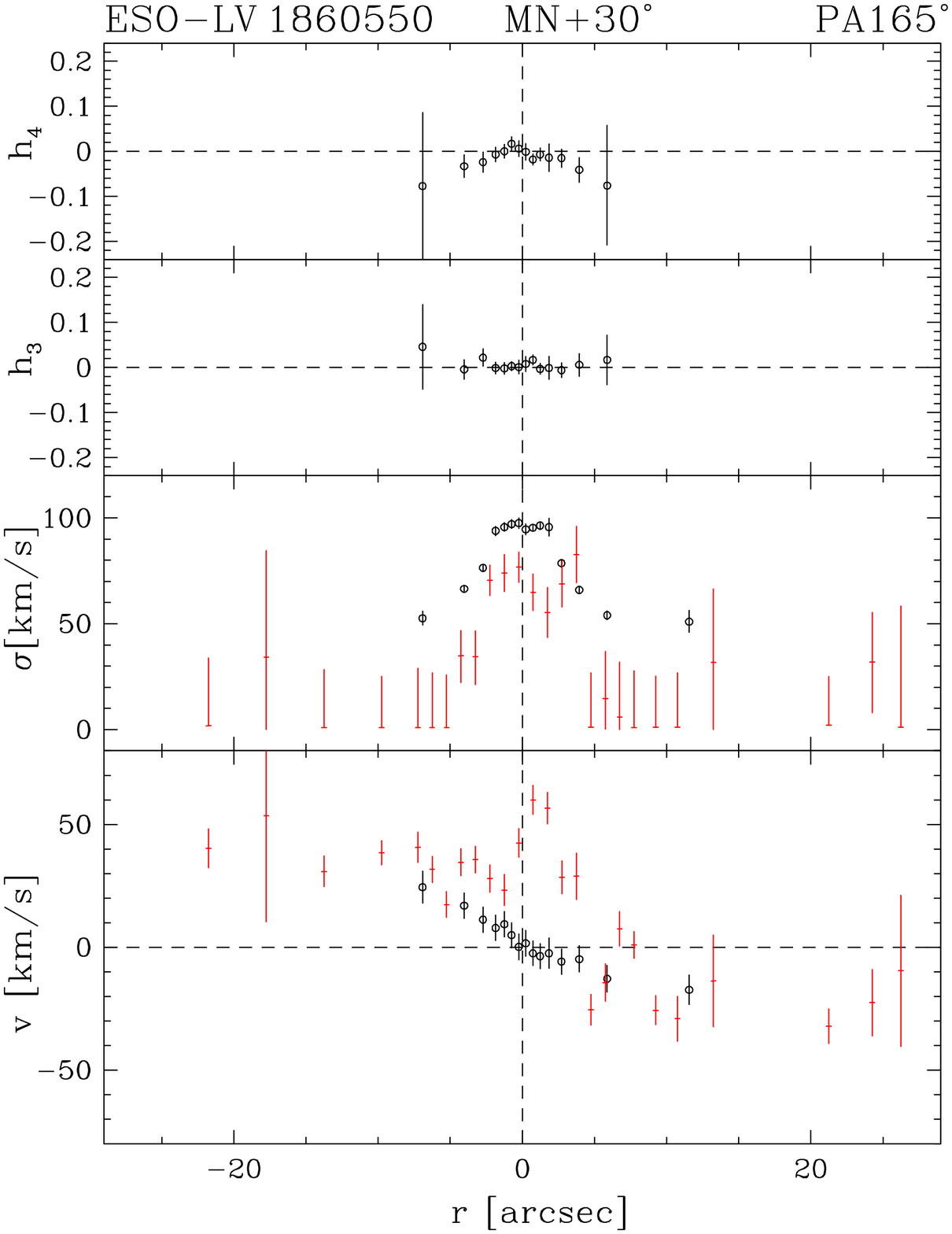,width=\linewidth}  
 \end{minipage}  
\caption{Kinematic parameters for stars (circles) and ionized-gas  
  (crosses) measured along the observed axes of the sample 
  galaxies. The radial profiles of the line-of-sight velocity (after 
  the subtraction of systemic velocity and with no correction for 
  inclination), velocity dispersion (corrected for instrumental 
  velocity dispersion), third and fourth order coefficient of the 
  Gauss-Hermite decomposition of the stellar LOSVD are shown (from 
  bottom to top). For each panel the galaxy name, location and 
  position angle of the slit are given.} 
\label{fig:kinematics_lsb}    
\end{center}  
\end{figure*}  
\begin{figure*}  
\addtocounter{figure}{-1}  
\begin{center}  
 \begin{minipage}[b]{.49\linewidth}  
  \centering\epsfig{figure=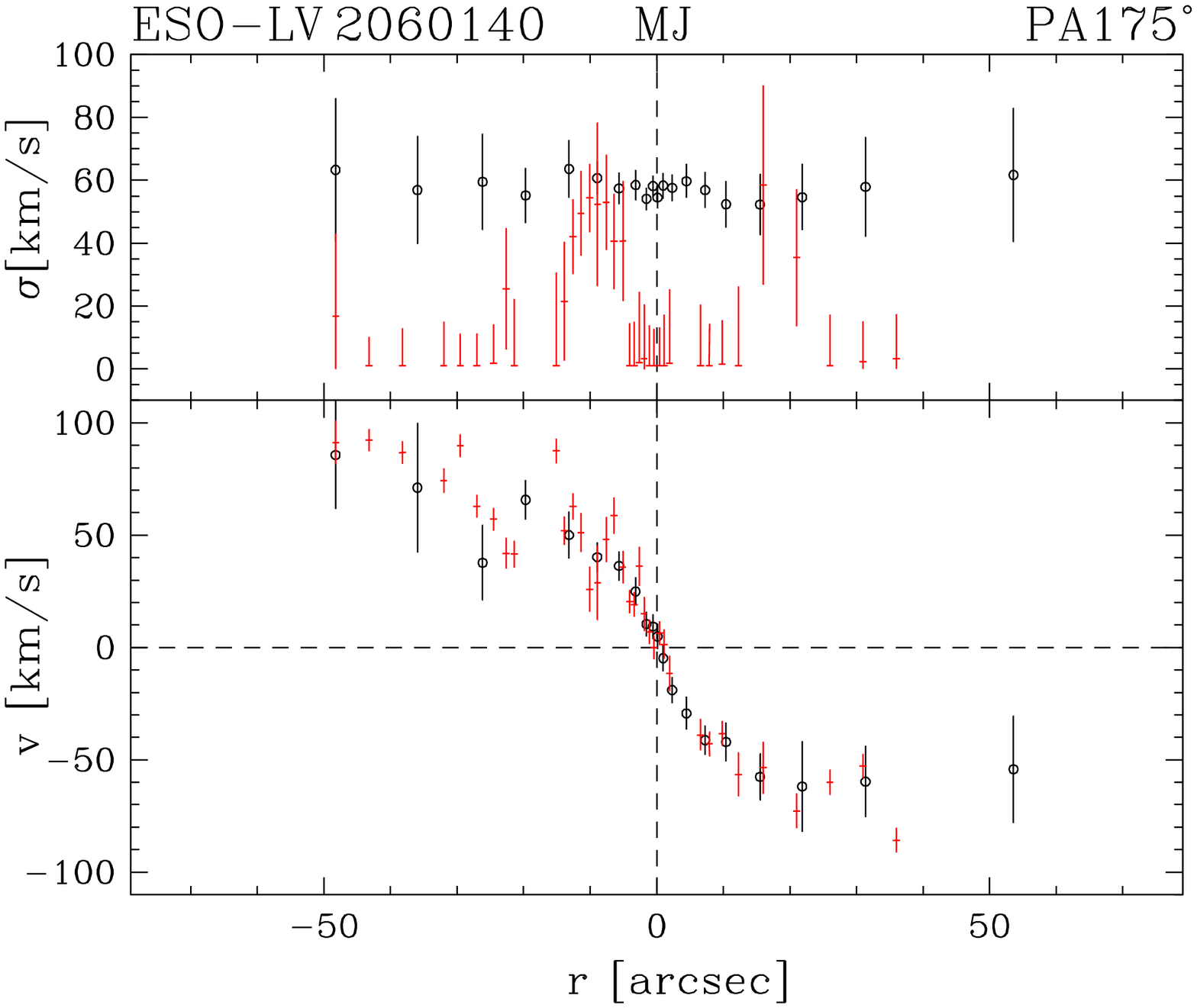,width=\linewidth}  
 \end{minipage} \hfill  
 \begin{minipage}[b]{.49\linewidth}  
  \centering\epsfig{figure=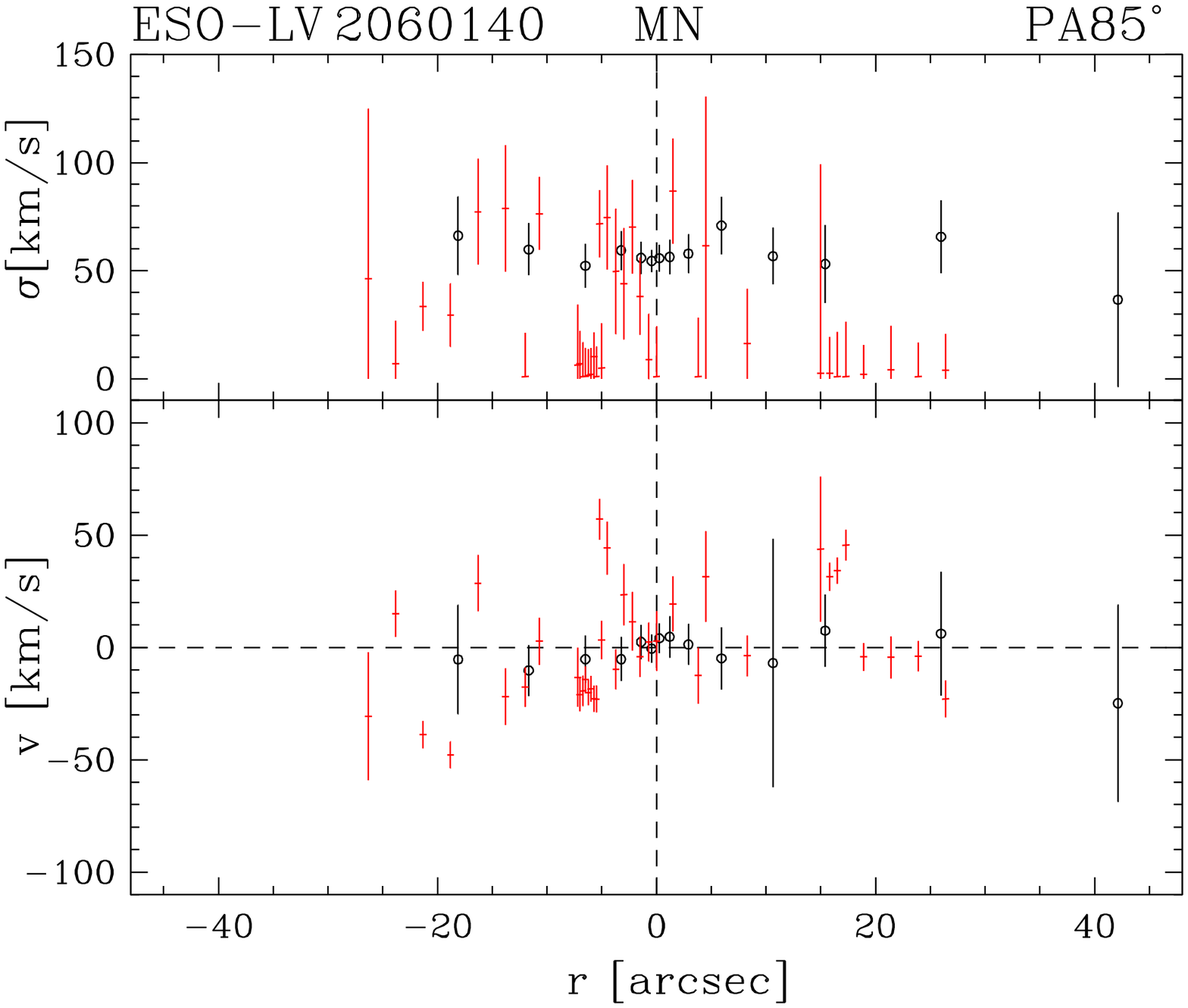,width=\linewidth}  
 \end{minipage}  
 \begin{minipage}[b]{.49\linewidth}  
  \centering\epsfig{figure=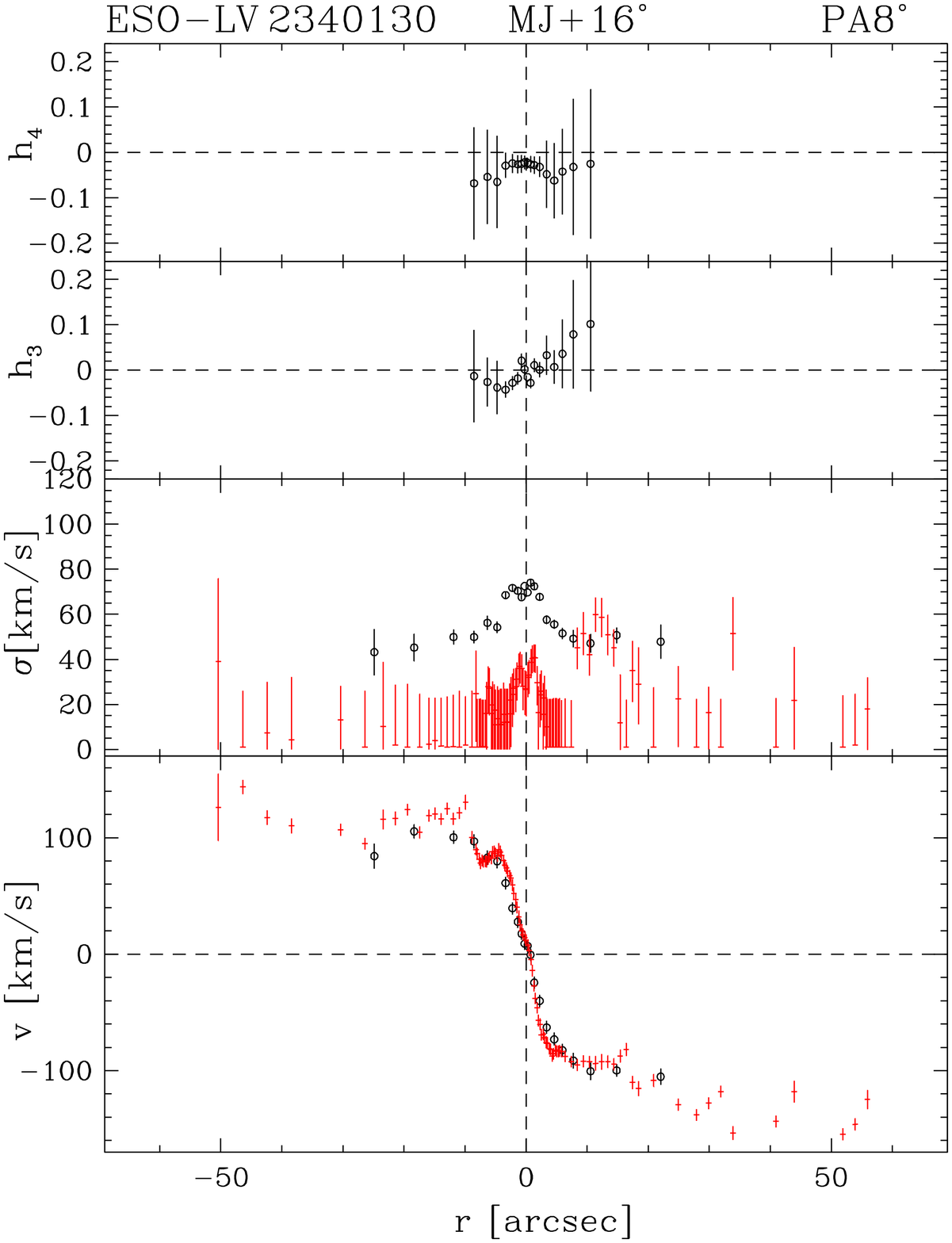,width=\linewidth}  
 \end{minipage} \hfill  
 \begin{minipage}[b]{.49\linewidth}  
  \centering\epsfig{figure=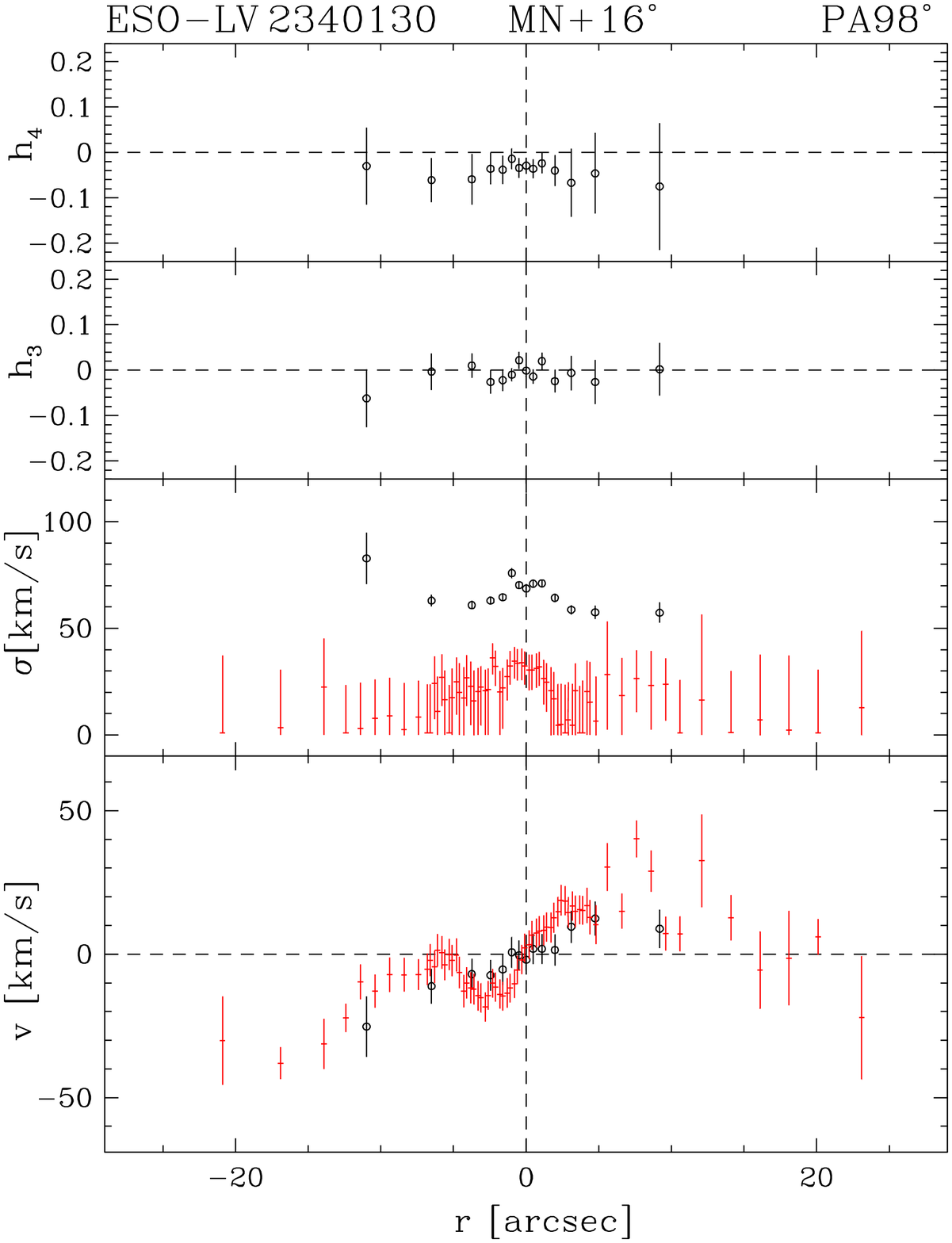,width=\linewidth}  
 \end{minipage}  
\caption{Continued}  
\end{center}  
\end{figure*}  
\begin{figure*}  
\addtocounter{figure}{-1}  
\begin{center}  
 \begin{minipage}[b]{.49\linewidth}  
  \centering\epsfig{figure=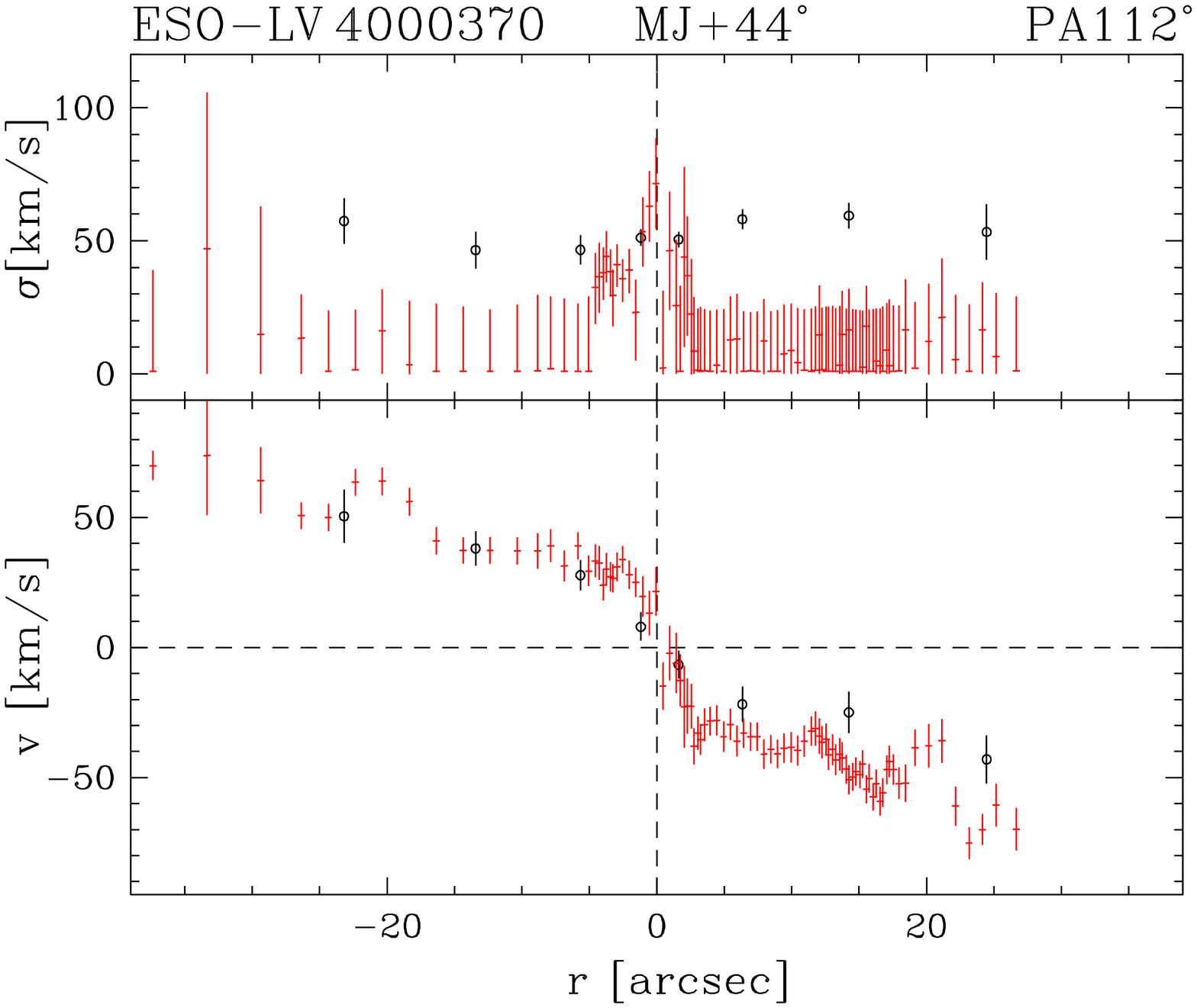,width=\linewidth}  
 \end{minipage} \hfill  
 \begin{minipage}[b]{.49\linewidth}  
  \centering\epsfig{figure=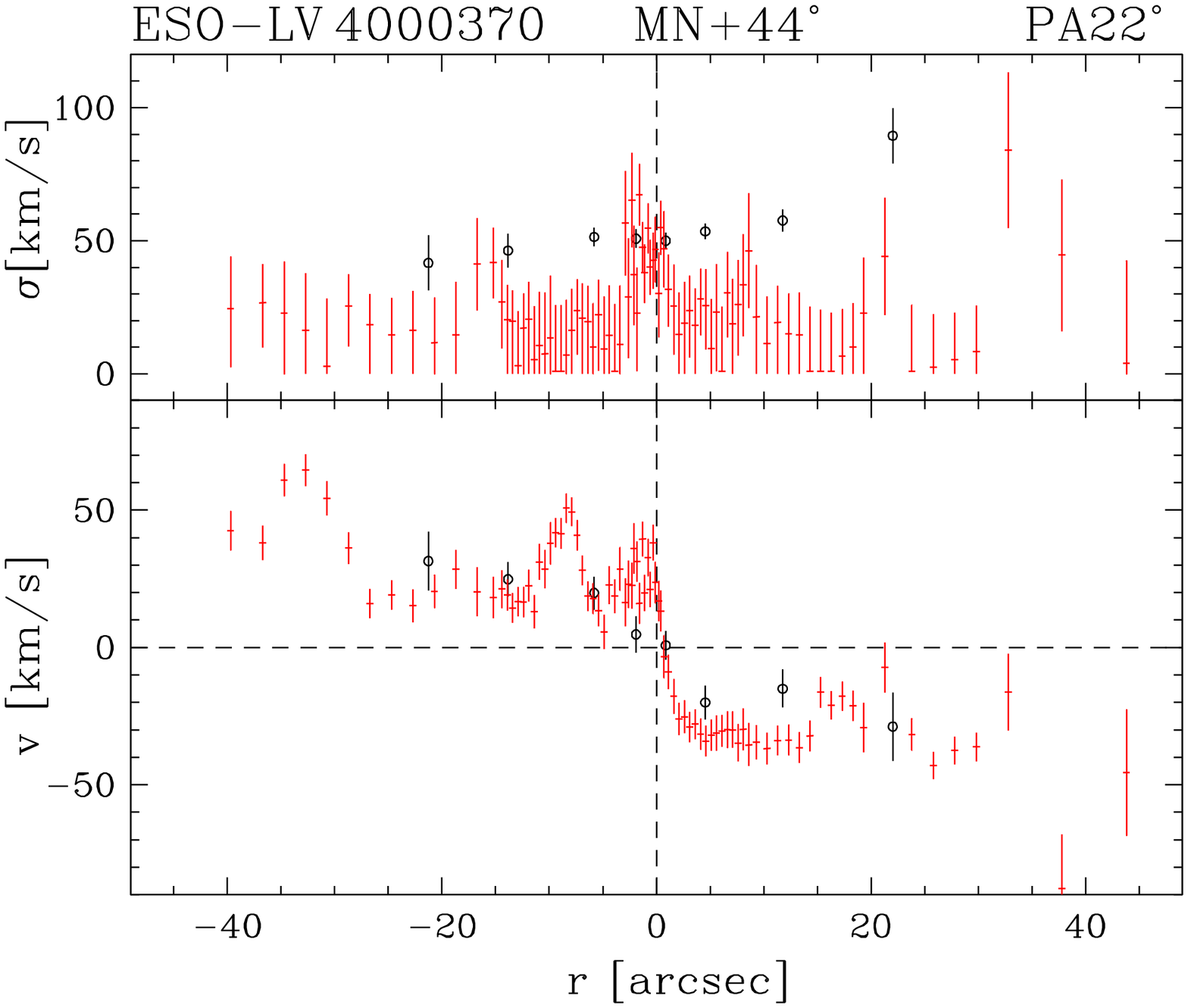,width=\linewidth}  
 \end{minipage}  
 \begin{minipage}[b]{.49\linewidth}  
  \centering\epsfig{figure=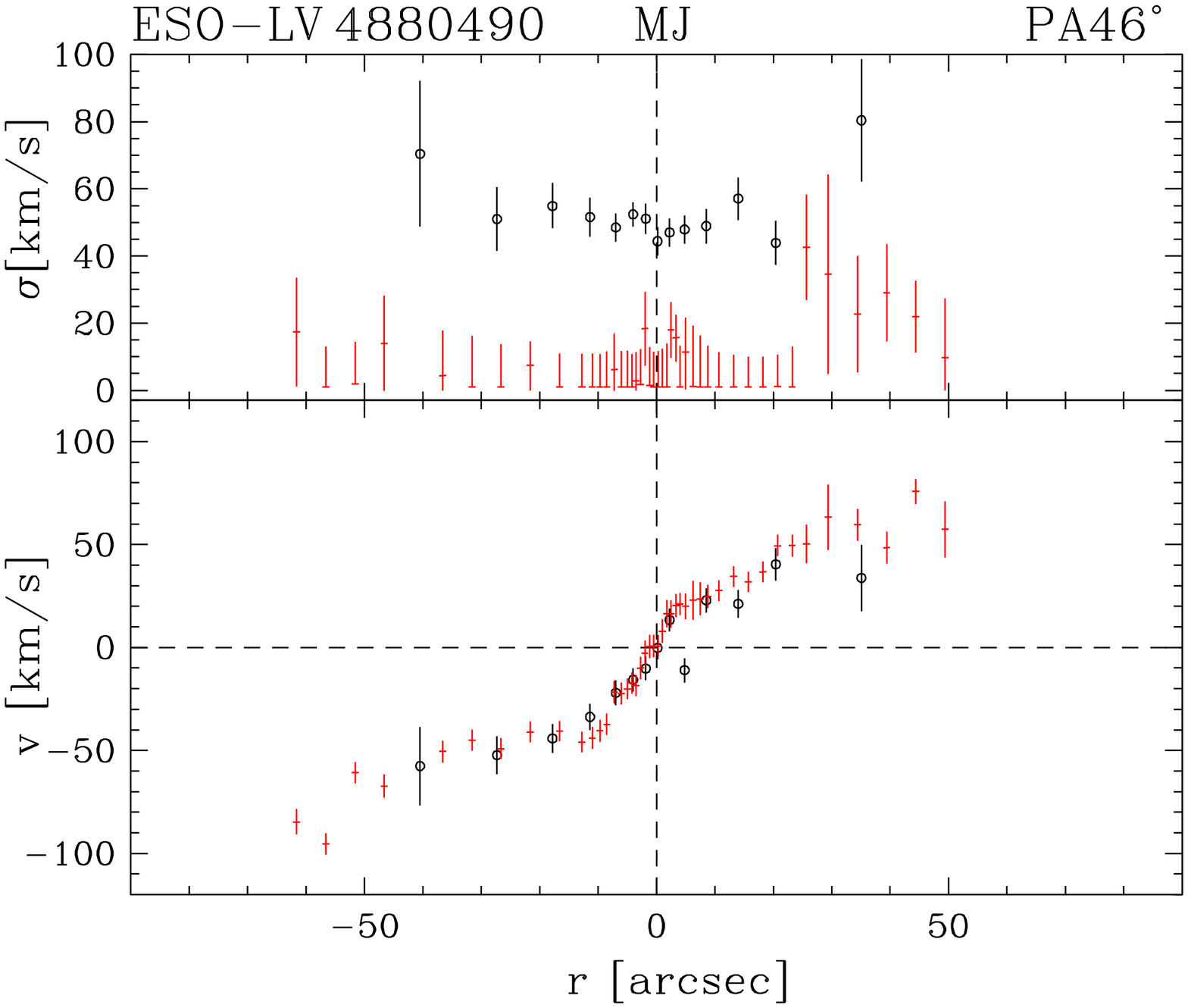,width=\linewidth}  
 \end{minipage} \hfill  
 \begin{minipage}[b]{.49\linewidth}  
  \centering\epsfig{figure=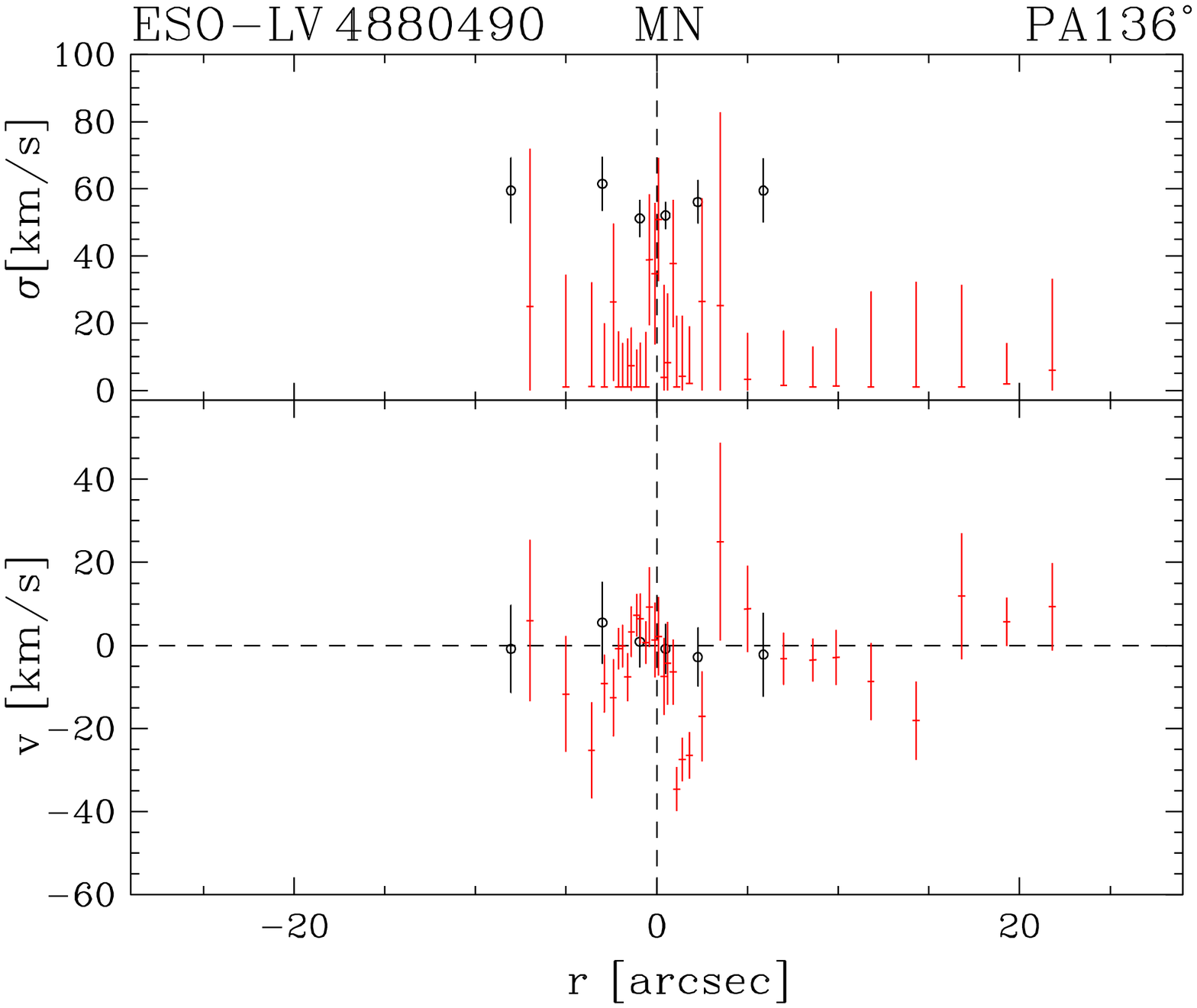,width=\linewidth}  
 \end{minipage}  
\caption{Continued}  
\end{center}  
\end{figure*}  
\begin{figure*}  
\addtocounter{figure}{-1}  
\begin{center}  
 \begin{minipage}[b]{.49\linewidth}  
  \centering\epsfig{figure=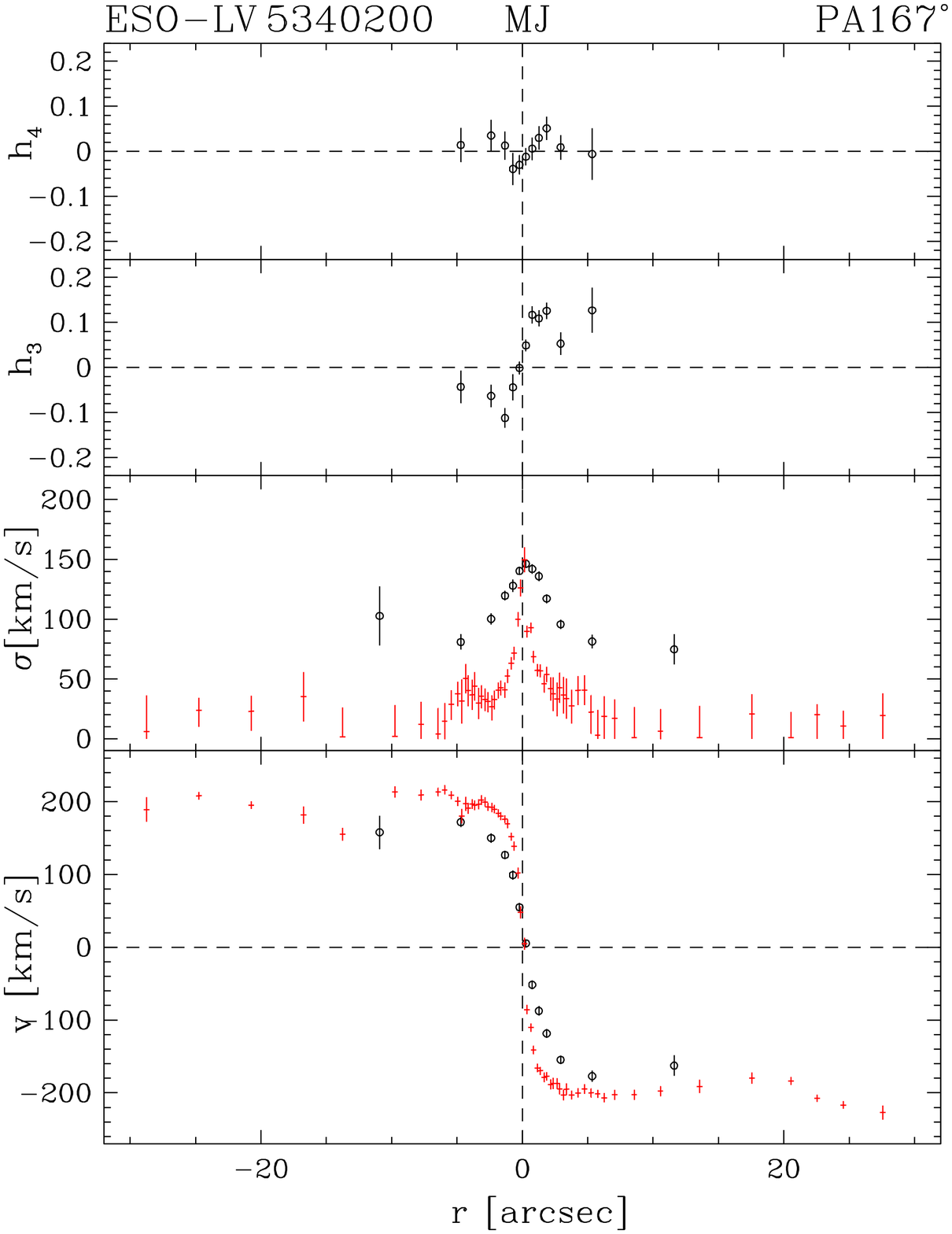,width=\linewidth}  
 \end{minipage} \hfill  
 \begin{minipage}[b]{.49\linewidth}  
  \centering\epsfig{figure=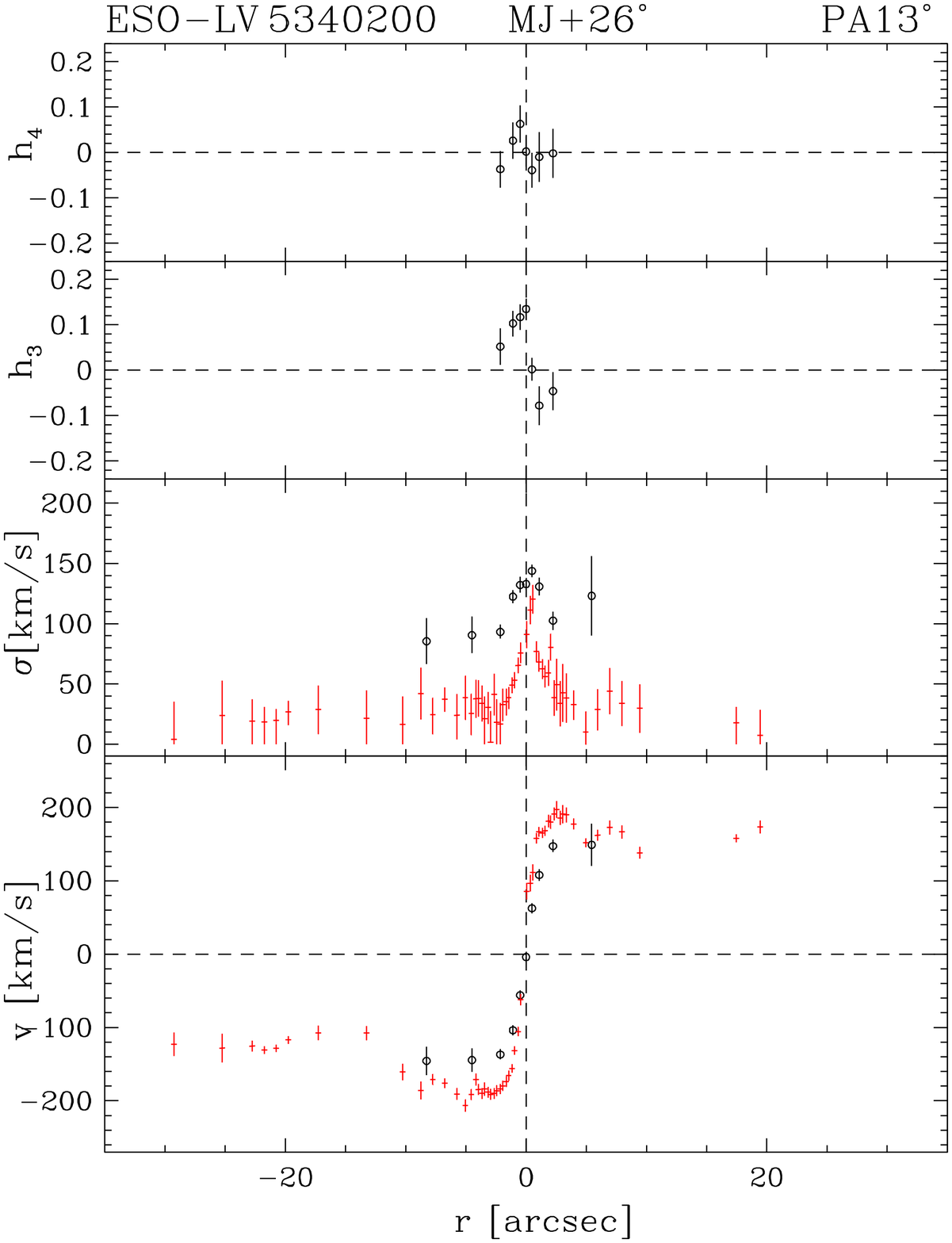,width=\linewidth}  
 \end{minipage}  
 \begin{minipage}[b]{.49\linewidth}  
  \centering\epsfig{figure=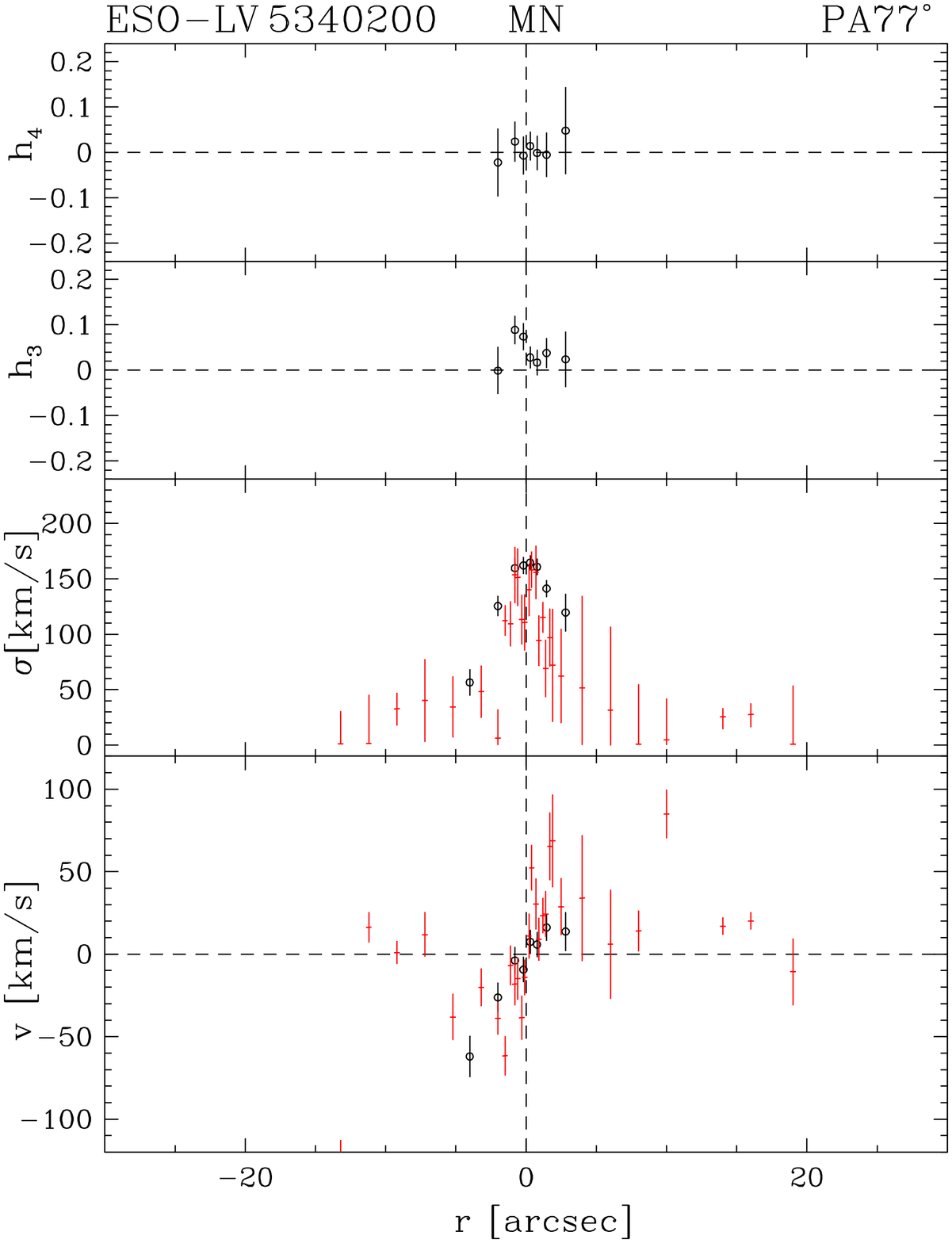,width=\linewidth}  
 \end{minipage} \hfill  
 \begin{minipage}[b]{.49\linewidth}  
  \centering\epsfig{figure=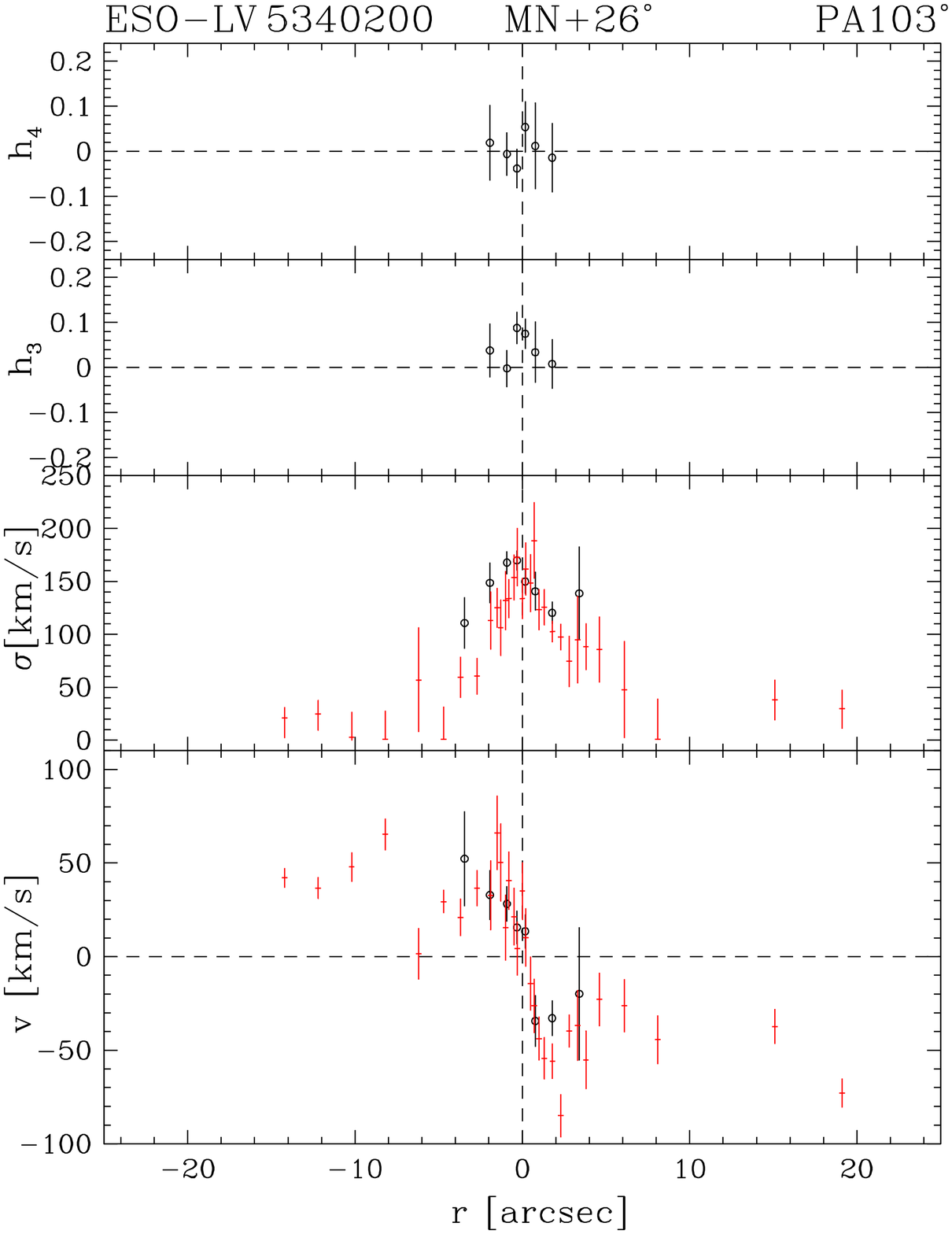,width=\linewidth}  
 \end{minipage}  
\caption{Continued}  
\end{center}  
\end{figure*}  
}

\newcommand{\placeFigseven}{
\begin{figure*}  
\begin{center}  
 \begin{minipage}[b]{.49\linewidth}  
  \centering\epsfig{figure=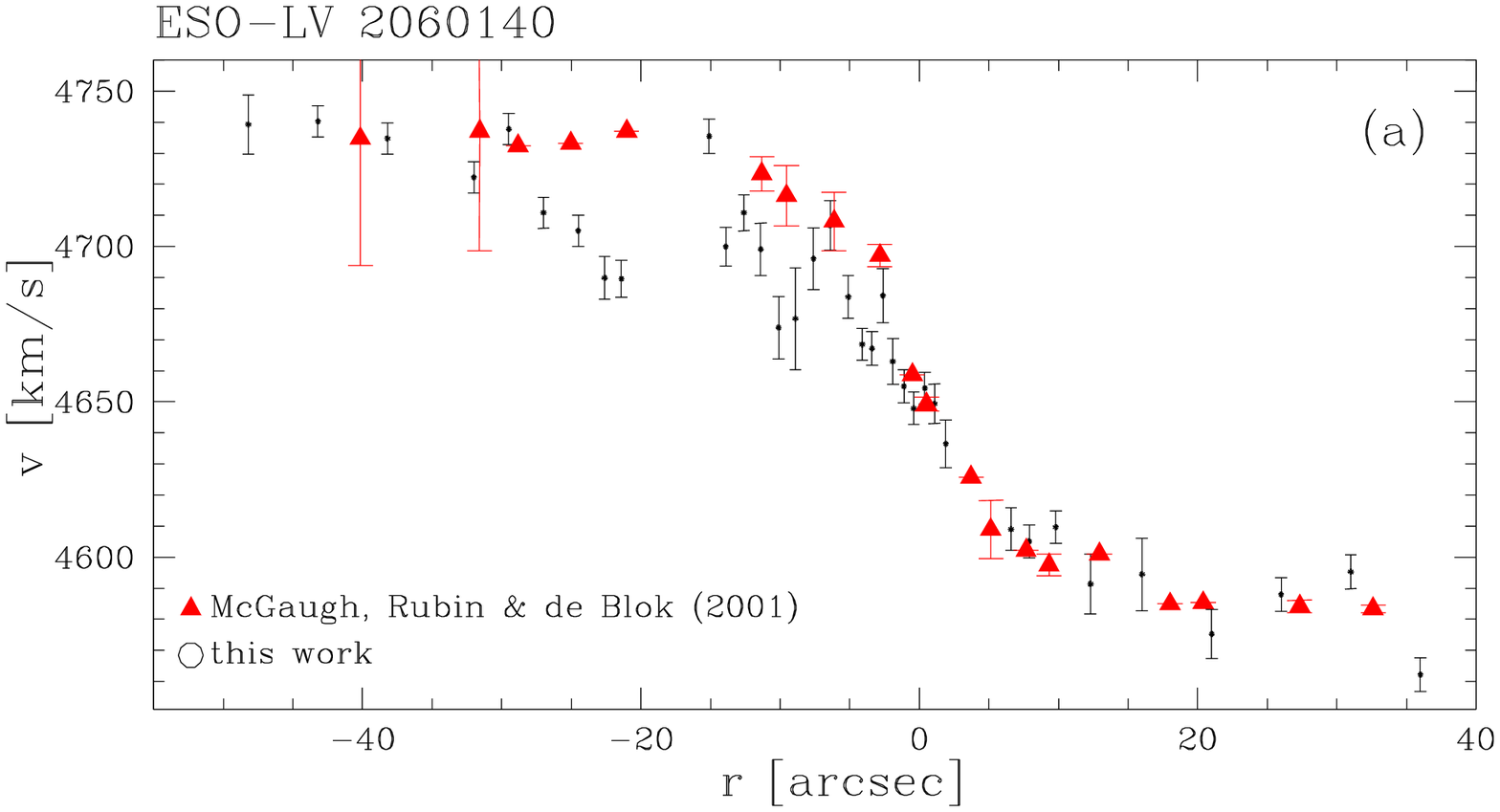,width=\linewidth}  
 \end{minipage} \hfill  
 \begin{minipage}[b]{.49\linewidth}  
  \centering\epsfig{figure=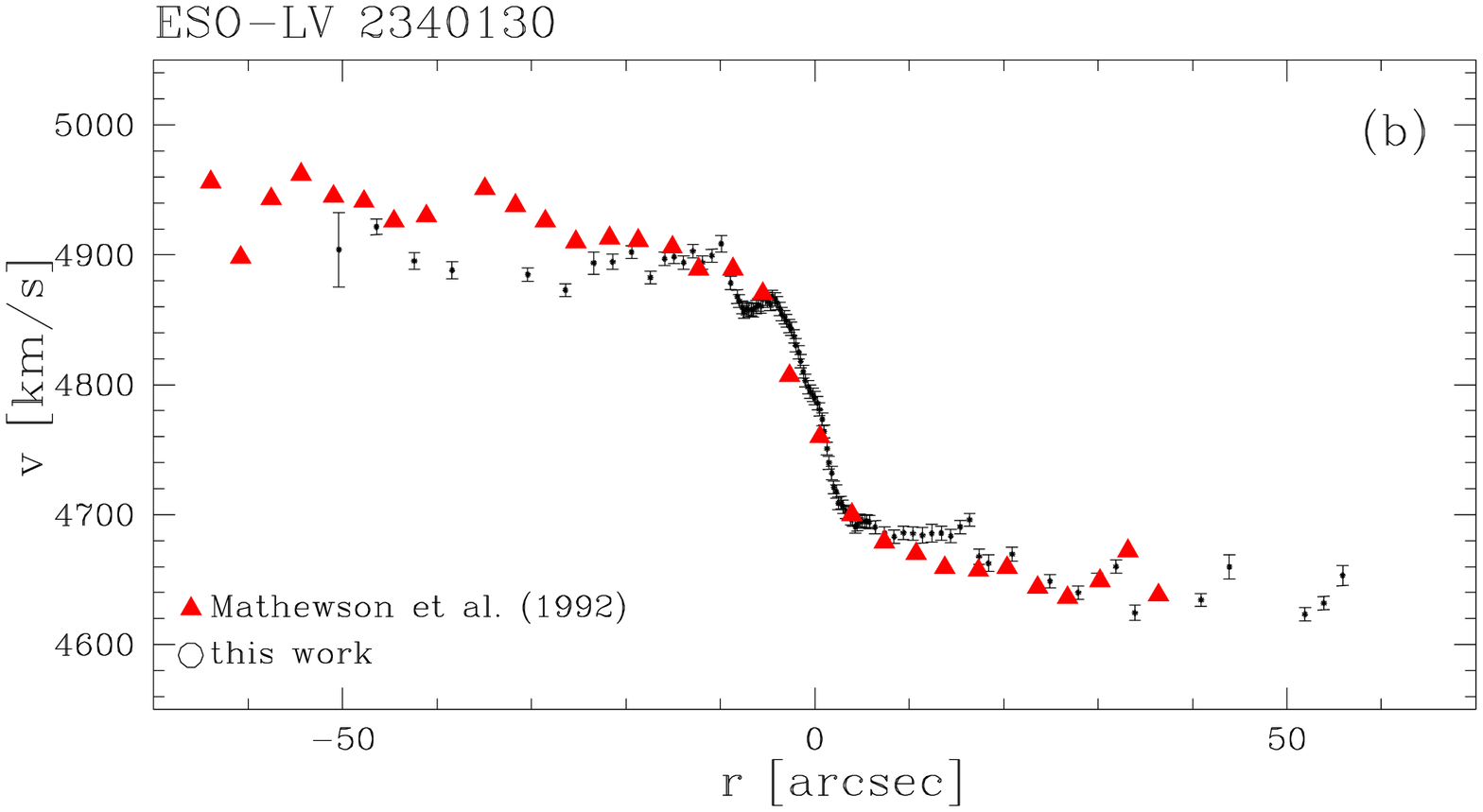,width=\linewidth}  
 \end{minipage}  
 \begin{minipage}[b]{.49\linewidth}  
  \centering\epsfig{figure=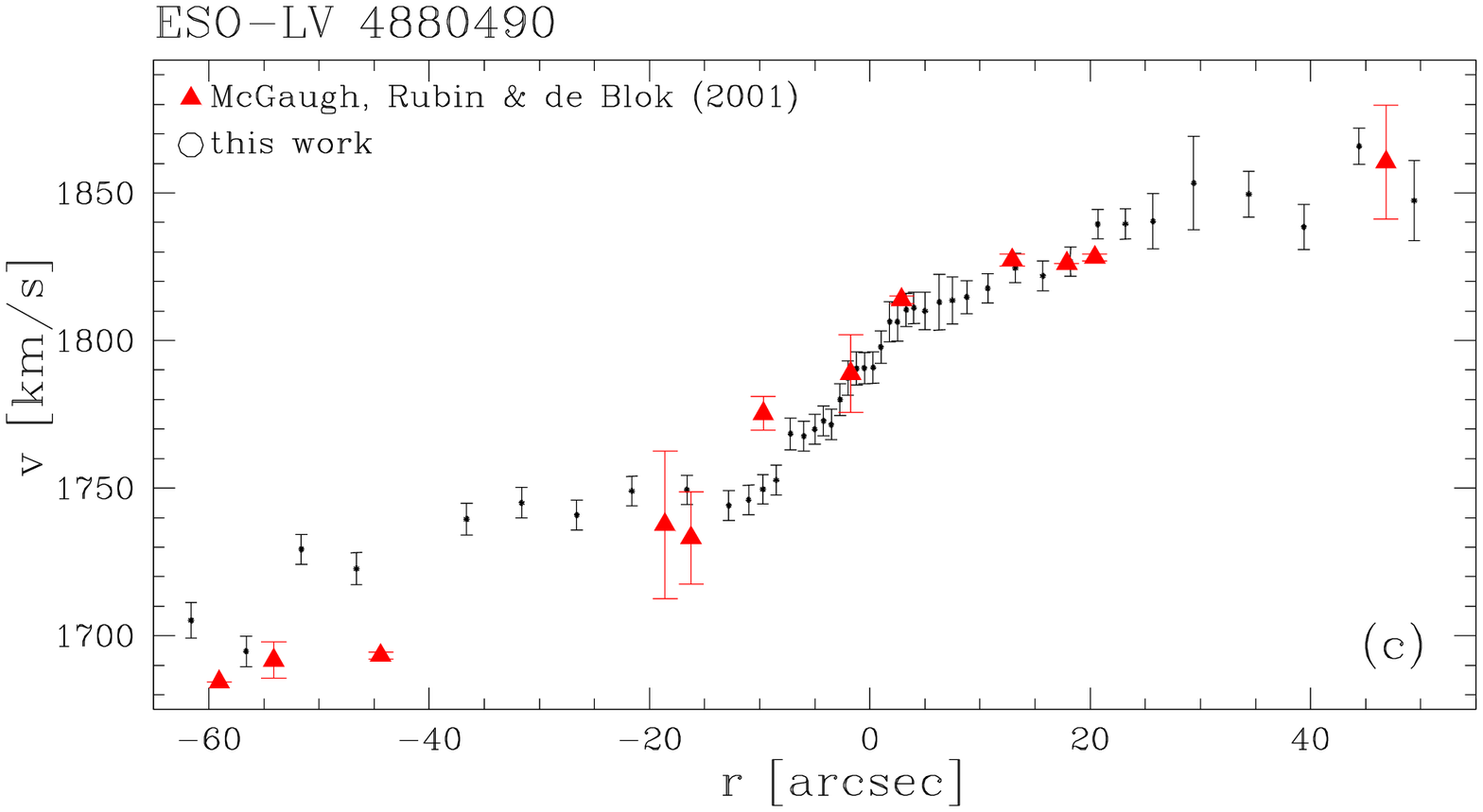,width=\linewidth}  
 \end{minipage} \hfill  
 \begin{minipage}[b]{.49\linewidth}  
  \centering\epsfig{figure=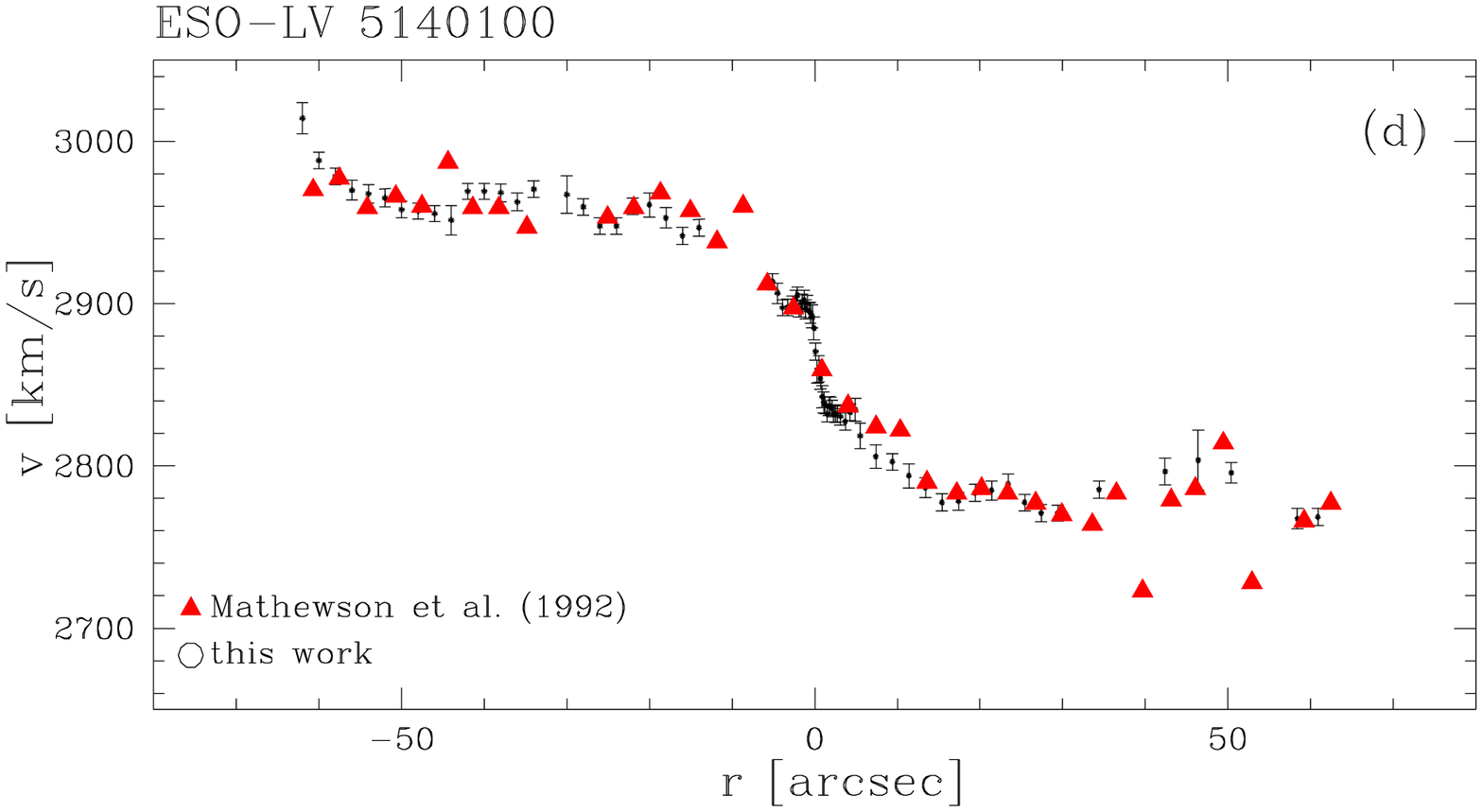,width=\linewidth}  
 \end{minipage}  
\caption{The line-of-sight velocities of the ionized gas derived in 
  this study (circles) for ESO-LV~2060140, ESO-LV~2340130,
  ESO-LV~4880490, and ESO-LV~5140100 [see App.~\ref{sec:hsb}] compared
  with those (triangles) obtained by \citet{Mathewson1992} and
  \citet{McGaugh2001}.}
\label{fig:kincomparison} 
\end{center} 
\end{figure*}  
}

\newcommand{\placeFigeight}{
\begin{figure*}  
\begin{center}  
 \begin{minipage}[b]{.49\linewidth} 
   \centering\epsfig{figure=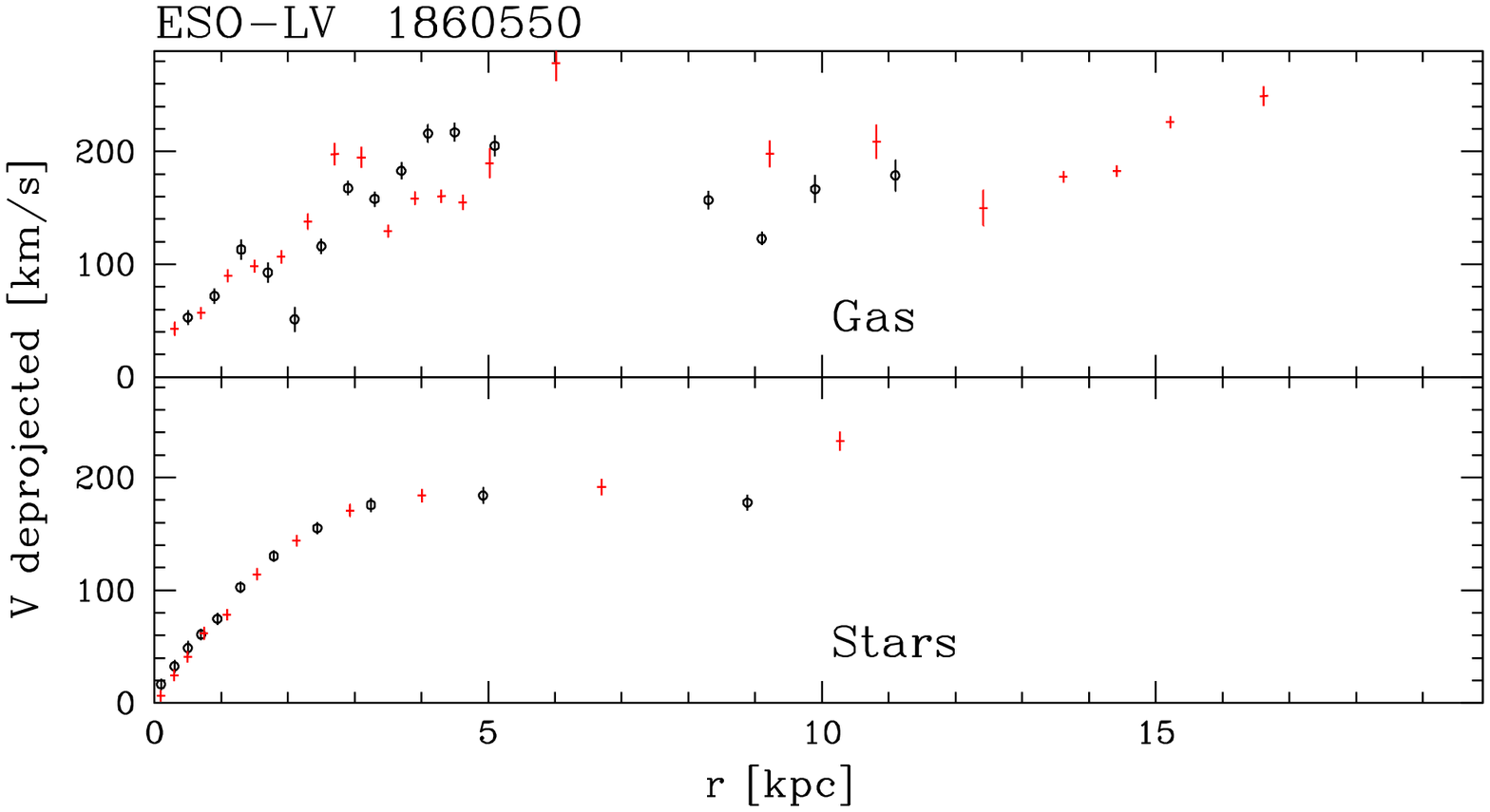,width=\linewidth} 
 \end{minipage} \hfill 
 \begin{minipage}[b]{.49\linewidth} 
   \centering\epsfig{figure=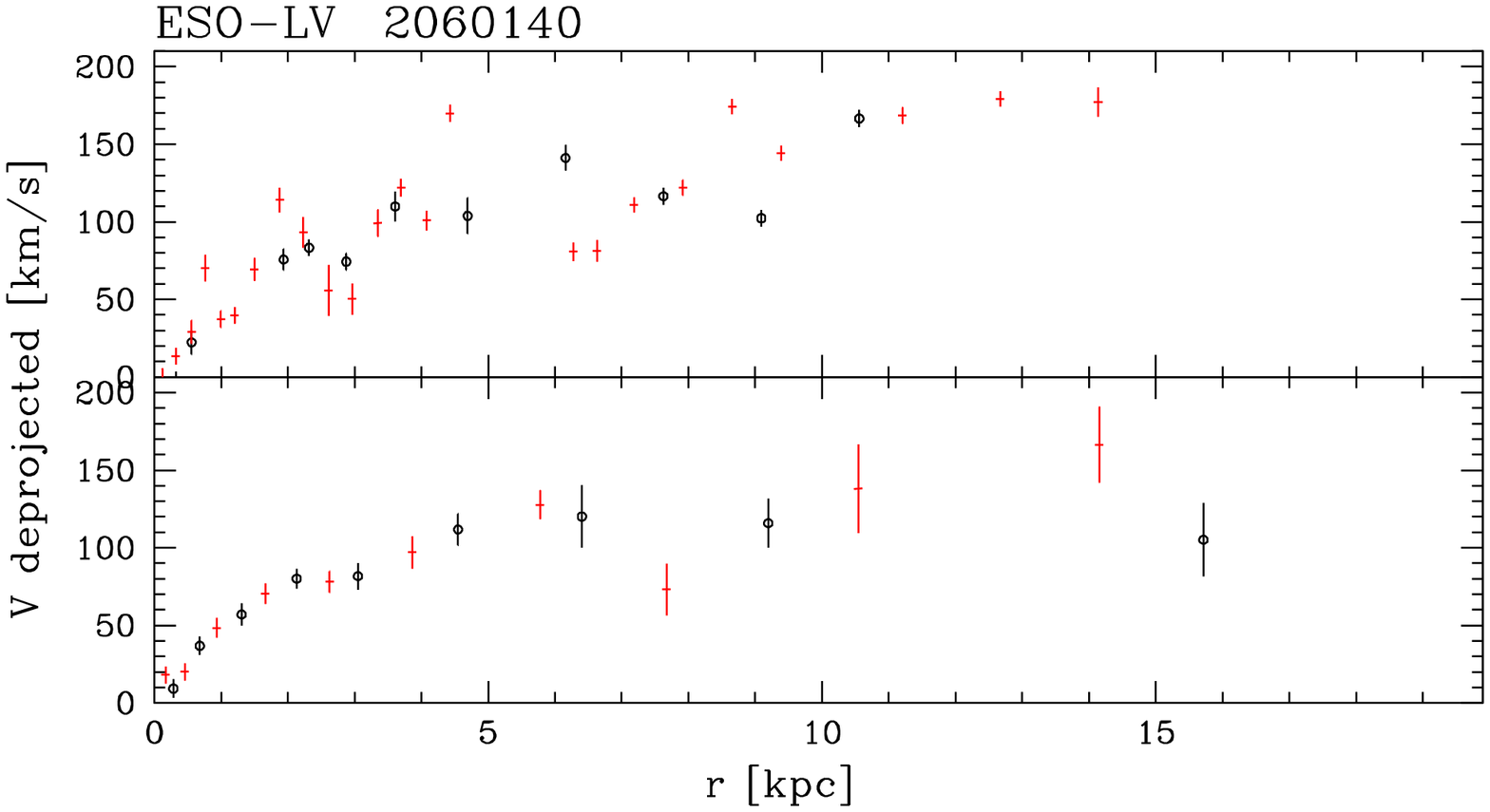,width=\linewidth}  
 \end{minipage} 
 \begin{minipage}[b]{.49\linewidth}  
   \centering\epsfig{figure=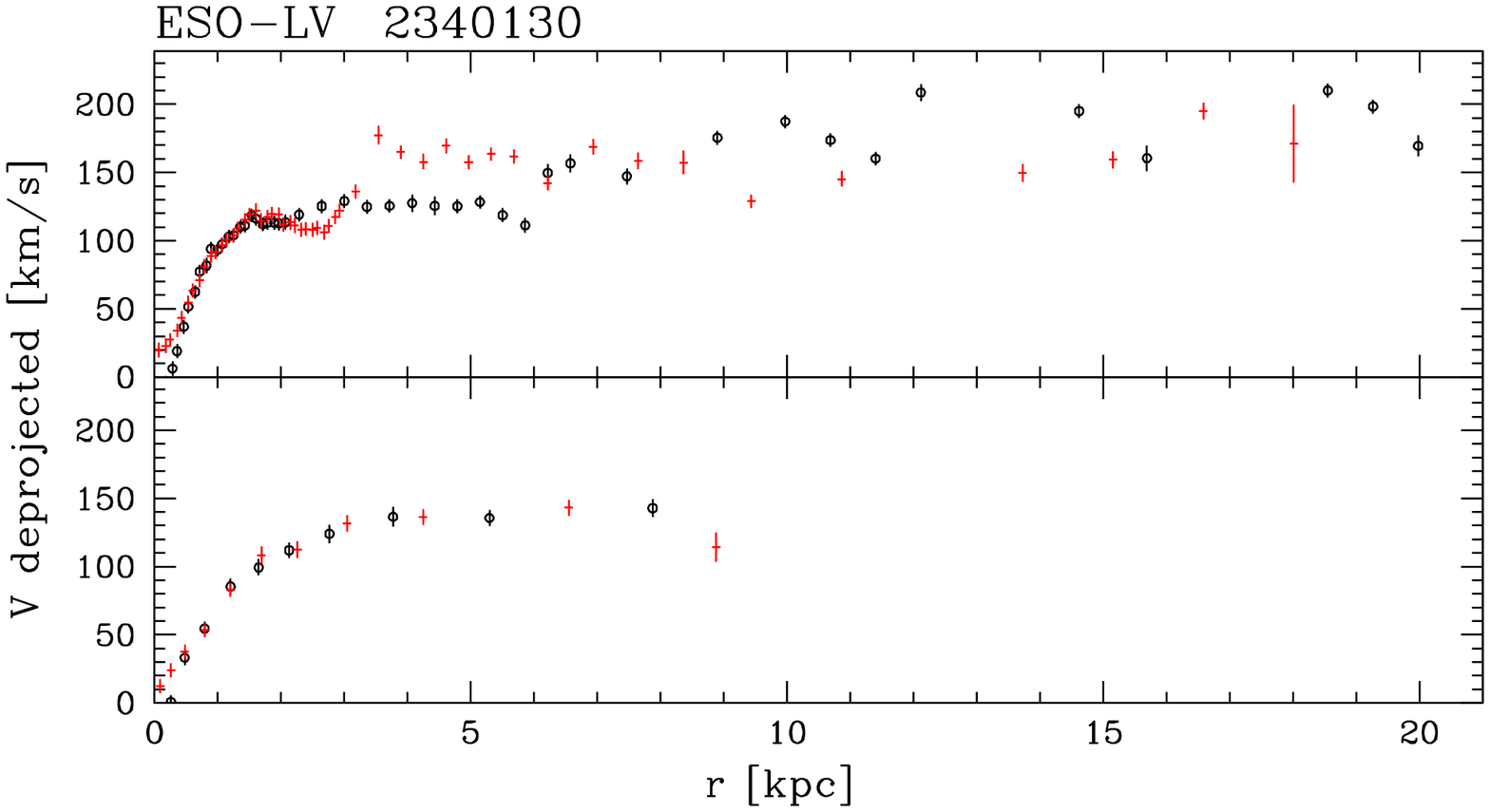,width=\linewidth}  
 \end{minipage} \hfill  
 \begin{minipage}[b]{.49\linewidth}  
   \centering\epsfig{figure=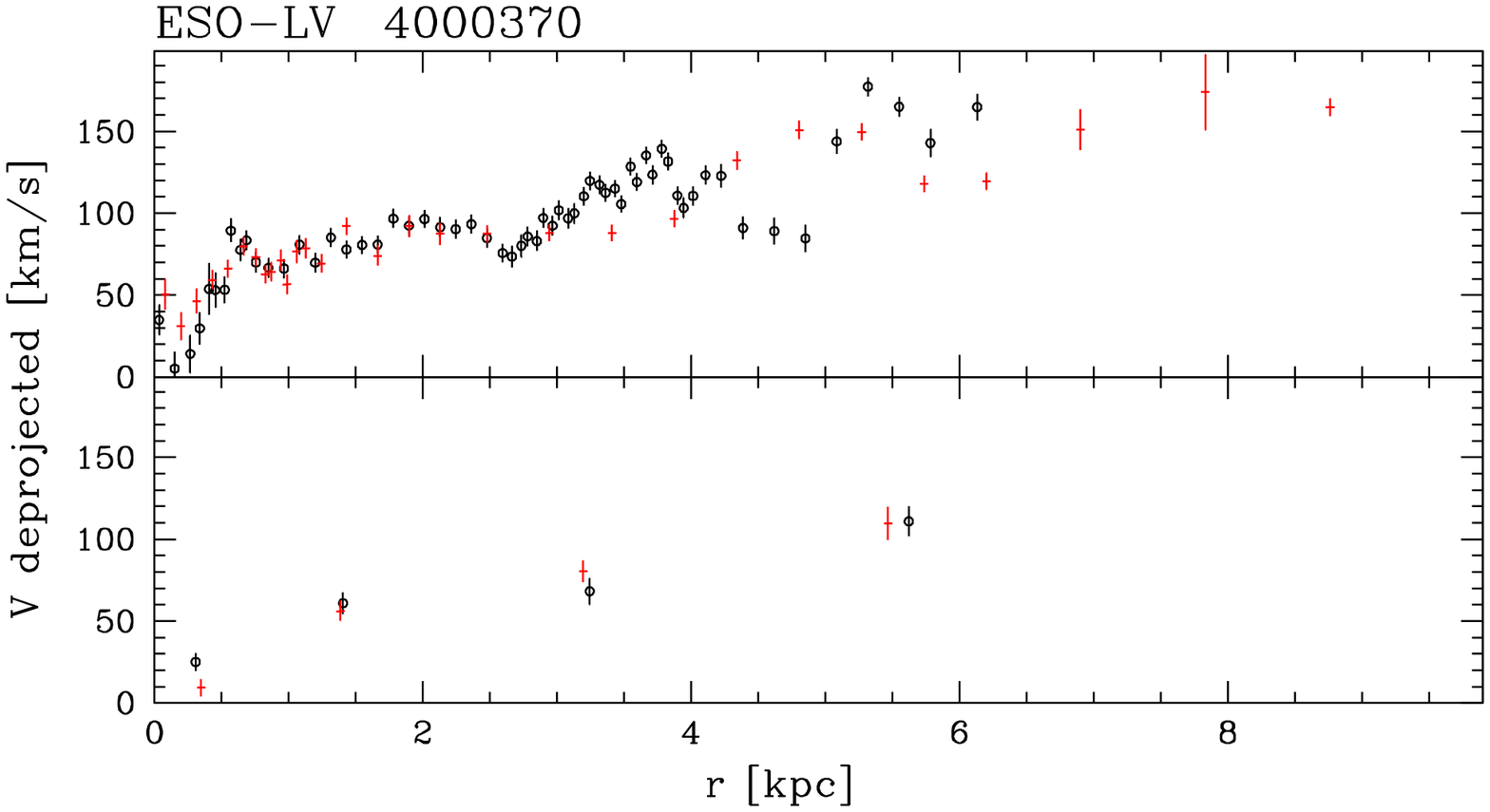,width=\linewidth}  
 \end{minipage}  
 \begin{minipage}[b]{.49\linewidth}  
   \centering\epsfig{figure=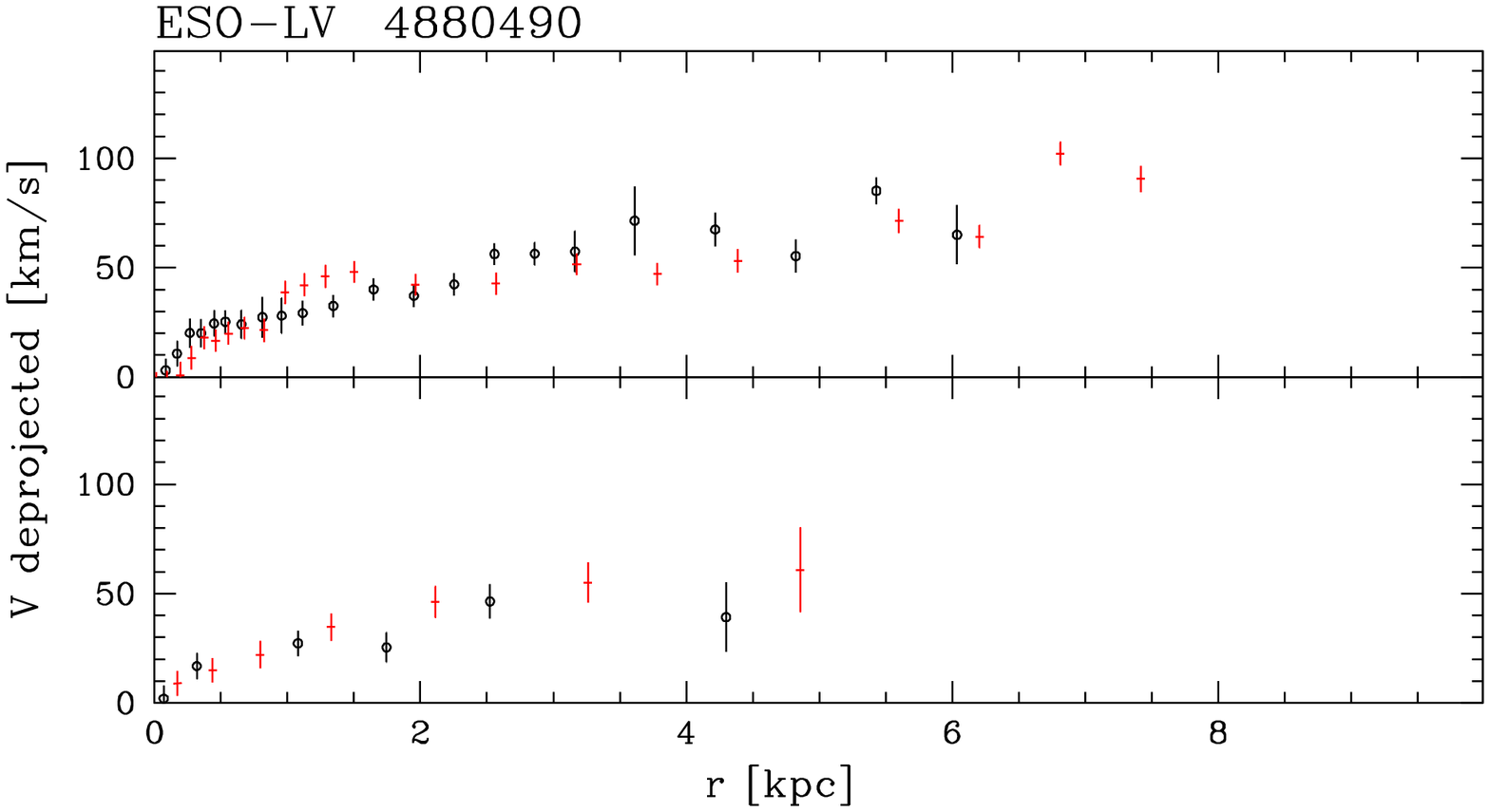,width=\linewidth}  
 \end{minipage} \hfill  
 \begin{minipage}[b]{.49\linewidth}  
   \centering\epsfig{figure=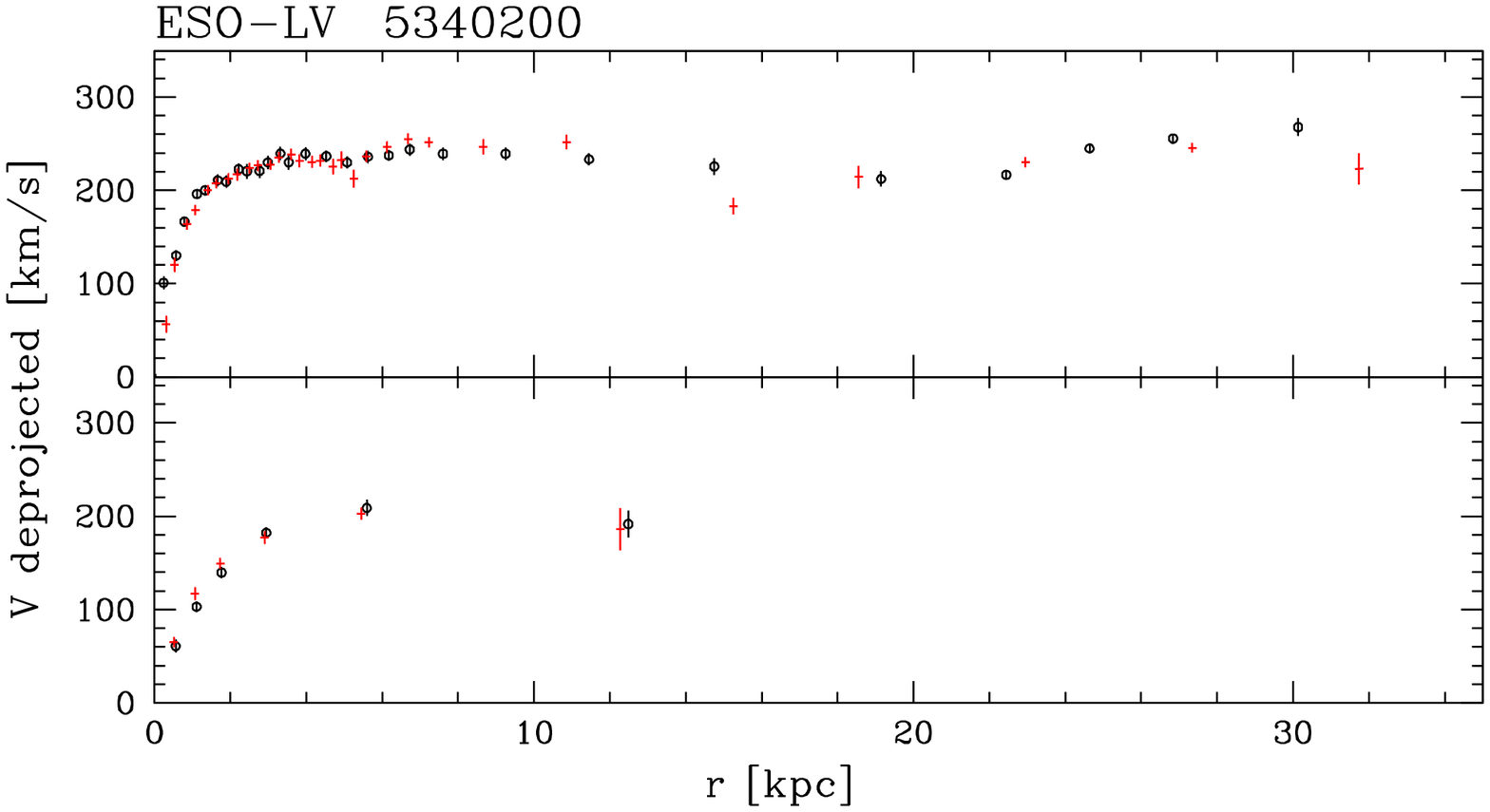,width=\linewidth} 
 \end{minipage}  
\caption{Ionized-gas and stellar rotation curves of the sample galaxies. 
  The velocities of the ionized gas (upper panel) and stellar
  component (lower panel) are shown after being folded around the
  symmetry centre and deprojected by taking into account the disc
  inclination and the misalignment between the observed axis and the galaxy
  major axis.}
\label{fig:fold_lsb}  
\end{center}  
\end{figure*}  
}

\newcommand{\placeFigAone}{
\begin{figure}  
\begin{center}  
\centering\epsfig{figure=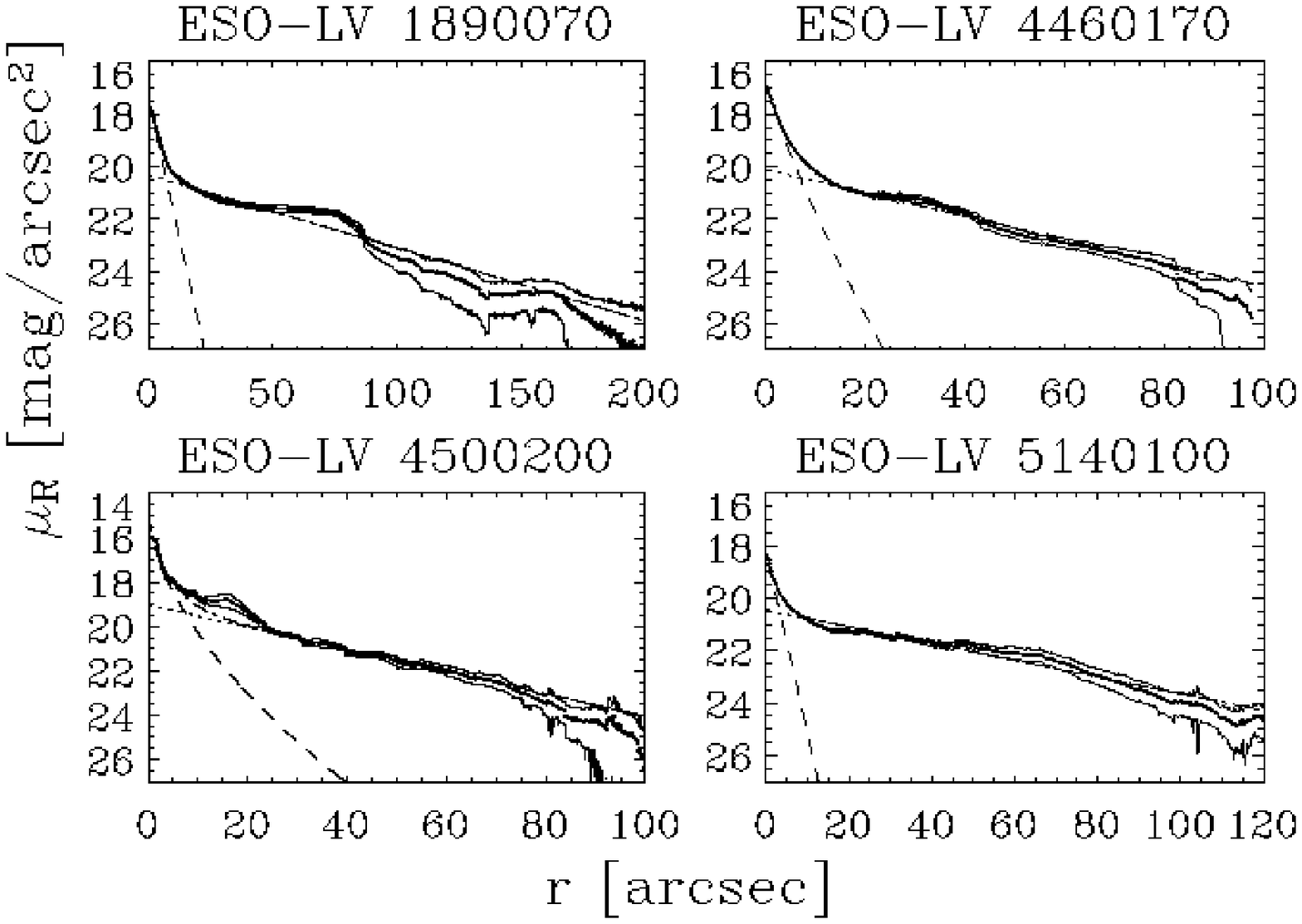, width=9cm}  
\caption{As in Fig.~\ref{fig:decomposition_lsb} but for the HSB galaxies.} 
\label{fig:decomposition_hsb}  
\end{center}  
\end{figure}  
}

\newcommand{\placeFigAtwo}{
\begin{figure*}  
\begin{center}  
 \begin{minipage}[b]{.49\linewidth}  
   \centering\epsfig{figure=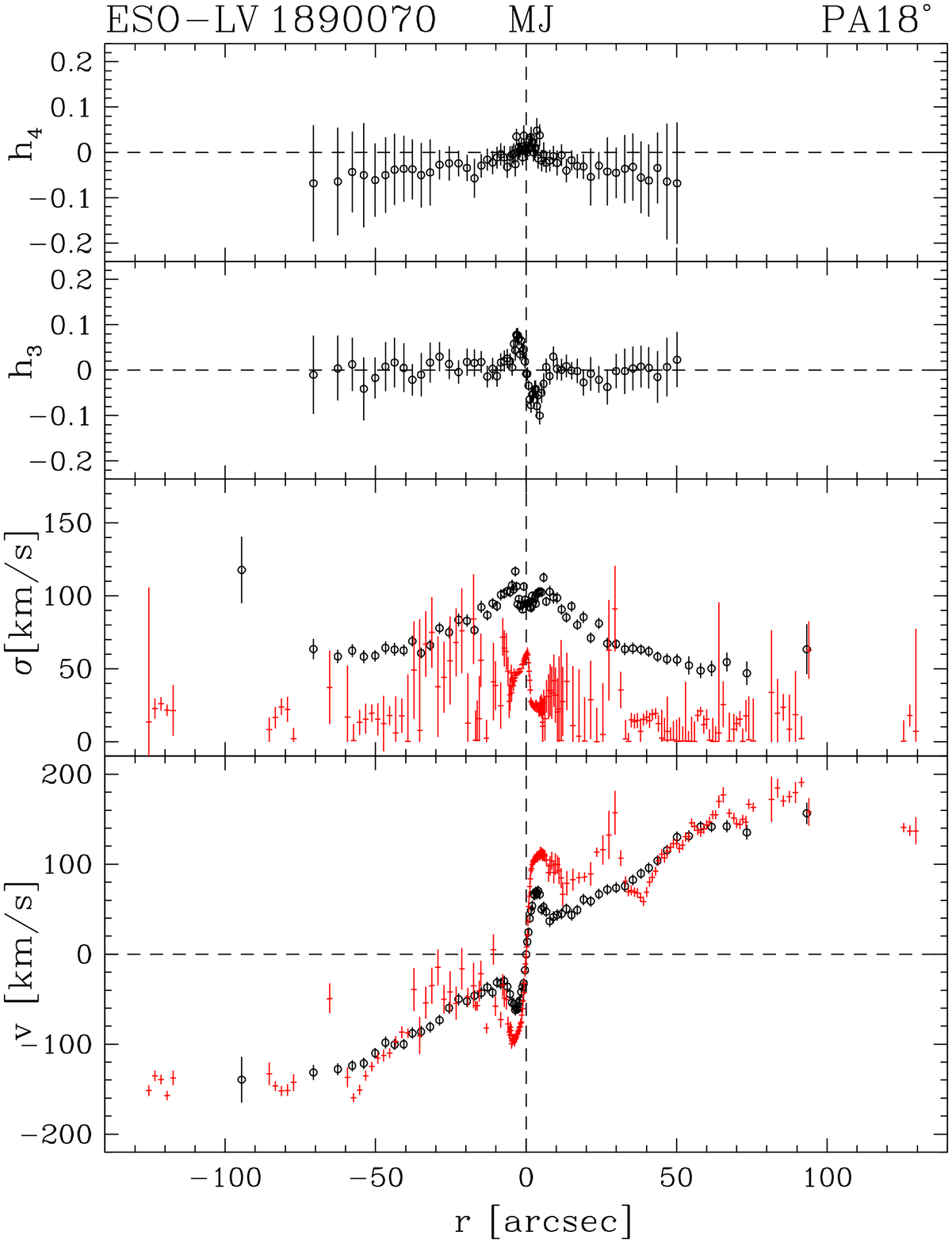,width=\linewidth}  
 \end{minipage} \hfill  
 \begin{minipage}[b]{.49\linewidth}  
   \centering\epsfig{figure=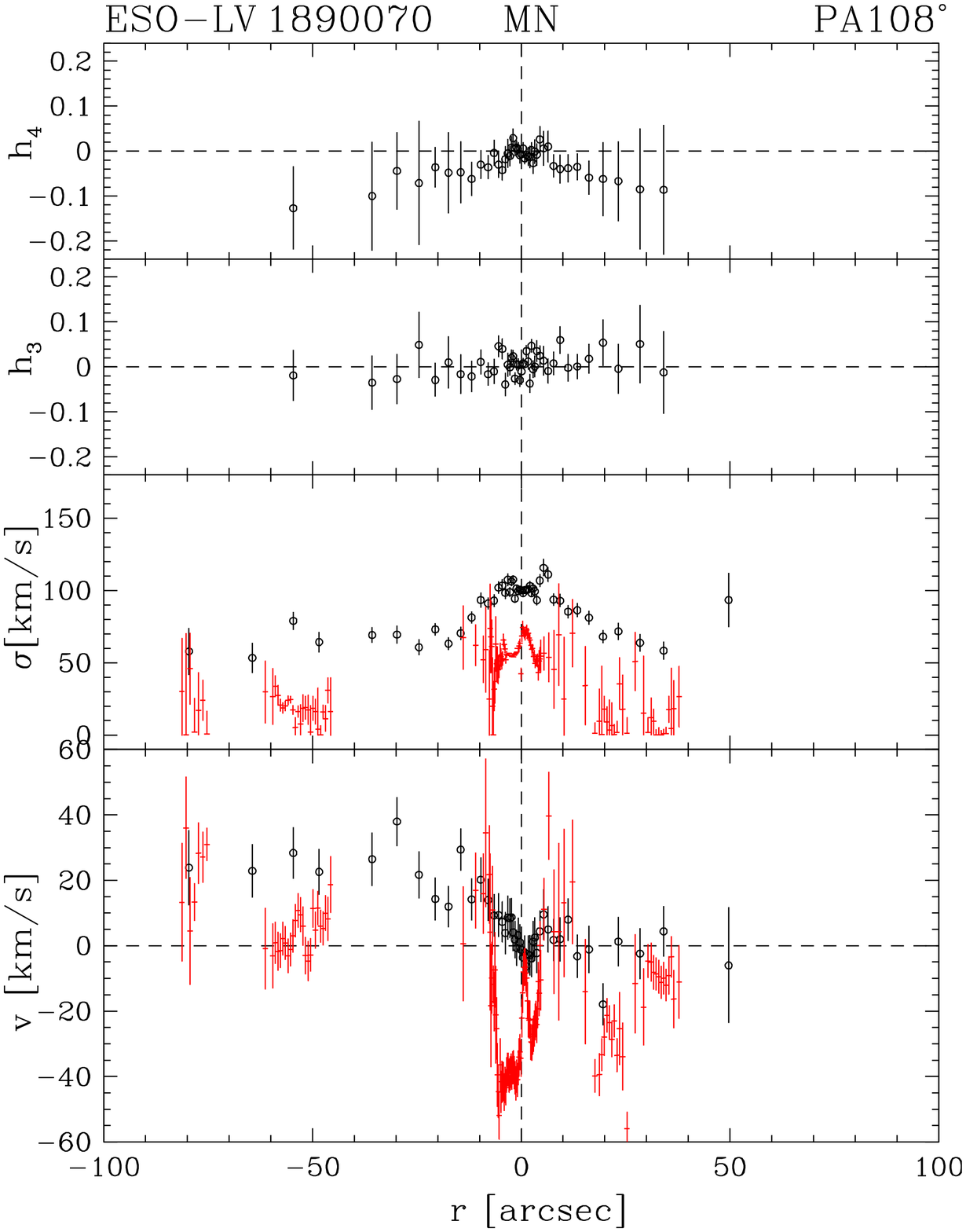,width=\linewidth}  
 \end{minipage}  
 \begin{minipage}[b]{.49\linewidth}  
   \centering\epsfig{figure=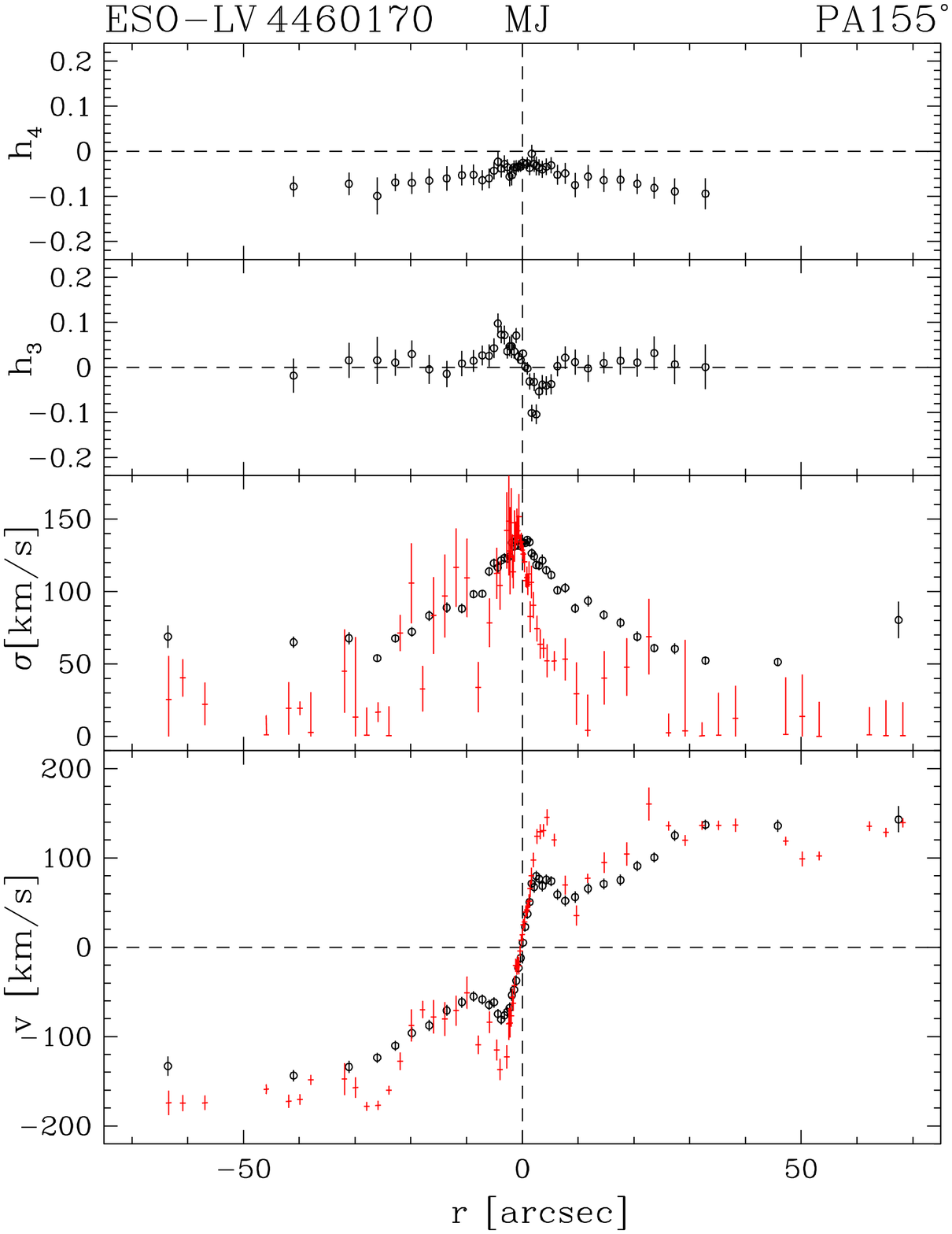,width=\linewidth}  
 \end{minipage} \hfill  
 \begin{minipage}[b]{.49\linewidth}  
   \centering\epsfig{figure=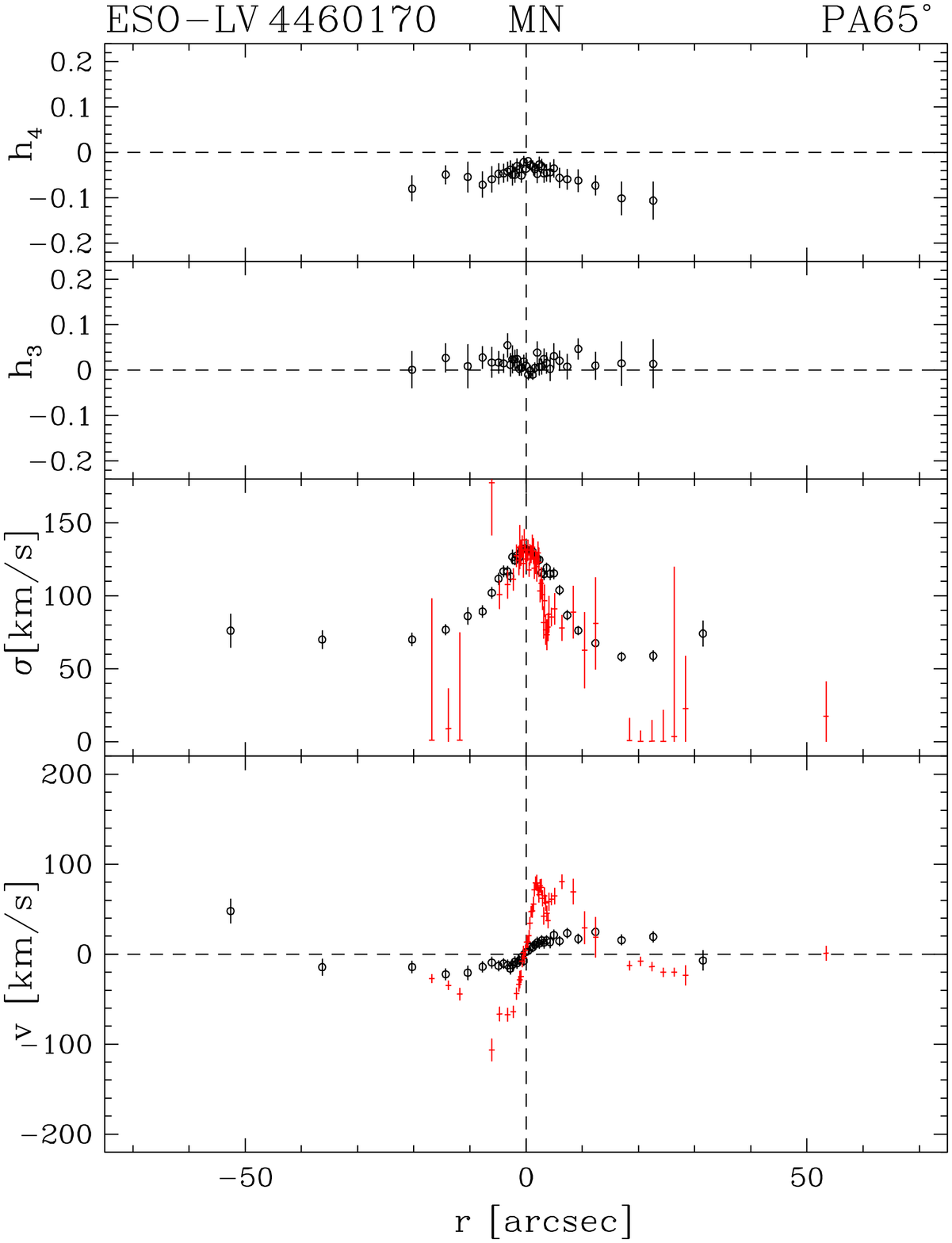,width=\linewidth}  
 \end{minipage} 
\caption{As in Fig.~\ref{fig:kinematics_lsb} but for the HSB galaxies.} 
\label{fig:kinematics_hsb}    
\end{center}  
\end{figure*}  
\begin{figure*}  
\addtocounter{figure}{-1}  
\begin{center}  
 \begin{minipage}[b]{.49\linewidth}  
   \centering\epsfig{figure=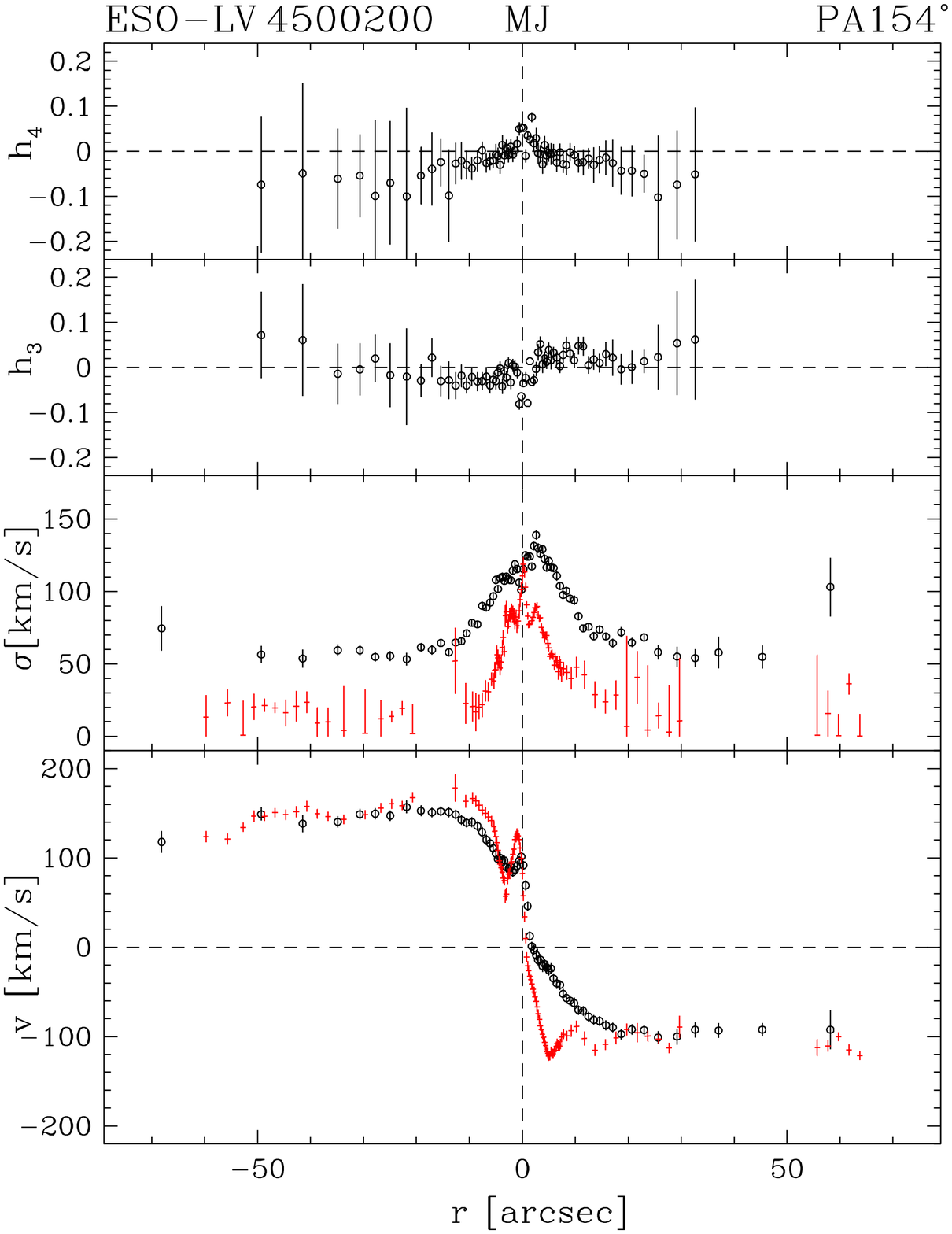,width=\linewidth}  
 \end{minipage} \hfill  
 \begin{minipage}[b]{.49\linewidth}  
   \centering\epsfig{figure=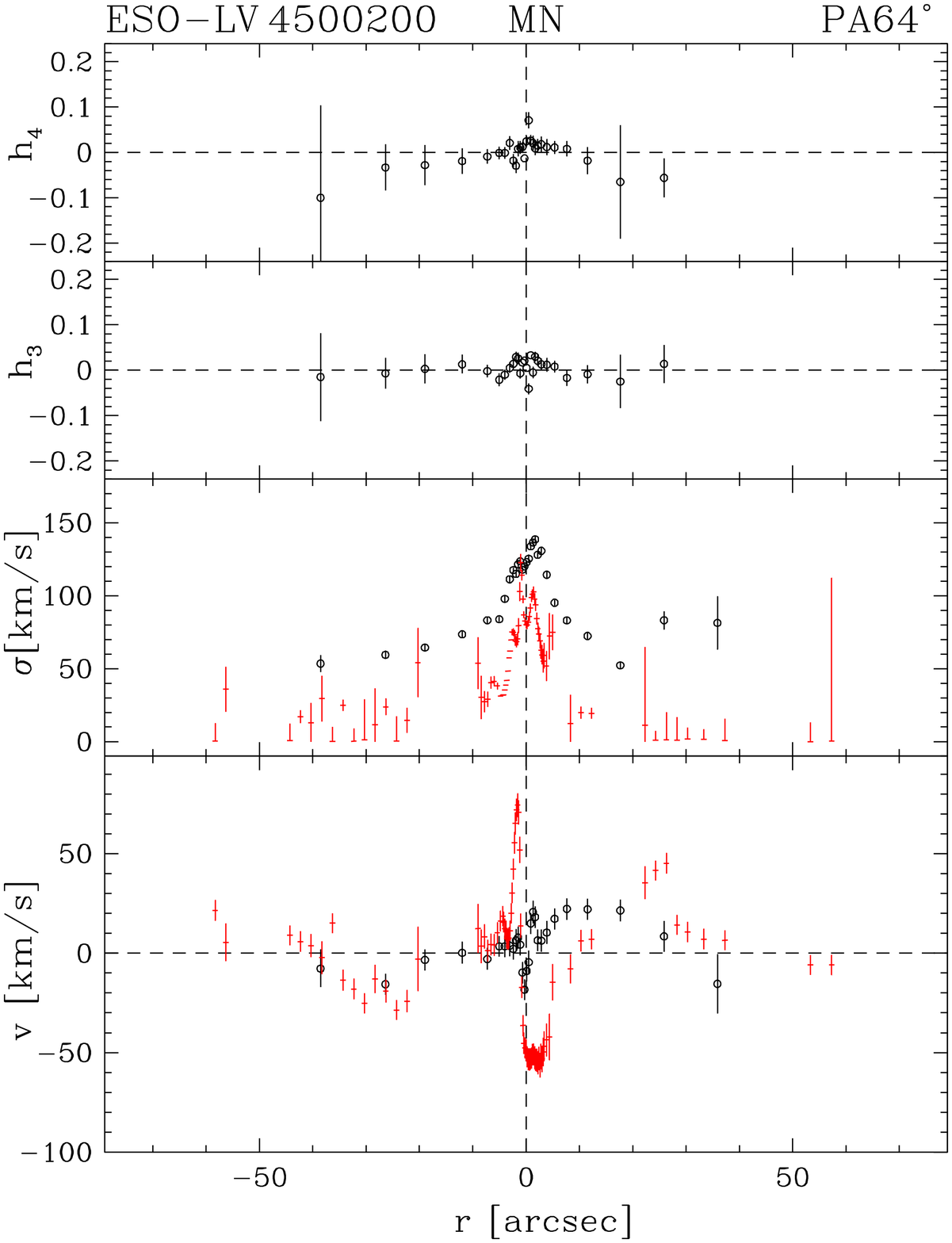,width=\linewidth}  
 \end{minipage}  
 \begin{minipage}[b]{.49\linewidth}  
  \centering\epsfig{figure=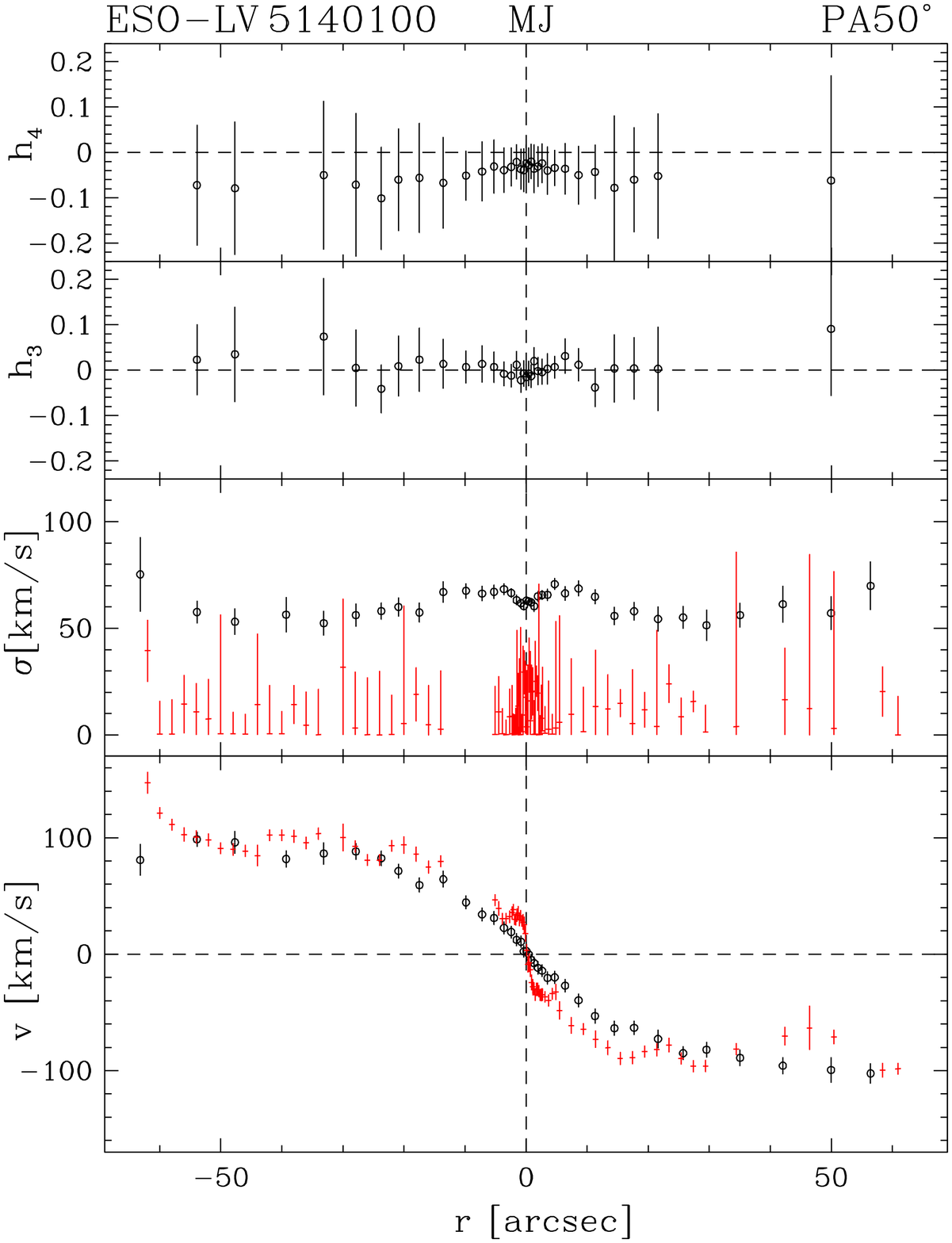,width=\linewidth}  
 \end{minipage} \hfill  
 \begin{minipage}[b]{.49\linewidth}  
  \centering\epsfig{figure=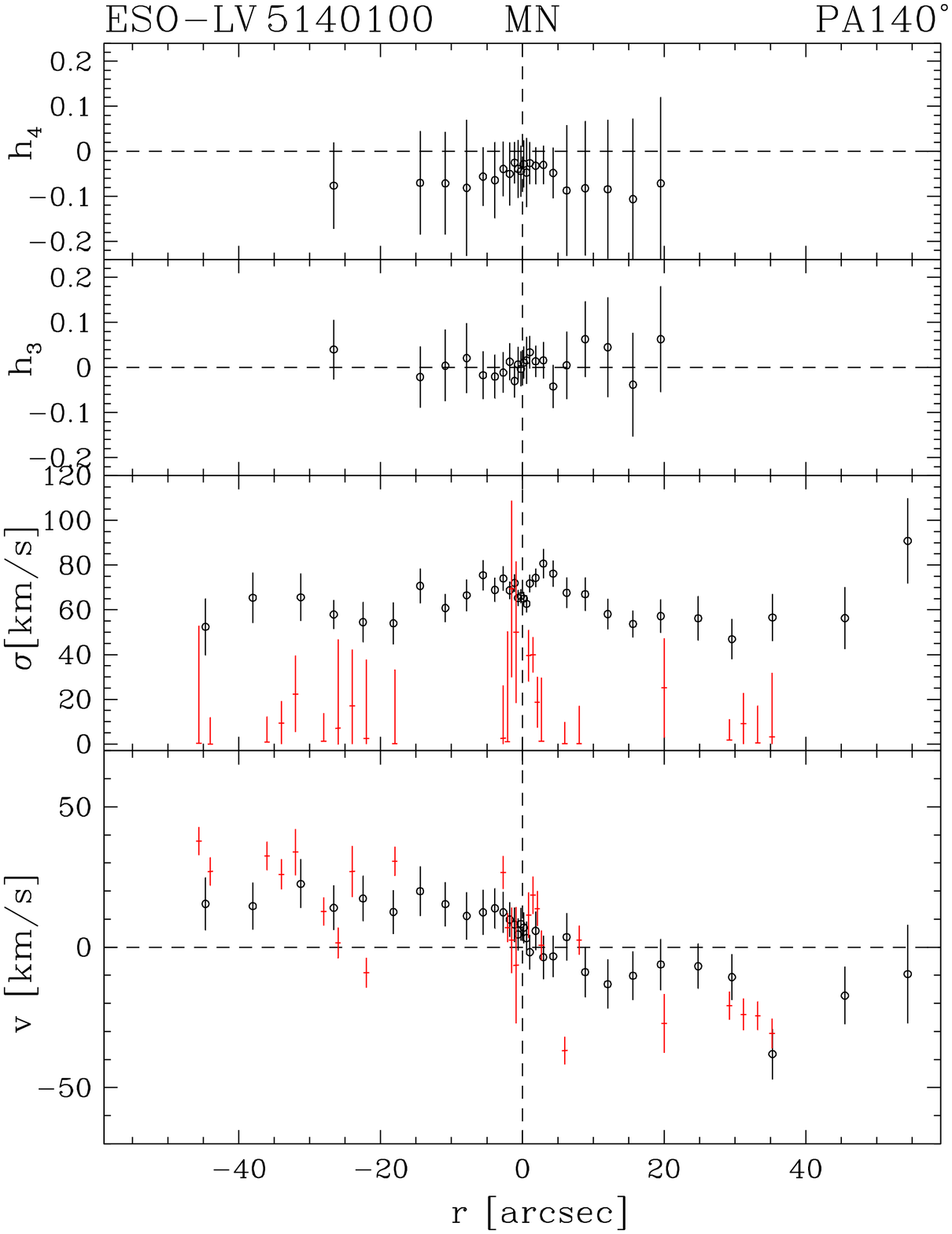,width=\linewidth}  
 \end{minipage}   
\caption{Continued}  
\end{center}  
\end{figure*}  
}

\newcommand{\placeFigAthree}{
\begin{figure*}  
\begin{center}  
 \begin{minipage}[b]{.49\linewidth} 
   \centering\epsfig{figure=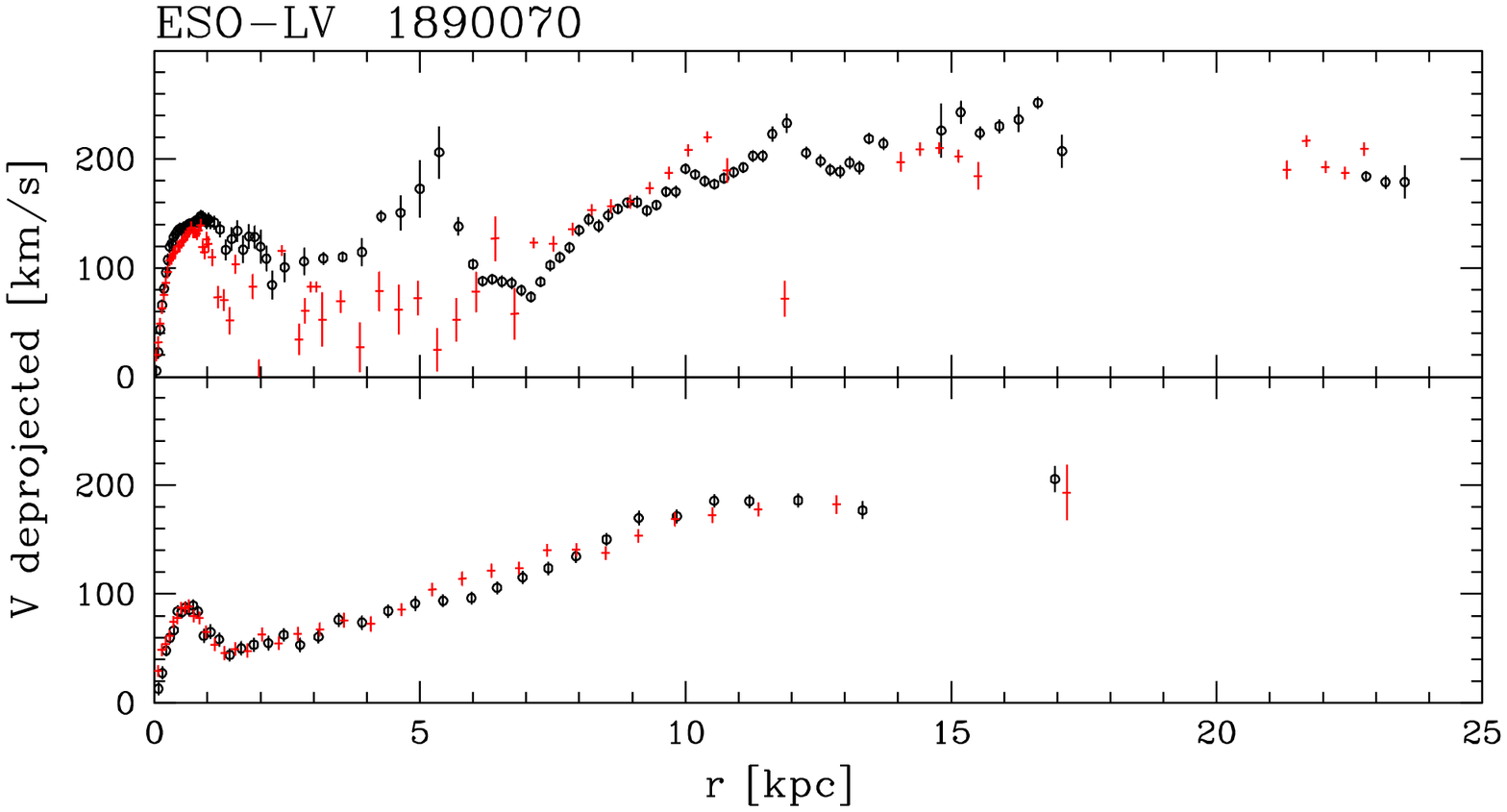,width=\linewidth} 
 \end{minipage} \hfill 
 \begin{minipage}[b]{.49\linewidth} 
   \centering\epsfig{figure=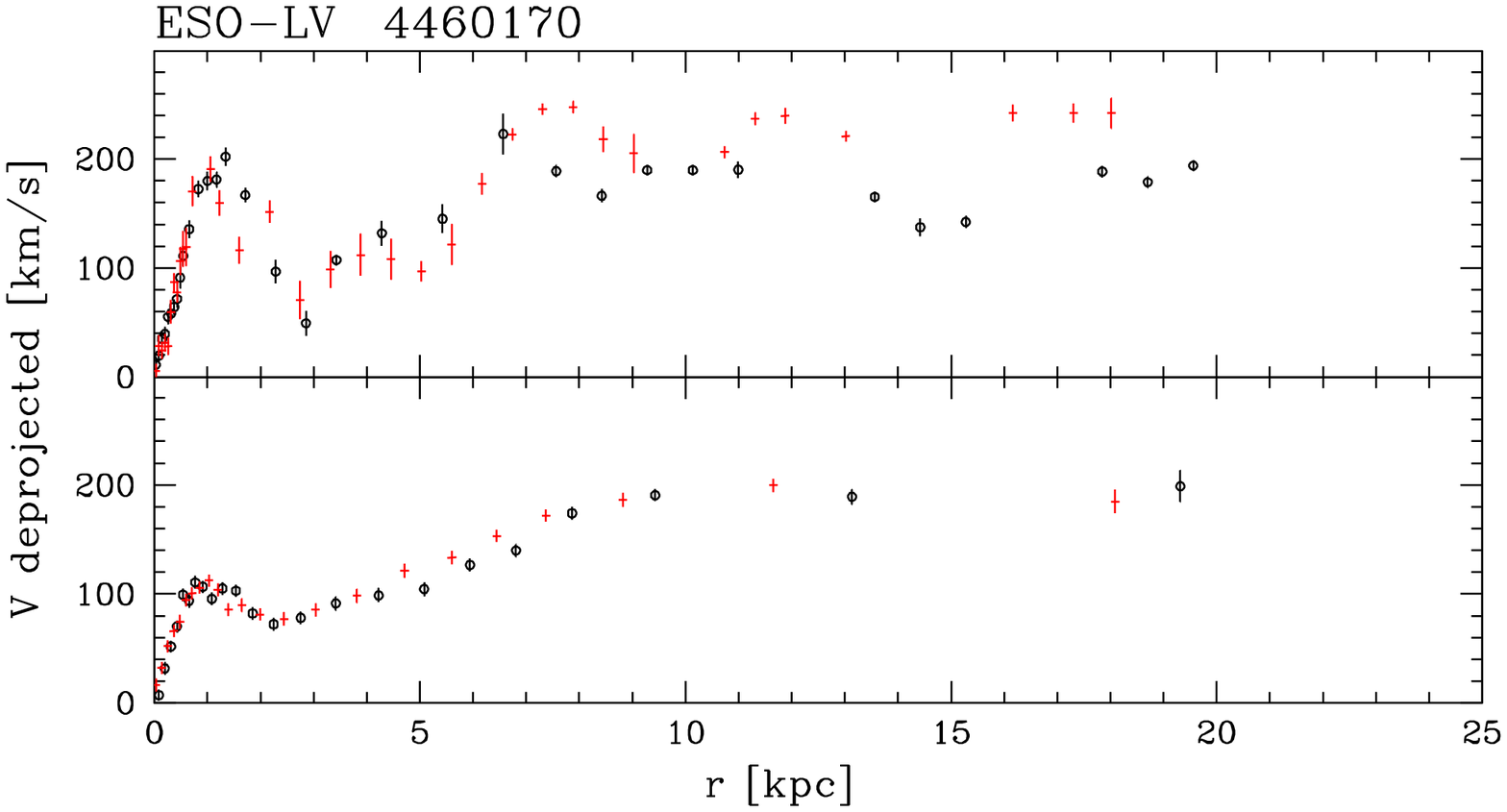,width=\linewidth}  
 \end{minipage} 
 \begin{minipage}[b]{.49\linewidth}  
   \centering\epsfig{figure=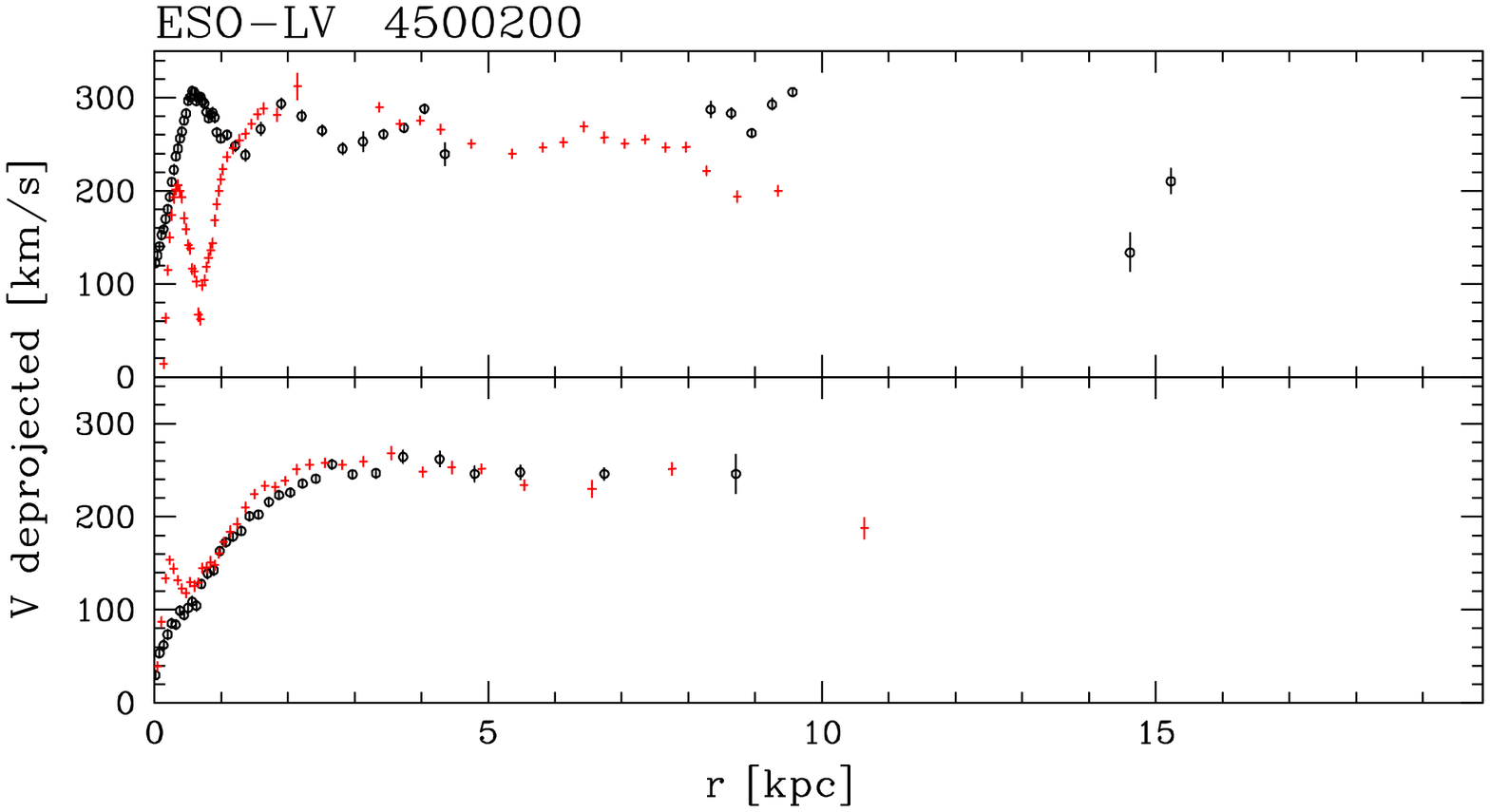,width=\linewidth}  
 \end{minipage} \hfill  
 \begin{minipage}[b]{.49\linewidth}  
   \centering\epsfig{figure=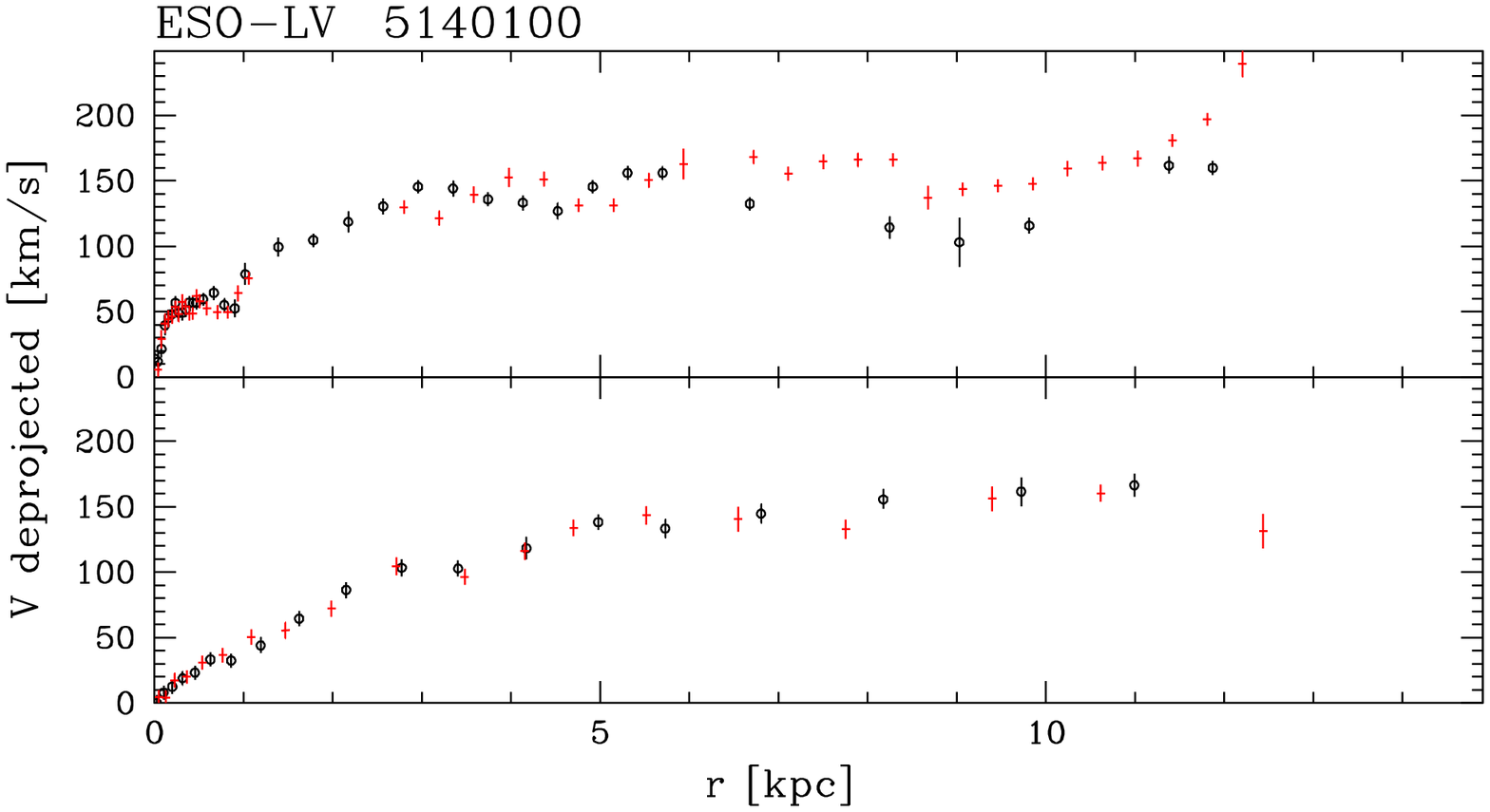,width=\linewidth}  
 \end{minipage}  
\caption{As in Fig.~\ref{fig:fold_lsb} but for the HSB galaxies.} 
\label{fig:fold_hsb}  
\end{center}  
\end{figure*}
}

\def\aj{AJ}                   
\def\araa{ARA\&A}             
\def\apj{ApJ}                 
\def\apjl{ApJ}                
\def\apjs{ApJS}               
\def\aap{A\&A}                
\def\aaps{A\&AS}              
\def\mnras{MNRAS}             
  
\def\kms{$\rm km\;s^{-1}$}    
\def\kmskpc{$\rm km\;s^{-1}\;kpc^{-1}$}       
\def\ha{H$\alpha$}   
\def\hb{H$\beta$}   
\def\oiii{[O~{\small III}]}    
\def\oisky{[O~{\small I}]$\,\lambda5577.3$}

\title[Kinematics of low surface brightness disc galaxies]{Structure and
dynamics of galaxies with a low surface brightness disc. I. The stellar and
ionised-gas kinematics\thanks{Based on observations collected at the
European Southern Observatory, Chile (ESO Prog. 067.B-0283,
069.B-0573, and 070.B-0171).}\thanks{Tables \ref{tab:starkin_lsb},
\ref{tab:gaskin_lsb}, \ref{tab:starkin_hsb}, and \ref{tab:gaskin_hsb}
are available only in electronic format}}
 
\author[A. Pizzella et al.]{A. Pizzella$^{1}$\thanks{E-mail:  
 alessandro.pizzella@unipd.it},   
 E. M. Corsini$^{1}$,  
 M. Sarzi$^{2}$,  
 J. Magorrian$^{3}$,  
 J. M\'endez-Abreu$^{1,5,6}$,  
 L. Coccato$^{4}$,  
\newauthor L. Morelli$^{1}$ 
 and F. Bertola $^{1}$\\  
$^1$ Dipartimento di Astronomia, Universit\`a di Padova, 
  vicolo dell'Osservatorio 3, I-35122 Padova, Italy.\\ 
$^2$ Centre for Astrophysics Research, University of Hertfordshire, 
College Lane, Hatfield, Herts AL10 9AB, United Kingdom.\\ 
$^3$ Theoretical Physics, Department of Physics, University of Oxford, 
1 Keble Road, Oxford OX1 3NP, United Kingdom.\\	 
$^4$ Max-Planck Institut f\"ur extraterrestrische Physik, 
  Giessenbachstrasse, D-85748 Garching, Germany.\\ 
$^5$ INAF-Osservatorio Astronomico di Padova, 
  vicolo dell'Osservatorio 5, I-35122 Padova, Italy.\\ 
$^6$ Universidad de La Laguna, Av. Astrof\'{\i}sico Francisco S\'anchez s/n, 
  E-38206 La Laguna, Spain.\\ 
}

\begin{document}  
  
\date{Accepted ... Received ...; in original ...}  
  
\pagerange{\pageref{firstpage}--\pageref{lastpage}} \pubyear{2008}  
  
\maketitle  
  
\label{firstpage}  
  
\begin{abstract}  
Photometry and long-slit spectroscopy are presented for a sample of 6
galaxies with a low surface brightness stellar disc   and a bulge  
The characterising parameters of the bulge and disc components
were derived by means of a two-dimensional photometric
decomposition of the images of the sample galaxies. Their
surface-brightness distribution was assumed to be the sum of the
contribution of a S\'ersic bulge and an exponential disc, with
each component being described by elliptical and concentric isophotes
of constant ellipticity and position angle.
The stellar and ionised-gas kinematics were measured along the
major and minor axis in half of the sample galaxies, whereas the other
half was observed only along two diagonal axes. Spectra along two
diagonal axes were obtained also for one of the objects with major and
minor axis spectra.
The kinematic measurements extend in the disc region out to a
surface-brightness level $\mu_R\approx24$ mag$\,$arcsec$^{-2}$,
reaching in all cases the flat part of the rotation curve.
The stellar kinematics turns out to be more regular and symmetric
than the ionised-gas kinematics, which often shows the presence of
non-circular, off-plane, and non-ordered motions. This   raises   the
question about the reliability of the use of the ionised gas as the
tracer of the circular velocity in the modeling of the mass
distribution, in particular in the central regions of low surface
brightness galaxies.

\end{abstract}  
  
  
\section{Introduction}  
 
Low surface brightness (LSB) galaxies are operatively
defined as galaxies with a central face-on surface brightness fainter
than 22.6 $B-$mag$\,$arcsec$^{-2}$. Therefore they are more difficult
to identify and study than their high surface brightness (HSB)
counterparts. Most but not all of them are dwarf galaxies
\citep{Schombert1988, Schombert1992, Impey1996}.
Giant LSB galaxies are fairly well described as exponential discs with
no significant bulge components \citep{Romanshin1983, McGaugh1994,
Sprayberry1995}. They fall within the range of luminosities defined by
HSB discs. At fixed luminosity, the LSB discs have lower central
surface brightnesses and larger scale lengths than the discs of HSB
galaxies, although none of them has structural parameters as extreme
as the prototype Malin 1 \citep{Bothun1987}. It is apparent that
LSB galaxies exist in a wide range of morphological types. Most LSB
galaxies discovered so far are bulgeless, but there are also
galaxies with LSB discs that have a significant bulge component
\citep{Beijersbergen1999}.

\placeTabone 

The amount and distribution of the dark matter (DM) in galaxies
are usually determined using the gas rotation curve and through the
study of an excess of galactic rotation compared to what can be
induced by the luminous matter, the contribution of which is generally
estimated by adjusting the mass-to-light ratio until the central
velocity gradient can be matched.
The structure of the DM halo is more directly revealed in galaxies
where the luminous component gives a nearly zero contribution to the
mass budget, because this makes mass modeling easier and the derived
DM distribution less uncertain. This is the case of the giant LSB
galaxies.
They are DM dominated at all radii when a stellar mass-to-light ratio
consistent with population synthesis modeling is adopted to fit the
observed rotation curves \citep{Swaters2000}. These results are in
agreement with the findings of \citet{Zwaan1995}, who realized that
LSB galaxies follow the same Tully-Fisher relation as HSB
galaxies. They argued that this implies that the mass-to-light ratio
of LSB galaxies is typically a factor of 2 larger than that of normal
galaxies of the same total luminosity and morphological type. For
these reasons, LSB galaxies were considered ideal targets for studying
the properties of DM halos and for testing if they have a central
cuspy power-law mass density distribution as predicted by cold dark
matter (CDM) cosmology \citep{Navarro1996, Navarro1997, Moore1998,
Moore1999}.
 
Different methods   are   used to measure the inner slope of the DM
density profile in LSB galaxies: fitting a mass model which takes into
account the contribution of stars, gas and DM to the rotation curve
\citep{vdBosch2000, vdBosch2001, Swaters2003a}, fitting a power law to
the rotation curve \citep{Simon2003}, and fitting a power law to the
density distribution derived by inverting the rotation curve
\citep{deBlok2001a, deBlok2002, Swaters2003a}.
For most galaxies the inner density slope is unconstrained. Although
DM halos with constant density cores generally provide better fits,  
cuspy haloes can sometimes be consistent with the data.  
Systematic errors of long-slit (e.g., not-perfect positioning of the
slit on the galaxy nucleus and errors on the position angle of the
disc) and radio (e.g., beam smearing) observations   may   lead to an
underestimate of the inner slope and contribute to the total
uncertainties on the interpretation of the observed kinematics
\citep{Swaters2003a,Swaters2004}   although only constant density cores
can reproduce some observations \citep{deBlok2001b} .  
 
High-resolution two-dimensional (2D) velocity fields obtained
with integral-field spectroscopy \citep{Bolatto2002, Swaters2003b,
Kuzio2006, Pizzella2008} remove most of systematic errors. But, they
are not conclusive. The analysis of these data in the inner regions of
most of the observed galaxies reveals that the gas is not moving onto
circular orbits in a disc coplanar to the stellar one
\citep[e.g.,][]{Swaters2003b, Rhee2004, Hayashi2006}.
It is a common phenomenon in the centres of HSB galaxies too. In fact,
 non-circular \citep[e.g.,][]{Gerhard1989, Berman2001},
pressure-supported \citep[e.g.,][]{Fillmore1986, Kormendy1989,
Bertola1995, Cinzano1999}, and off-plane \citep[e.g.,][]{Corsini1999,
Pignatelli2001, Corsini2003} gas motions are often observed in
unbarred HSB galaxies for which dynamical modeling allows a direct
comparison between the gas and circular velocity. Recently, \citet{Christlein2008}
found kinematic anomalies in the outer disk region of nearby disc galaxies.

This poses a question about the reliability of central mass
distribution derived in both LSB and HSB galaxies from the analysis of
the gas rotation curves. To address this issue in LSB galaxies a
crucial piece of information, which has been missed so far, is needed:
the stellar kinematics.
This is the first of series of paper aimed at investigating the mass
distribution of galaxies with LSB discs by means of dynamical
models based on both stellar and gaseous kinematics. The paper is
organized as follows.
The galaxy sample is presented in \S~\ref{sec:sample}. The photometric
and spectroscopic observations are described in
\S~\ref{sec:observations}. The structural parameters of the bulge and
disc of the sample galaxies are derived by analysing their 2D surface
brightness distribution in \S~\ref{sec:decomposition}. The stellar and
ionised-gas kinematics are measured from the long-slit spectra out to
0.5--1.5 optical radii with high signal-to-noise ratio in
\S~\ref{sec:kinematics}. The results and conclusions are given in
\S~\ref{sec:conclusions}.
Dynamical modeling of the data is deferred to a companion paper
(Magorrian et al, in preparation)

\section{Galaxy sample}  
\label{sec:sample}  
 
We set out to find LSB discs in spiral galaxies without
necessarily excluding the presence of a bulge. 
  We included both barred and unbarred galaxies  
An initial sample of 10
target galaxies was initially selected on the basis of the central
disc surface-brightness estimate derived from The Surface Photometry
Catalog of the ESO-Uppsala Galaxies \citep[ESO-LV]{ESOLV}.
An exponential law was fitted to the surface-brightness radial profile
at large radii, where the light distribution of the galaxy is expected
to be dominated by the disc contribution.  The fit was done using a
non-linear least-squares minimization method. It was based on the
robust Levenberg-Marquardt method \citep[e.g.][]{Press1996}
implemented by \citet{More1980}. The actual computation was done using
the MPFIT algorithm implemented by C.~B. Markwardt under the IDL
environment\footnote{The updated version of this code is available on
http://cow.physics.wisc.edu/~craigm/idl/idl.html}.
All the sample galaxies are classified as spiral galaxies with a
morphological type ranging from Sa to Sm \citep[RC3]{RC3}. The
galactic latitudes were chosen so as to minimize the foreground
extinction ($|b|>15^\circ$, RC3). The inclinations were selected to be
larger than $30^\circ$ (ESO-LV) to allow a reliable measurement of the
stellar and ionised-gas kinematics. Finally, the galaxies were
chosen to be morphologically undisturbed and have a clean stellar
foreground to not complicate the photometric and spectroscopic
analysis.

This sample was subsequently culled to 6 galaxies following a
scrupulous 2D photometric decomposition of our own images (see
\S~\ref{sec:decomposition}), which led to find that 4 targets were in
fact hosting HSB discs. The basic properties, photometry, and
kinematics of these HSB spiral galaxies are given in
App.~\ref{sec:hsb}.
In the rest of the paper we discuss only the 6 galaxies which host a
LSB disc. Our selection criteria are similar to those of
\citet{Beijersbergen1999}. The result of this selection is that the
galaxies in our sample host HSB bulges with a central face-on
surface-brightness brighter than 22.6 $B-$mag arcsec$^{-2}$  
and may have a significant bulge component as compared to typical LSB
galaxy samples. The presence of a bulge is unavoidable when we sample
both early and late type spirals  .
The sample galaxies and their basic properties are listed in Table
\ref{tab:sample_lsb}.
 
\placeTabtwo 
\placeTabthree
\placeTabfour 

\section{Observations and data reduction}  
\label{sec:observations} 
 
\subsection{Broad-band imaging and long-slit spectroscopy}  
\label{sec:setup}  
 
The photometric and spectroscopic data of the sample galaxies were
obtained at the Very Large Telescope (VLT) of the European Southern
Observatory (ESO) at Paranal Observatory during three different runs
between 2001 and 2003. The observations were carried out in service
mode on May 05 -- July 03, 2001 (run 1), April 13 -- June 03, 2002
(run 2), and October 30, 2002 -- February 04, 2003 (run 3).
 
The Focal Reducer Low Dispersion Spectrograph 2 (FORS2) mounted the 
volume-phased holographic grism GRIS\_1400V18 with 1400 grooves 
mm$^{-1}$ and the 1 arcsec $\times$ 6.8 arcmin slit. 
In run 1 the detector was a Site SI-424A CCD with $2048\times2048$
pixels of $24\times24$ $\mu$m$^2$. The wavelength range from 4750 to
5800 \AA\ was covered with a reciprocal dispersion of 0.51\AA\
pixel$^{-1}$ and a spatial scale of 0.20 arcsec pixel$^{-1}$.
In run 2 and 3 the detector was a mosaic of two MIT/LL CCDID-20 $\rm
2k \times 4k$ CCDs   with a gap of about 4 arcsec along the
spatial direction.  . Each CCD had $2048\times4096$ pixels of
$15\times15$ $\mu$m$^2$. The wavelength range from 4630 to 5930 \AA\
was covered with a reciprocal dispersion of 0.65\AA\ pixel$^{-1}$ and
a spatial scale of 0.250 arcsec pixel$^{-1}$ after a $2\times2$ pixel
binning.
 
Major and minor-axis spectra were obtained for half of the sample
galaxies, whereas the other half was observed only along two diagonal
axes. Spectra along two diagonal axes were obtained also for
ESO-LV~5340200, which has major and minor axis spectra as well.  The
integration time of the galaxy spectra ranged between 2400 and 3050
s. The log of the spectroscopic observations as well as total
integration times and slit position angle are given in Table~2.
%
At the beginning of each exposure, the galaxy was centred on the
slit. The acquisition images were obtained with no filter in
run~1. The Site CCD efficiency curve roughly corresponded to a broad
$V$ band. In run 2 and 3 the acquisition images were obtained with the
Gunn $z-$band filter ($\lambda_{\rm eff}=9100$ \AA , $\rm FWHM=1305$
\AA). The total integration times are listed in Table
\ref{tab:imalog_lsb}. In run 2 and 3 an offset of 10 arcsec along the
direction of the slit was applied between repeated exposures in order
to deal with the gap between the two CCDs in the mosaic spectrum.
 
In run 1 and 3 a number of spectra of giant stars with spectral type 
ranging from G to K were obtained to be used as templates in measuring 
stellar kinematics. Integration times and spectral type the template 
stars as well as the log of the observations are given in Table 
\ref{tab:templog}. 
Additionally, at least one flux standard star per night was observed 
to calibrate the flux of the spectra. The standard calibration frames 
(i.e., biases, darks, and flat-field spectra) as well as the spectra of 
the comparison arc lamp were taken in the afternoon before each 
observing night. 
The mean value of the seeing FWHM during the spectroscopic exposures
as measured by fitting a 2D Gaussian to the guide star ranged from 0.8
to 2.7 arcsec.
 
\placeFigone
\placeFigtwo

\subsection{Reduction of the photometric data}  
\label{sec:photometry}  
 
All the images were reduced using standard MIDAS\footnote{MIDAS is
developed and maintained by the European Southern Observatory.}
routines. First, the images were bias subtracted and then they were
corrected for pixel-to-pixel intensity variations by using a mean flat
field for each night.
The sky-background level was removed by fitting a second-order 
polynomial to the regions free of sources in the images.  
The different frames of each galaxy were rotated, shifted, and aligned
to an accuracy of a few hundredths of a pixel using common field stars
as reference. After checking that their point spread functions (PSFs)
were comparable, the frames were combined to obtain a single image.
The cosmic rays were identified and removed using a sigma-clipping 
rejection algorithm. 
2D Gaussian fits to the field stars in the resulting images yielded
the final FWHM measurement of seeing PSF listed in
Table~\ref{tab:imalog_lsb}.
Finally, the mean residual sky level was measured in each combined and 
sky-subtracted image, and the error on the sky subtraction was 
estimated. 
The median value of the residual sky level was determined in a large 
number of $5\times5$ pixel areas. These areas were selected in empty 
regions of the frames, which were free of objects and far from the 
galaxy to avoid the contamination of the light of field stars and 
galaxies as well as of the target galaxy itself. The mean of these 
median values was zero, as expected. For the error in the sky 
determination we adopted half of the difference between the maximum 
and minimum of the median values obtained for the sampled areas. 
 
For each galaxy, we derived a `luminosity growth curve' by measuring
the integrated magnitudes within circular apertures of increasing
radius by means of the IRAF\footnote{IRAF is distributed by the
National Optical Astronomy Observatories which are operated by the
Association of Universities for Research in Astronomy under
cooperative agreement with the National Science Foundation.} task
ELLIPSE within the STSDAS package.
Photometric calibration was performed by fitting this growth curve to
that given in $R-$band for the same circular apertures by ESO-LV. The
surface-brightness radial profiles of the sample galaxies are shown in
Fig.~\ref{fig:decomposition_lsb}.
According to the color gradients measured by \citet{Beijersbergen1999}
in a sample of bulge-dominated LSB galaxies, we estimated that the
accuracy of the absolute calibration was better than 0.3 mag
arcsec$^{-2}$. The $R-$band radial of ESO-LV~2060140 by
\citet{Beijersbergen1999} was used to test the accuracy of our
calibration (Fig. \ref{fig:imacomparison}).
 
\placeTabfive

\subsection{Reduction of the spectroscopic data}   
\label{sec:spectroscopy}   
 
All the spectra were bias subtracted, flat-field corrected, cleaned of 
cosmic rays, corrected for bad columns, and wavelength calibrated 
using standard MIDAS routines. 
 
The flat-field correction was performed by means of a quartz lamp, 
which were normalized and divided into all the spectra, to correct for 
pixel-to-pixel sensitivity variations and large-scale illumination 
patterns due to slit vignetting. 
The cosmic rays were identified by comparing the photon counts in each 
pixel with the local mean and standard deviation and eliminated by 
interpolating over a suitable value. The residual cosmic rays were 
eliminated by manually editing the spectra. 
 
The wavelength calibration was performed by means of the MIDAS package 
XLONG. Each spectrum was re-binned using the wavelength solution 
obtained from the corresponding arc-lamp spectrum. We checked that the 
wavelength re-binning had been done properly by measuring the 
difference between the measured and predicted wavelengths of about 20 
of un-blended arc-lamp lines which were distributed over the whole 
spectral range of a wavelength-calibrated spectrum. The resulting rms 
about the dispersion solution was 0.10 \AA\ corresponding to an 
accuracy in the wavelength calibration of 6 \kms\ at 5280 \AA . 
The instrumental resolution was derived as the mean of the Gaussian 
FWHMs measured for the same un-blended arc-lamp lines we measured for 
assessing the accuracy of wavelength calibration. The mean FWHM of the 
arc-lamp lines was $2.04\pm0.08$ \AA , $2.11\pm0.07$ \AA , and 
$2.10\pm0.07$ \AA\ for run 1, 2, and 3, respectively. In run 1 the 
$1.0$-arcsec wide slit was projected onto 6 pixels. In run 2 and 3 it 
was projected onto 4 pixels. Therefore, the sampling of the arc-lamp 
lines was lower than in run 1. Nevertheless, FWHM values of the 
different runs are in agreement within the errors.  They correspond to 
an instrumental velocity dispersion $\sigma_{\rm inst} = 50$ \kms\ at 
5280 \AA . 

All the spectra were corrected for the misalignment between the CCD
mosaic and slit following \citet{Bender1994}. The spectra obtained for
the same galaxy along the same axis were coadded using the centre of
the stellar continuum as a reference. This allowed to improve the
signal-to-noise ratio ($S/N$) of the resulting 2D spectrum.
The accuracy in the wavelength calibration and instrumental velocity
dispersion as a function of position along the slit were derived from
the brightest night-sky emission line in the observed spectral range
of the coadded 2D spectra of the galaxies.
This was the \oisky\ emission line. Its 
velocity curve and velocity dispersion profile were measured along the 
full slit extension as done in \citet{Corsini1999}. The central 
wavelength and FWHM of the night-sky line were evaluated fitting at 
each radius a Gaussian to the emission line and a straight line to its 
adjacent continuum. Then the values were converted to velocity and 
velocity dispersion, respectively. The velocity curve was fitted by 
quadratic polynomial by assuming the rms of the fit to be the 
$1\sigma$ velocity error. No velocity gradient was found and the fit 
rms was 5 \kms .  
The instrumental velocity dispersion derived from the FWHM of the 
\oisky\ line increased from 46 \kms\ at the slit centre to 53 
\kms\ at the slit edges.    
Both the error on the wavelength calibration and instrumental velocity
dispersion measured from the \oisky\ line in the coadded spectra of
the galaxies are consistent with those derived from the lamp lines in
the arc spectra.
 
A one-dimensional spectrum was obtained for each kinematic template 
star as well as for each flux standard star. The spectra of the 
kinematic templates were de-redshifted to laboratory wavelengths. 
In the galaxy and stellar spectra the contribution of the sky was 
determined by interpolating along the outermost $60$ arcsec at the two 
edges of the slit, where the target light was negligible, 
and then subtracted. The galaxy spectra always extended less than $90$ 
arcsec from the centre. A sky subtraction better than $1\%$ was achieved. 

\section{Structural parameters}     
\label{sec:decomposition} 

The structural parameters of the sample galaxies were derived by
applying a 2D bulge-disc photometric decomposition to their images. To
this aim the GASP2D algorithm by \citet{Mendez2008} was used.
 
The galaxy surface-brightness distribution was assumed to be the sum 
of the contributions of a bulge and a disc component. 
The S\'ersic law \citep{Sersic1968} was adopted to describe the surface 
brightness of the bulge component. It is parametrized by $r_{\rm e}$, 
$I_{\rm e}$ and $n$ which are respectively the effective radius, 
surface brightness at $r_{\rm e}$, and a shape parameter describing 
the curvature of the radial profile. The bulge isophotes are ellipses 
centred on galaxy centre $(x_0,y_0)$, with constant position angle 
PA$_{\rm b}$ and constant axial ratio $q_{\rm b}$. 
The surface brightness distribution of the disc component was assumed 
to follow an exponential law \citep{Freeman1970} with $h$ and $I_0$ 
the scale length and central surface brightness, respectively.  The 
disc isophotes are ellipses centred on $(x_0,y_0)$, with constant 
position angle PA$_{\rm d}$ and constant axial ratio $q_{\rm d}$. 
 
Since the fitting algorithm of GASP2D is based on a $\chi^2$ 
minimization, it was important to adopt initial trials for free 
parameters as close as possible to their actual values.  To this aim 
the ellipse-averaged radial profiles of surface brightness, 
ellipticity, and position angle were analysed by following the 
prescriptions by \citet{Mendez2008}. 
Isophote-fitting with ellipses, after masking foreground stars, 
background galaxies, and residual bad columns, was carried out using 
ELLIPSE. In all cases, first ellipses were fitted by allowing their 
centres to vary. Within the errors, no variation in the ellipse 
centres was found for all the galaxies studied in this paper. The 
final ellipse fits were done at fixed ellipse centres. The 
ellipse-averaged profiles of surface brightness, ellipticity, and 
position angle were measured. For all the sample galaxies the 
ellipse-averaged profiles of surface brightness are shown in 
Fig. \ref{fig:decomposition_lsb}. 
The trial values of $I_{\rm e}$, $r_{\rm e}$, $n$, $I_0$, and $n$ 
were obtained by performing a standard photometric decomposition on the 
ellipse-averaged surface-brightness profile. 
The trial values of PA$_{\rm b}$ and $q_{\rm b}$ were found by 
interpolating at $r_{\rm e}$ the ellipse-averaged position-angle and 
ellipticity profiles, respectively.  The trial values of PA$_{\rm d}$ 
and $q_{\rm d}$ were obtained by fitting a constant to the outer 
portion of the ellipse-averaged position-angle and ellipticity 
profiles, respectively. The coordinates of the maximum galaxy surface 
brightness were taken as trial values of the coordinates $(x_0,y_0)$ 
of the galaxy centre. 
 
\placeTabsix

Starting from these initial, trial parameters the bulge-disc
model of the surface brightness was fitted iteratively by GASP2D to
the pixels of the galaxy image to derive the photometric parameters of
the bulge ($I_{\rm e}$, $r_{\rm e}$, $n$, PA$_{\rm b}$, and $q_{\rm
b}$) and disc ($I_0$, $h$, PA$_{\rm d}$, and $q_{\rm d}$) and the
position of the galaxy centre $(x_0,y_0)$. The MPFIT algorithm based
on a robust Levenberg-Marquardt method of non-linear least-squares
minimization was used in the IDL environment.
Each image pixel has been weighted according to the variance of its 
total observed photon counts due to the contribution of both galaxy 
and sky, and determined assuming photon noise limitation and taking 
the detector read-out noise into account. 
The seeing effects were taken into account by convolving the model 
image with a circular Gaussian PSF with the FWHM measured using the 
stars in the galaxy image (Table \ref{tab:imalog_lsb}). The convolution 
was performed as a product in Fourier domain before the least-squares 
minimization. 
 
The parameters derived for the structural components of the sample
galaxies are collected in Table \ref{tab:parameters_lsb}. No correction
for galaxy inclination was applied. The result of the photometric
decomposition of the surface brightness distribution of the sample
galaxies is shown in Fig. \ref{fig:decomposition_lsb}.
 
The formal errors obtained from the $\chi^2$ minimization method are 
not representative of the real errors in the structural parameters 
\citep{Mendez2008}. Therefore, the errors given in Table 
\ref{tab:parameters_lsb} were obtained through a series of Monte Carlo 
simulations. A set of 3000 images of galaxies with a S\'ersic bulge 
and an exponential disc was generated. The structural parameters of 
the artificial galaxies were randomly chosen among the following 
ranges 
 
\begin{equation} 
0.5 \leq r_{\rm e} \leq 3\ {\rm kpc} \qquad   
0.5 \leq q_{\rm b} \leq 0.9 \qquad  
0.5 \leq n \leq 6 
\end{equation} 
for the bulges, and 
 
\begin{equation} 
1 \leq h \leq 6\ {\rm kpc} \qquad   
0.5 \leq q_{\rm d} \leq 0.9 
\end{equation} 
for the discs.  
The artificial galaxies also satisfied the following conditions 
 
\begin{equation} 
q_d \leq q_b \qquad   
11 \leq M_R \leq 16\ {\rm mag}. 
\end{equation} 
The simulated galaxies were assumed to be at a distance of 45 Mpc, 
which corresponds to a scale of 222 pc arcsec$^{-1}$. The adopted 
pixel scale, CCD gain, and read-out-noise were respectively 0.25 
arcsec pixel$^{-1}$, 2.52 e$^-$ ADU$^{-1}$, and 3.87 e$^-$ to mimic 
the instrumental setup of the photometric observations. Finally, a 
background level and photon noise were added to the artificial images 
to yield a signal-to-noise ratio similar to that of the observed ones. 
The images of artificial galaxies were analysed with GASP2D as if they
were real. The errors on the fitted parameters were estimated by
comparing the input $p_{\rm in}$ and measured $p_{\rm out}$
values. The artificial and observed galaxies were divided in bins of
0.5 magnitudes. The relative errors $1-p_{\rm in}/p_{\rm out}$ on the
parameters of the artificial galaxies were assumed to be normally
distributed. In each magnitude bin the mean and standard deviation of
relative errors of artificial galaxies were adopted as the systematic
and typical errors for the observed galaxies.
 
The structural parameters of ESO-LV~2060140 were obtained from
multi-band photometry by \citet{Beijersbergen1999}. They assumed the
bulge and disc to have both an exponential surface brightness radial
profile and a one-dimensional photometric decomposition was
applied. The $R-$band bulge ($\mu_{\rm 0,b}=20.19$ mag arcsec$^{-2}$,
$h_{\rm b}=1.62$ arcsec) and disc parameters ($\mu_{\rm 0,d}=21.57$
mag arcsec$^{-2}$, $h_{\rm b}=17.75$ arcsec) are in good agreement
with those we derived (Table \ref{tab:parameters_lsb}), although we
adopted for the bulge a S\'ersic profile with $n=1.64$ ($\mu_{\rm
0,b}=20.35$ mag arcsec$^{-2}$, $h_{\rm b}=0.52$ arcsec).
 
  In Table~\ref{tab:discs_lsb} we report the face-on $B-$band
central surface brightness of the discs. It was derived from $\mu_0$
(Tab.\ref{tab:parameters_lsb}) corrected for galaxy inclination
(Tab.~\ref{tab:sample_lsb}) and was computed by adopting the mean
$(B-R)$ color given by ESO-LV for radii larger than the galaxy
effective radius in order to minimize the contamination due to the
bulge light.

The six sample galaxies hosting a LSB disc were identified among the
10 originally selected galaxies, according to the structural
parameters derived from the 2D photometric decomposition
(Tab.~\ref{tab:discs_lsb}). 
All the bulges have a S\'ersic index $n\la2$, and few of them
(ESO-LV~2060140, ESO-LV~4000370) have an apparent flattening similar
to that of the disc (Tab.~\ref{tab:discs_lsb}). According to
\citet{Kormendy2004}, these disc-like features are the photometric
signatures of pseudobulges.

\placeFigthree
\placeTabseven

\section{Stellar and ionised-gas kinematics}   
\label{sec:kinematics}   
  
\subsection{Stellar kinematics and central velocity dispersion} 
 
The stellar kinematics was measured from the galaxy absorption 
features present in the wavelength range between 5050 and 5550 \AA , 
and centred on the Mg line triplet ($\lambda\lambda\,5164,5173,5184$ 
\AA) by applying the Fourier Correlation Quotient method 
\citep[FCQ]{Bender1990} as done in \citet{Bender1994}. 
The spectra were re-binned along the dispersion direction to a natural 
logarithmic scale, and along the spatial direction to obtain 
$S/N\geq20$ per resolution element. In a few spectra the $S/N$ 
decreases to 10 at the outermost radii. The galaxy continuum was 
removed row-by-row by fitting a fourth to sixth order polynomial as in 
\citet{Bender1994}. 
  The star SAO~137330 was adopted as kinematic template for
runs 1 and 2, and SAO~99192 for run 3.   These stars were selected
to have the narrower auto-correlation function (Table
\ref{tab:templog}) and to minimize the mismatch with the galaxy
spectra (Fig. \ref{fig:tempmism}).
 
For each galaxy spectrum the line-of-sight velocity distribution 
(LOSVD) was derived along the slit and its moments, namely the radial 
velocity $v_\star$, velocity dispersion $\sigma_\star$, and values of 
the coefficients $h_3$ and $h_4$ were measured. At each radius, they 
have been derived by fitting the LOSVD with a Gaussian plus third- and 
fourth-order Gauss-Hermite polynomials $H_3$ and $H_4$, which describe 
the asymmetric and symmetric deviations of the LOSVD from a pure 
Gaussian profile \citep{vdMarel1993,Gerhard1993}. 
The errors on the LOSVD moments were derived from photon statistics 
and CCD read-out noise, calibrating them by Monte Carlo simulations as 
done by \citet{Gerhard1998}.  
A large number of artificial spectra was generated for each measured 
set of LOSVD moments by first convolving the spectrum of the stellar 
template with the measured LOSVD and then adding different 
realizations of photon and read-out noises. The artificial spectra 
were analysed with FCQ as if they were real. The errors on the fitted 
parameters were estimated by comparing the input $p_{\rm in}$ and 
measured $p_{\rm out}$ values. The relative errors ($1-p_{\rm 
in}/p_{\rm out}$) were assumed to be normally distributed, with mean 
and standard deviation corresponding to the systematic and typical 
error on the relevant parameter, respectively. 
These errors do not take into account possible systematic effects due 
to template mismatch or the presence of dust and/or faint emission. 
 
\placeFigfour
  
The stellar velocity dispersion measured at large radii for many 
sample galaxies is comparable the instrumental velocity dispersion 
($\sigma_\star \simeq \sigma_{\rm inst} 
\approx 50$ \kms). The reliability of such measurements  
was assessed by means of a second series of Monte Carlo simulations. 
Artificial spectra were generated by convolving the spectrum of the 
stellar template with a LOSVD with $\sigma_\star = 20, 30, ..., 80, 
100, 120$ \kms, $h_3 = 0$ and $h_4 = 0, -0.1$, or $h_3 = 0, -0.1$ and 
$h_4 = 0$ and adopting different values of $S/N = 10, 20, ..., 100$. 
The artificial spectra were measured with FCQ. Finally, the values of 
$\Delta h_3 = h_{\rm 3, in}-h_{\rm 3, out}$, $\Delta h_4 = h_{\rm 4, 
in}-h_{\rm 4, out}$ were computed. 
The correlation of $|\Delta h_3|$ and $|\Delta h_4|$ with 
$\sigma_\star$ and $S/N$ is shown in Fig. \ref{fig:sigmaSNlimit}. The 
measured values of $h_3$ and $h_4$ were considered unreliable when 
$|\Delta h_3|>0.1$ and/or $|\Delta h_4|>0.1$. For this range of 
$\sigma_\star$ and $S/N$, only $v_\star$ and $\sigma_\star$ were 
measured and it was assumed $h_3=h_4=0$. 
  For three of our targets, namelly ESO-LV 2060140, ESO-LV
4000370 and ESO-LV 4880490, $h_3$ abd $h_4$ parameters were not
measured due to the low signal-to-noise ratio.  
Moreover, the stellar velocity dispersion was systematically 
underestimated for $\sigma_\star\leq30$ \kms . This was found by 
measuring $\Delta\sigma_\star = \sigma_{\rm \star, in}-\sigma_{\rm 
\star, out}$ in artificial spectra with $|\Delta h_3|\leq0.1$  
and $|\Delta h_4|\leq0.1$ (Fig. \ref{fig:sigmatest}).  
Such a small value of $\sigma_\star$ is far below the velocity
dispersions measured for the sample galaxies indicating that
these measurements are genuine and are not strong affected by
instrumental broadening. This is not the case of the small velocity
dispersions ($\simeq25$ \kms ) measured for the faint foreground stars
($S/N\simeq20$) which happened to fall into the slit when some of the
galaxy spectra were obtained. 
These results are consistent with previous findings
by \citet{Joseph2001,Pinkney2003}.

The measured stellar kinematics of all the sample galaxies are 
reported in Table \ref{tab:starkin_lsb} and plotted in 
Fig. \ref{fig:kinematics_lsb}. The central velocity dispersion of the 
stellar component $\sigma_{\rm c}$ was derived by extrapolating the 
velocity dispersion radial profile to $r=0$ arcsec. This was done by 
fitting the eight innermost data points with an empirical 
function. The best-fitting function was selected to be either 
exponential, or Gaussian, or constant. No aperture correction was 
applied to $\sigma_{\rm c}$, as discussed by 
\citet{Baes2003} and \citet{Pizzella2004}. The derived values  
for $\sigma_{\rm c}$ of our sample galaxies are given 
Table \ref{tab:sample_lsb}. 

 
\placeFigfive
\placeTabeight
 
\subsection{Ionised-gas kinematics, circular velocity, and minor-axis rotation} 
 
The ionised-gas kinematics was measured by the simultaneous Gaussian
fit of the \hb\ and \oiii$\lambda\lambda 4959,5007$ emission lines
present in the galaxy spectra. The galaxy continuum was removed from
the spectra as done for measuring the stellar kinematics. A Gaussian
was fitted to each emission line, assuming they have the same
line-of-sight velocity $v_{\rm g}$, and velocity dispersion
$\sigma_{\rm g}$. An additional absorption Gaussian was considered in
the fitting function to take into account the presence of the \hb\
absorption line.  Far from the galaxy centre, adjacent spectral rows
were average to increase the signal-to-noise ratio ($S/N>10$) of
both the emission lines.
The fitting routine developed by \citet{Pizzella2004} is based on the 
MPFIT algorithm to perform the non-linear least-squares minimization 
in the IDL environment.  
 
\placeFigsix
\placeFigseven
We checked for some galaxies that the error in the kinematic parameter 
determination derived by Monte Carlo simulations did not differ 
significantly from the formal errors given as output by the 
least-squares fitting routine. We therefore decided to assume the 
latter as the errors on the gas kinematics. For each galaxy we derived 
the heliocentric systemic velocity, $V_\odot$, as the velocity of the 
centre of symmetry of the ionised-gas rotation curve (Table 
\ref{tab:sample_lsb}). 
The measured ionised-gas kinematics of all the sample galaxies are 
reported in Table \ref{tab:gaskin_lsb} and plotted in 
Fig. \ref{fig:kinematics_lsb}. 
The ionised-gas kinematics along the major axis of some of the sample 
galaxies have been measured by other authors. A comparison with these 
data sets was performed to assess the accuracy and reliability of our 
measurements (Fig. \ref{fig:kincomparison}). 
The \ha\ rotation curves obtained for ESO-LV~2060140 
\citep[Fig. \ref{fig:kincomparison}a]{McGaugh2001}, ESO-LV~2340130 
\citep[Fig. \ref{fig:kincomparison}b]{Mathewson1992}, ESO-LV~4880490 
\citep[Fig. \ref{fig:kincomparison}c]{McGaugh2001}, and ESO-LV~5140100 
\citep[see App.~\ref{sec:hsb},][Fig. \ref{fig:kincomparison}b]{Mathewson1992},
are in agreement within the errors with those measured by us after
applying a small offset in their systemic velocities ($\Delta V\leq
30$ \kms). As a general comment, we can notice that our data are
characterized by a finer spatial sampling. This allows to reveal
kinematic details that were not evident in previous observations. The
extension of the kinematic data are comparable. It has to be noted
that our measurements are based on the \hb\ and \oiii\ emission lines,
while the data available in the literature are based on the \ha\
emission line.

For all the sample galaxies, the ionised-gas rotation curve measured
along the major axis (or along diagonal axis which was closer to the
major axis) was obtained by folding the observed line-of-sight
velocities around the galaxy centre and systemic velocity after
averaging the contiguous data points. The curve was deprojected
by taking into account the inclination of the disc (Table
\ref{tab:sample_lsb}) and the misalignment between the observed axis
and the position angle of the galaxy major axis (Table 2).
The galaxy circular velocity $V_{\rm c}$ was derived by averaging the
outermost values of the flat portion of deprojected rotation curve
(see \citet{Pizzella2005} for a detailed descrption). Their $V_{\rm
c}$ are listed in Table \ref{tab:sample_lsb}.
For five out of six sample galaxies the ionised-gas kinematics
were measured either along the minor axis (ESO-LV~2060140,
ESO-LV~4880490, ESO-LV~5340200) or along a diagonal axis close to the
minor one (by $30^\circ$ and $16^\circ$ for ESO-LV~1860550 and
ESO-LV~2340130, respectively).
The corresponding rotation curves are either irregular and asymmetric
(ESO-LV~1860550, ESO-LV~2060140, ESO-LV~4880490) or characterized by a
well-defined velocity gradient (ESO-LV~2340130, ESO-LV~5340200).
Significant velocity values are unexpected along the galaxy minor axis
if the gas is moving in circular orbits in a disc coplanar to the
stellar one. The asymmetry of rotation curve of ESO-LV~1860550 ($\rm
PA = 165^\circ$) and the reversal of the velocities measured for
ESO-LV~2340130 ($\rm PA = 98^\circ$) can not be attributed to the
small misalignment ($\rm \Delta PA = 30^\circ$ and $16^\circ$ )
between the slit and the galaxy minor axis. They are suggestive of a
more complex gas distribution.
As a consequence, the ionised-gas velocity measured along the major
axis is not tracing the circular velocity in the inner regions of the
galaxy. To properly address this issue it is crucial to map the entire
velocity field of gas. This is the case of ESO-LV~1860550,
ESO-LV~4000370, and ESO-LV~5340200. The 2D velocity field of their
gaseous component has been recently measured by
\citet{Pizzella2008}. They found the presence of non-ordered gas
motions in ESO-LV~1860550, of a kinematically-decoupled component in
ESO-LV~4000370, and of a misalignement between the photometric and
kinematic major axis of ESO-LV~5340200.

\placeFigeight

\section{Summary and Conclusions}  
\label{sec:conclusions} 
  
We have presented and investigated the structural parameters and
kinematic properties of 6 galaxies with a LSB disc, for which we
obtained deep photometric and spectroscopic VLT observations.
Initially chosen following selection criteria similar to those of
\citet{Beijersbergen1999} and using published photometric profiles, we
have subsequently confirmed the presence of a LSB disc in our sample
galaxies by carrying out a careful 2D photometric bulge-disc
decomposition.
  Our sample galaxies differs as compared to typical LSB
galaxy samples for the possible prsence of a significant bulge component  

Our photometric analysis further revealed that our LSB galaxies have
bulges that can are well described by a S\'ersic profile with an index
$n\la2$, with few of them displaying an apparent flattening rather
similar to that of their surrounding disc. According to
\citet{Kormendy2004}, these disc-like features are the photometric
signature of a pseudobulge.
 
We obtained spectroscopic observations along the major and minor axis
in half of our sample galaxies (ESO-LV~2060140, ESO-LV~4880490 and
ESO-LV~5340200,), while we observed the other half along two diagonal
axes. Spectra along two diagonal directions were obtained also for
ESO-LV~5340200.
In all targets, the depth of our data allowed us to measure the
stellar and ionised-gas kinematics out to a surface-brightness level
$\mu_R\approx24$ mag$\,$arcsec$^{-2}$, comfortably reaching the outer
disc regions where the rotation curve flattens.
 
We derived central values for the stellar velocity dispersion of
$\sigma_c$ and measured the circular velocity in the outskirts of our
sample galaxy $V_{\rm c}$ by following the flat portion of the gas
rotation curve.
\citet{Ferrarese2002} found a correlation between $\sigma_c$ and
$V_{\rm c}$ and interpreted it as evidence for a connection between
the mass of the central supermassive black hole and that of the DM
halo \citep[see also][]{Baes2003, Buyle2006}.
Using the $\sigma_c$ and $V_{\rm c}$ measurements presented in this
paper, \citet{Pizzella2005} found that galaxies hosting LSB discs
follow a different $V_{\rm c}-\sigma_{\rm c}$ relation with respect to
their HSB counterparts. In fact, they show either higher $V_{\rm c}$
or lower $\sigma_{\rm c}$. This behaviour has been recently explained
in terms of morphology or, equivalently, total light concentration by
\citet{Courteau2007} and \citet{Ho2007}.

We have shown that the vast majority of our sample galaxies show
non-zero gas velocities along or close to their minor axes.
Significant velocities are unexpected along the galaxy minor axis if
the gas is moving in circular orbits in a disc coplanar to the stellar
one. They can be explained as due to non-ordered motions or by
invoking a warp and/or non-circular motions of the gaseous
component. As a consequence, the ionised-gas velocity measured along
the major axis is not tracing the circular velocity in the inner
regions of the galaxy.
Deviations from perfect circular gas motions are by no means limited
to the minor axis direction in our sample galaxies. Along the major
axis of our targets (or the closest diagonal direction) the
ionised-gas velocity curve is characterized by an higher degree of
asymmetry and irregularity with respect to the stellar one. As a
consequence, to derive the gas rotation curve it is often necessary to
smooth the data over large radial bins before folding the gas
velocities around the centre of symmetry. This is not the case for the
stars, since their   measured velocities   
fold quite well and do not show
sharp wiggles and bumps.
The rotation curves of the ionised gas and stars are plotted in
Fig. \ref{fig:fold_lsb} for all the sample galaxies. Data were folded
around the centre of symmetry and deprojected by taking into account
the inclination of the disc (Table \ref{tab:sample_lsb}) and the
misalignment between the observed axis and the position angle of the
galaxy major axis (Table 2). 
Globally the gas rotation curves of the sample galaxies are more
asymmetric, irregular, and affected by scatter than the corresponding
stellar ones.
The presence of a more disturbed gas motions and a regular stellar
kinematics is particularly clear in ESO-LV~1860550 and ESO-LV~2340130.
Nevertheless, locally both gas and stellar rotation can both appear
pretty smooth, in particular toward the centre (e.g., ESO-LV~2340130
and ESO-LV~5340200). On the other hand, the most obvious scatter of
the gas data points at larger radii is instrinsic and neither due to
low $S/N$ nor to the underlying \hb\ asborption line, since this is
observed only at small radii (where the kinematics is constrained also
by \oiii\ emission doublet).

While it is easy to spot non-circular, off-plane, and non-ordered
velocities along the minor axis, their presence along major axis may
remain undetected. This can affect the correct estimate of the central
velocity gradient, posing severe limitations to the use of the ionised
gas as the tracer of the circular velocity to derive the mass content
and distribution in the central region of galaxies. To address this
issue, it is crucial to measure the kinematics along different axes
and ideally to have integral-field data.
The 2D velocity field of the ionised gas in LSB galaxies has been
mapped only recently by means of integral-field spectroscopy
\citep{Kuzio2006, Coccato2008, Pizzella2008}.
\citet{Pizzella2008} reported the detection in ESO-LV~1890070 and
ESO-LV~4000370 of strong non-ordered motions of the ionised gas, which
prevent to disentangle between core and cuspy radial profiles of the
DM mass density.\\

The main conclusion of this work is that the stellar kinematics of
galaxies with LSB discs is less affected than the ionised-gas
kinematics one by large-scale asymmetries and small-scale
irregularities. The gas velocities have been found to deviate from the
from the circular velocity in particular and more often in the inner
regions of our sample galaxies. Unless the regularity of the
ionised-gas motions can be establish with integral-field data, our
findings strongly suggest that stars are more suitable to trace the mass
distribution of galaxies with a LSB disc, in particular when the inner
slope of the dark matter halo is the central question that one wishes
to answer.

Building a complete dynamical model to address the question of the
mass distribution of our sample galaxies is beyond the scope of this
paper, and will be presented in a companion paper
(Magorrian et al, in preparation)

\section*{Acknowledgments}  
 
\noindent 
This work was made possible through grants PRIN 2005/32 by Istituto 
Nazionale di Astrofisica (INAF) and CPDA068415/06 by Padua 
University. JMA receives financial support by INAF, LM acknowledges 
financial support from grant CPDR061795/06 by Padua University.  EMC 
acknowledges the University of Hertfordshire and University of Oxford 
for hospitality while this paper was in progress. 
 
\bibliographystyle{mn2e} 

\appendix
\section{HSB galaxies}
\label{sec:hsb}

\placeTabAone

ESO-LV~1890070, ESO-LV~4460170, ESO-LV~4500200, and ESO-LV~5140100
(Tab.~\ref{tab:sample_hsb}) were originally selected according to the
criteria given in \S~\ref{sec:sample}.
Their photometric (Tab.~\ref{tab:imalog_hsb}) and spectroscopic
(Tab.~\ref{tab:speclog_hsb}) observations were carried out and data
analysed together with the sample galaxies, as discussed in the paper.

In spite of the selection criteria, ESO-LV~1890070, ESO-LV~4460170,
ESO-LV~4500200, and ESO-LV~5140100 do not host a LSB disc, according
to the structural parameters derived from the 2D photometric
decomposition (Tab.~\ref{tab:parameters_hsb}).
The central surface brightness of the discs derived from ESO-LV
photometry was overestimated with respect to that from the available
CCD images (Tab.~\ref{tab:discs_hsb}). This is due to the limited
radial extension of ESO-LV data and to the shallow surface-brightness
radial profile observed at the largest measured radii. The flattening
of the light profile is actually due to the presence of a bar in
ESO-LV~1890070, ESO-LV~4460170, and ESO-LV~4500200 (as
found by subtracting the photometric model from the observed image of
the galaxy), and by the downbending of the disc radial profile in
ESO-LV~5140100 (Fig.~\ref{fig:decomposition_hsb}).

The stellar and ionised-gas kinematics of the HSB galaxies is plotted
in Fig.~\ref{fig:kinematics_hsb} and given in Tables
\ref{tab:starkin_hsb} and \ref{tab:gaskin_hsb}. These data allowed to 
derive the bulge velocity dispersion $\sigma_{rm c}$ and disc circular
velocity $V_{\rm c}$ (Tab.~\ref{tab:sample_hsb}), which were used by
\citet{Pizzella2005} to study the $V_{\rm c}-\sigma_{\rm c}$ relationship 
in HSB and LSB galaxies. The rotation curves of the 
ionised gas and stars are plotted in Fig.~\ref{fig:fold_hsb}. 

All the galaxies show non-zero velocities for the ionised
gas and stars along the minor axes.
The minor-axis velocity gradients measured for ESO-LV~1890070,
ESO-LV~4460170, and ESO-LV~4500200 can be explained as due to the
non-circular motions induced by the bar.
Moreover ESO-LV~1890070 and ESO-LV~5140100 host a dynamically cold
component in their nucleus.
The $\sigma_\star$ radial profile of ESO-LV~1890070 along both the 
major and minor axis are characterized by a central minimum. A similar 
phenomenon was observed by \citet{Emsellem2001} in a number of barred 
galaxies and interpreted as the signature of a nuclear stellar disc or 
bar.  The presence of such a fast-rotating nuclear component in 
ESO-LV~1890070 is also supported by the sharp stellar velocity 
gradient ($\sim 300\,$\kmskpc ) measured along the galaxy major axis. 
A central drop was observed in the $\sigma_\star$ radial profile of 
ESO-LV~5140100 too. But this is an unbarred galaxy according to the 
results of the photometric decomposition. Therefore, the lower central 
value of $\sigma_\star$ has to be attributed to the small, 
low-luminosity, and exponential bulge of the galaxy.


\placeTabAthree
\placeTabAtwo

\placeFigAone
\placeTabAfour 
\placeTabAfive

\placeTabAsix
\placeTabAseven

\placeFigAtwo
\placeFigAthree

\bsp

\label{lastpage}  
 
\end{document}